\documentclass[aps,prd,twocolumn,amssymb,eqsecnum,showpacs,showkeyes,secnumarabic,graphics,floatfix,nofootinbib,tightenlines,longbibliography,superscriptaddress,numbers,sort&compress]{revtex4-1}

\usepackage{graphicx}
\usepackage{epsfig} 
\usepackage{bm}
\usepackage{cancel}  
\usepackage{dcolumn}  
\usepackage{amsmath} 
\usepackage{amssymb}
\usepackage{hhline} 
\usepackage[dvipsnames]{xcolor}
\usepackage[utf8]{inputenc}
\usepackage{cancel}
\usepackage[dvipsnames]{xcolor}  
\usepackage{float}
\usepackage{rotating}

\usepackage[breaklinks=true,colorlinks=true,
linkcolor=blue,urlcolor=blue,citecolor=MidnightBlue,
bookmarks=true,bookmarksopenlevel=2]{hyperref}
\usepackage{mathrsfs} 
\usepackage{subcaption}
\usepackage[titletoc]{appendix}
\usepackage{physics}
\usepackage{ragged2e}
\usepackage{hyperref}
\usepackage{changepage}
\usepackage{tikz}
\usepackage{natbib}

\newlength{\extralength}
\renewcommand{\theequation}{\arabic{equation}}

\def\apj{{Astroph.\@ J.\ }}

\def\aj{{Astron.\@ J.\ }}

\def \beq  {\begin{equation}}
\def \eeq  {\end{equation}}
\def \ber  {\begin{eqnarray}}
\def \eer  {\end{eqnarray}}
\usepackage{titlesec}
\titleformat{\paragraph}[runin]
{\normalfont\normalsize\itshape}{\theparagraph}{1em}{}
\titlespacing*{\paragraph} {0pt}{0pt}{0pt}

\begin{document}

\renewcommand\thesection{\Roman{section}} 
\numberwithin{equation}{section}         
\renewcommand\theequation{\thesection.\arabic{equation}} 

\newcommand{\newc}{\newcommand}

\newc{\be}{\begin{equation}}
\newc{\ee}{\end{equation}}
\newc{\ba}{\begin{eqnarray}} 
\newc{\ea}{\end{eqnarray}}
\newc{\bea}{\begin{eqnarray*}}
\newc{\eea}{\end{eqnarray*}}
\newc{\D}{\partial}
\newc{\ie}{{\it i.e.} }
\newc{\eg}{{\it e.g.} }
\newc{\etc}{{\it etc.} }
\newc{\etal}{{\it et al.}}
\newc{\lcdm}{$\Lambda$CDM }
\newc{\lcdmnospace}{$\Lambda$CDM}
\newcommand{\nn}{\nonumber}
\newc{\ra}{\Rightarrow}
\newc{\omm}{$\Omega_{m}$ }
\newc{\ommnospace}{$\Omega_{m}$}
\newc{\fs}{$f\sigma_8$ }
\newc{\fsz}{$f\sigma_8(z)$ }
\newc{\fsnospace}{$f\sigma_8(z)$}
\newc{\plcdm}{Planck/$\Lambda$CDM }
\newc{\plcdmnospace}{Planck15/$\Lambda$CDM}
\newc{\wlcdm}{WMAP7/$\Lambda$CDM }
\newc{\wlcdmnospace}{WMAP7/$\Lambda$CDM}
\newcommand{\fss}{{\rm{\it f\sigma}}_8}
\newcommand{\LP}[1]{\textcolor{red}{[{\bf LP}: #1]}}

\title{Metastable Cosmological Constant and Gravitational Bubbles: Ultra-Late-Time Transitions in Modified Gravity}
\author{Dimitrios Efstratiou}\email{d.efstratiou@uoi.gr
}
\affiliation{Department of Physics, University of Ioannina, GR-45110, Ioannina, Greece}
\author{Leandros Perivolaropoulos}\email{leandros@uoi.gr}
\affiliation{Department of Physics, University of Ioannina, GR-45110, Ioannina, Greece}

\date {\today}  

\begin{abstract}
The observed cosmological constant may originate as the minimum value  $U_{min}$ of a scalar field potential, where the scalar field is frozen due to a large mass $m$. If this vacuum is metastable, it may decay to a true vacuum either at present or in the future. Assuming its decay rate $\Gamma$ is comparable to the Hubble expansion rate $H_0$, we estimate the scale of true vacuum bubbles and analyze their evolution. We find that their initial formation scale is sub-millimeter, and their tension causes rapid collapse if $m \gtrsim 1.7 \cdot 10^{-3}\, eV$. For smaller masses, the bubbles expand at the speed of light. We extend our analysis to scalar-tensor theories with non-minimal coupling, finding that the nucleation scale of gravitational constant bubbles remains consistent with the sub-millimeter regime of General Relativity (GR). The critical mass scale remains around $10^{-3}\,eV$, similar to the minimally coupled case. A theoretical estimate at redshift $z_{obs} \sim 0.01$ suggests an observable bubble radius of $\sim 50$ Mpc, implying a gravitational transition triggered $\sim 300$ Myr ago, with a present-day size approaching $100$ Mpc. Additionally, we explore mass ranges ($m < 10^{-3}\,eV$) and coupling parameter $\xi$ ranges ($10^{-8}\,eV^{2-n} - 10^{-1}\,eV^{2-n}$) that lead to a variation in the gravitational constant $\Delta G/G_N$ within the $1\%-7\%$ range. This analysis is conducted in the framework of non-minimal coupling theories, where the coupling function takes the form $F(\phi)=1/\kappa - \xi \phi^n$, with $\kappa=8\pi G_N$ and $2 \leq n \leq 9$. Finally, we review various local physics or/and transition based proposed solutions to the Hubble tension, including ultra-late-time transitional models ($z \sim 0.01$), screened fifth-force mechanisms, and the $\Lambda_{\rm s}$CDM model, which features a transition at $z \sim 2$. We discuss observational hints supporting these scenarios and the theoretical challenges they face.

\end{abstract}
\maketitle 

\section{Introduction}\label{sec1}

The $\Lambda$ Cold Dark Matter ($\Lambda$CDM) model has emerged as the prevailing cosmological model due to its remarkable successes in explaining a wide range of observations. One of the key strengths of the \lcdm model is its simplicity. It is based on the assumption that the Universe is composed of cold dark matter particles (CDM), which provide the gravitational scaffolding for the formation of large-scale structures  which is responsible for structure formation \cite{BERTONE2005279,1981AJ.....86.1825B,1970ApJ...160..811F,1980ApJ...238..471R,1970ApJ...159..379R,Zwicky:1933gu,1937ApJ....86..217Z} and a cosmological constant $\Lambda$ \cite{Carroll_2001}\cite{doi:10.1146/annurev.aa.30.090192.002435}, associated with vacuum energy that drives the accelerated expansion of the Universe. This straightforward framework allows for a concise and elegant description of the Universe's evolution.

The \lcdm model also exhibits exceptional consistency with numerous observations. It successfully accounts for cosmic microwave background (CMB) radiation \cite{Page:2003fa}, the large-scale distribution of galaxies, and the clustering patterns of matter in the Universe \cite{Bernardeau:2001qr}\cite{Bull:2015stt}. Additionally, it accurately predicts the abundances of light elements produced during the early stages of the Universe, known as Big Bang nucleosynthesis (BBN) \cite{Cyburt:2015mya, Iocco:2008va,Schramm:1997vs,Steigman:2007xt}. Moreover, the model's predictions align with the observations of the accelerating expansion of the universe \cite{Perlmutter_1999}\cite{Riess:1998cb}, which were awarded the Nobel Prize in Physics in 2011.

 Despite its exceptional effectiveness in explaining cosmology in simple words, the validity of \lcdm has recently been subjected to scrutiny \cite{Abdalla_2022,Anchordoqui:2021gji,Buchert:2015wwr,DiValentino:2021izs,Schmitz:2022hsz,Schoneberg:2021qvd}. All this ``turmoil'' in the Cosmology society and the questioning of the sovereignty of the concordance cosmological model is motivated by a variety of theoretical problems and observational difficulties of \lcdm.

The most crucial theoretical issues facing $\Lambda$CDM are the fine-tuning \cite{Burgess:2013ara,Martin:2012bt,RevModPhys.61.1} and coincidence problems \cite{Velten:2014nra}\cite{fitch}. The fine-tuning or cosmological constant $\Lambda$ problem is connected to the fact that there is a sizable gap between observations and the predicted from the theory values of the cosmological constant which reach at least 60 orders of magnitude \cite{Martin:2012bt}\cite{Copeland:2006wr,Sola:2013gha,Weinberg:2008zzc}. The nature of the second problem is related to the coincidental, approximately equal values of observed energy densities $\Omega_\Lambda$ and $\Omega_m$ nowadays despite their totally different evolution properties.

Apart from the theoretical problems, there are some observational issues that appear in astrophysical and cosmological data. More specifically there are signals that appear to be in some tension with the standard model as specified by the Planck18 parameter values \cite{Planck:2018nkj}\cite{Planck:2018vyg}. This tension could be equal to $2\sigma$ or larger.

Two important tensions are the growth tension ($S_8$) ($2 - 3\sigma$) and the Hubble tension (5$\sigma$). In the first, direct measurements of the growth rate of cosmological perturbations cluster counts (CC)~\cite{Rozo:2009jj,Rapetti:2008rm,SPT:2014wkb,Ruiz:2014hma}, weak lensing (WL)~\cite{Schmidt:2008hc,kids1,Hildebrandt:2016iqg,Joudaki:2017zdt,DES:2017myr,DES:2017qwj,DES:2018ufa,Kohlinger:2017sxk} and redshift-space distortions (RSD)~\cite{2013MNRAS.429.1514S,Macaulay:2013swa,Johnson:2015aaa,Nesseris:2017vor,Kazantzidis:2018rnb} indicate a lower growth rate than that indicated by Planck-\lcdm (lower matter density) \cite{LIGOScientific:2018mvr, PhysRevD.96.063517,Joudaki:2017zdt}. In the second, the combined local direct measurements of the Hubble constant are in $5\sigma$ tension with indirect measurements of $H_0$ with CMB. This tension can be larger if combinations of local measurements are used \cite{DiValentino:2021izs}\cite{Wong:2019kwg}. The best-fit value given by the Planck/\lcdm is $H_0= 67.4 \pm 0.5\, km \, s^{-1}\, Mpc^{-1}$ \cite{Planck:2018vyg}. The local measurements, using Cepheid calibrators by the SH0ES Team, give a higher value at $H_0=73.04 \pm 1.04 \pm 0.5\, km \, s^{-1}\, Mpc^{-1}$. For a more analytical review of the standard model tensions, you can check \cite{Perivolaropoulos:2021jda}.

Many different solutions have been put out in an effort to address the Cosmological Constant Problem (fine-tuning) and the Cosmic Coincidence Problem. One of the potential answers is that a scalar field could fill the role of dark energy. The quintessence models \cite{Jennings_2010,Ferreira:1997hj,PhysRevD.37.3406,Wetterich:1987fm} are included in this specific category of models. A quintessence model is a scalar field minimally coupled to gravity. A great advantage of these models is the possible connection with inflation (quintessential inflation) an incredibly brief period of time during which the early Universe was rapidly expanding \cite{deHaro:2021swo}. Despite its advantages, there appears to be a major drawback. A small scalar field mass is required to describe the acceleration of the Universe ($m_\phi \lesssim H_0 \approx 10^{-33}\, eV$). In general, it is challenging to explain how such an ultralight mass may fit into the energy scales found in particle physics \cite{Carroll:1998zi}.

 In a special scenario of quintessence on the phenomenological front, the Universe is currently expanding due to a long-lasting (quintessence) false vacuum state, which may be a holdover from an earlier inflationary period \cite{Peebles:1987ek}\cite{Carvalho:2006fy}. In this regard, Landim and Abdalla have put out an intriguing scenario called metastable dark energy (MDE) \cite{Landim:2016isc}. The quantum tunneling from an unstable false to a stable true vacuum state is the basis of the model. As opposed to early times, as it occurs in the old inflationary model first proposed by Guth \cite{Guth:1980zm}\cite{Guth:1982pn}, the metastable state decay process occurs in the current low energy Universe. Due to the low temperature of the vacuum-matter phase, this transition from false to true vacuum can be discussed in the context of the semi-classical approach developed by Coleman and his collaborators \cite{Coleman:1977py}\cite{Callan:1977pt}. Interestingly, an extension of the MDE model has been explored in which the metastable vacuum undergoes a transition into dark matter, potentially producing axion-like particles as a viable dark matter candidate \cite{deSouza:2024sfl}. However, it has been shown that this process would not lead to a complete transition to a dark matter-dominated phase, even in the distant future.
 
 Some advantages of the MDE model is its consistency with most data as it behaves like \lcdm and its connection with inflation by assuming that the present false vacuum energy density is a leftover from an early inflationary epoch. Also, it is motivated by the string landscape  \cite{Lima:2020nwt}. Moreover, such a model has the potential to resolve the Hubble tension (an ultra-late transition event has been suggested as a solution) \cite{Perivolaropoulos:2021bds}. For a comprehensive review of various dark energy models and modified gravity theories, including \( f(R) \) gravity, scalar field theories, and holographic dark energy, see \cite{Bamba:2012cp}.

The expected scales for such an event are the following: 1. The energy scale of the false vacuum $\epsilon$ should represent the value of the observed cosmological constant ($\Lambda$) 2. The vacuum lifetime of such an ultra-late phase transition must be comparable to the age of the Universe ($t_d=H_0^{-1}$), for one bubble nucleation event the decay rate per four-volume must be set equal to $H_0^{-4}$  \cite{PhysRevD.46.2384}\cite{Krauss:2013vya}.

The sound horizon scale \(r_s\) at recombination (standard ruler) and the standardized bolometric absolute magnitude \(M_B\) of Type Ia supernovae (SnIa standard candles) have been used as probes for the measurement of cosmological distances and thus for the measurement of the Hubble constant \(H_0\), the most fundamental parameter of cosmology. The best-fit values of \(H_0\) obtained using the two distance calibrators are at 5$\sigma$ discrepancy with each other, as we have already referred to.

Each type of measurement makes specific assumptions whose possible violation would lead to significant systematic errors that would invalidate the accuracy of the corresponding measurement. Measurements based on SnIa standard candles assume the validity of the distance ladder approach and in particular that physical laws and environmental effects around calibrated SnIa are the same in all the three rungs of the distance ladder. Measurements based on the sound horizon scale used as a standard ruler assume that the physical laws before recombination are consistent with the standard cosmological model and that the expansion history of the universe \(H(z)\) is consistent with the standard $\Lambda$CDM model and its parameters determined by the Planck18 \citep{Planck:2018vyg}.

In accordance with the above three types of assumptions there are three classes of models for the resolution of the Hubble tension:
\begin{enumerate}

\item \textbf{Early time models}: These models assume physics beyond the standard model to decrease the sound horizon scale at recombination
\begin{equation}
r_s = \int_{z_\text{rec}}^\infty \frac{dz\, c_s(z)}{H(z; \Omega_{0b}h^2, \Omega_{0\gamma}h^2, \Omega_{0CDM}h^2)}
\end{equation}
to a lower value induced by Early Dark Energy (EDE), modified gravity or dark radiation\citep{Poulin:2018cxd, Kamionkowski:2022pkx, Simon:2022adh, Braglia:2020bym, Niedermann:2020dwg, Smith:2020rxx, Braglia:2020auw, Brax:2013fna, Adi:2020qqf, Clifton:2011jh, Lin:2018nxe, DiValentino:2015bja, Rossi:2019lgt, Braglia:2020iik, Abellan:2023tbi, Seto:2021xua, Sakstein:2019fmf, Vagnozzi:2020zrh,Ballardini:2023mzm}. One notable proposal is the AdS-EDE model \citep{Ye:2020btb}, which introduces an Anti-de Sitter (AdS) phase in EDE around recombination. This mechanism has been shown to further alleviate the Hubble tension by lifting the inferred $H_0$ value while maintaining consistency with CMB, BAO, and SNIa constraints. Thus, these models exploit the degeneracy of \(r_s\) with \(H_0\) in the context of the observable angular scale of the CMB sound horizon \(r_s\)
\begin{equation}\label{thetasigma}
\theta_s = \frac{r_s H_0}{\int_0^{z_\text{rec}} \frac{dz}{E(z)}}
\end{equation}
to lead to a higher value of \(H_0\) while respecting the best fit Planck18/$\Lambda$CDM form of \(E(z) \equiv H(z)/H_0\) between recombination and present time.

The main problem of this class is that despite fine-tuning of theoretical models that support them, they are only able to decrease the statistical significance of the Hubble tension and they cannot fully eliminate it. In addition, they favor a higher value of the matter density parameter \(\Omega_{0m}\) thus worsening the \(S_8\) tension \citep{Alestas:2021nmi, Jedamzik:2020krr, Vagnozzi:2023qvp, Vagnozzi:2021gjh}. However, it has been shown that the increase in $\omega_m = \Omega_{0m} h^2$ in these models is a consequence of the CMB+BAO background compatibility, while $\Omega_{0m}$ itself does not shift significantly compared to $\Lambda$CDM \citep{Ye:2020oix}.

\item \textbf{Late time models (H(z) deformation)}: These models assume that there is deformation of the Hubble expansion history \(H(z)\) with respect to the Planck18/$\Lambda$CDM prediction at late cosmological times (\(z \lesssim 2\))\citep{Li:2019yem, Pan:2019hac, Panpanich:2019fxq, DiValentino:2019ffd, DiValentino:2019jae, Li:2019san, Clark:2020miy, DiValentino:2020naf, Alestas:2021luu, Alestas:2020mvb, Press:2007}. This deformation shifts the value of \(H_0\) measured by the sound horizon scale standard ruler to a value consistent with the distance ladder measurement by effectively changing the denominator of Eq \eqref{thetasigma}. However, as discussed in the present analysis even this model has difficulty to fit simultaneously BAO and SnIa data.

The main problem of this class of models is that \(E(z) \equiv H(z)/H_0\) deformations are highly constrained by BAO and SnIa data in redshifts larger than \(z \simeq 0.1\) \citep{Alestas:2021nmi, Keeley:2022ojz, Chen:2020vsk, Anchordoqui:2021gji, Mau:2022gsl, Cai:2022dkh, Heisenberg:2022gqk, Vagnozzi:2021tjv, Davari:2022qpf, Gomez-Valent:2023bku}. Thus, this class of models has been put in disfavor during the last few years. The exception constitutes the $\Lambda_{\rm s}$CDM model which shifts the deformation to high redshifts (\(z \simeq 2\)) and makes it abrupt so that it has a minimal effect on late data while maintaining consistency with the distance to the CMB sound horizon at last scattering \citep{Akarsu:2023jzv, Akarsu:2021fol}.

An alternative approach to modifying the late-time expansion history is through extensions of Horndeski gravity that include the Galileon term \(G_3\). Such models introduce additional interactions in the scalar field Lagrangian, which can alter the cosmic evolution at both early and late times. Specifically, certain models with non-standard kinetic terms—while maintaining no-ghost conditions—modify the expansion history significantly at \(z \lesssim 2\), making them a relevant candidate for addressing late-time cosmological tensions \cite{Ferrari:2023qnh}.

\item \textbf{Ultralate time models}: These models \citep{Marra:2021fvf, Alestas:2020zol, Perivolaropoulos:2021bzz} assume that there is a physics or environmental change between the second and third rungs of the distance ladder i.e., that the Cepheid calibrated SnIa are not the same as the Hubble flow SnIa either due to a change of the physical laws or due to some change in their structure or environment (dust, metallicity etc). Thus, \(M_B\) in the third rung of the distance ladder (Hubble flow rung \(z \in [0.01, 0.1]\)) is lower than the value measured in the second rung. Thus, these models explore the degeneracy between \(H_0\) and \(M_B\) in the context of the Hubble flow observable
\begin{equation}
\mathcal{M} \equiv M_B + 5\log \left( \frac{c/H_0}{Mpc} \right) + 25
\end{equation}
to lead to a lower \(H_0\) measurement in the context of the distance ladder measurements.

The main problem of this class of models is fine-tuning. There is no current clear theoretical motivation for the assumed physics transition at such ultralate redshifts (\(z \sim 0.01\)) even though it is straightforward to construct such models using a degree of fine-tuning \citep{Perivolaropoulos:2022khd}. On the other hand, the physics transitions hypothesis is easily testable using a wide redshift data and in fact there are some hints for such an effect in both Cepheid and Tully-Fisser data \citep{Perivolaropoulos:2022vql, Alestas:2021xes}.
\end{enumerate}

In this paper we focus on ultra-late time transitions in the framework of scalar-tensor theories of gravity \cite{Brans:1961sx, Jordan:1959eg,Damour:1993id,Xie:2007gq, EspositoFarese:2000ij}. Scalar-tensor theories of gravity are a class of theories that generalize Einstein's theory of General Relativity (GR) by introducing a scalar field that couples to gravity. These theories are motivated by various considerations, such as attempts to unify gravity with other fundamental forces, to explain dark energy and cosmic inflation, and to address the shortcomings of GR at different scales.

In scalar-tensor theories, the gravitational interaction is mediated by both a tensor field (the metric \( g_{\mu\nu} \)) and a scalar field (\( \phi \)). The action for such theories can generally be written as follows:

\begin{equation}
    S = \int d^4x \sqrt{-g} \left[ \frac{F(\phi)}{2} R - \frac{1}{2} g^{\mu\nu} \partial_\mu \phi \partial_\nu \phi - U(\phi) + \mathcal{L}_m \right],
\end{equation}
where \( F(\phi) \) is a function of the scalar field that determines the coupling between the scalar field and the Ricci scalar \( R \). The term \( U(\phi) \) represents the potential of the scalar field, and \( \mathcal{L}_m \) denotes the Lagrangian density of the matter fields.

Variation of the action with respect to the metric \( g_{\mu\nu} \) and the scalar field \( \phi \) produces the following field equations:

\begin{equation}
  \begin{aligned}
    F(\phi) G_{\mu\nu} &= T_{\mu\nu} + \nabla_\mu \nabla_\nu F(\phi) - g_{\mu\nu} \Box F(\phi) \\
    &\quad+ \frac{1}{2} g_{\mu\nu} \left( \nabla_\alpha \phi \nabla^\alpha \phi + 2U(\phi) \right) - \nabla_\mu \phi \nabla_\nu \phi,  
  \end{aligned}
\end{equation}

\begin{equation}
    \Box \phi - U'(\phi) + \frac{1}{2} F'(\phi) R = 0,
\end{equation}
where \( G_{\mu\nu} \) is the Einstein tensor, \( T_{\mu\nu} \) is the energy-matter tensor, \( \nabla_\mu \) denotes the covariant derivative, \( \Box = \nabla^\mu \nabla_\mu \) is the d'Alembertian operator, and \( F'(\phi) \) and \( U'(\phi) \) are the derivatives of \( F(\phi) \) and \( U(\phi) \) with respect to \( \phi \).

Scalar-tensor theories can be formulated in two different frames: the Jordan frame and the Einstein frame \cite{Faraoni:1999hp}. These frames are related by a conformal transformation of the metric.

\begin{itemize}
    \item \textbf{Jordan Frame:} The scalar field is non-minimally coupled to the Ricci scalar as shown in the action above.
    \item \textbf{Einstein Frame:} A conformal transformation \( \tilde{g}_{\mu\nu} = F(\phi) g_{\mu\nu} \) transforms the action into a form where the Ricci scalar is minimally coupled, simplifying the gravitational equations but complicating the matter coupling.
\end{itemize}

In the Einstein frame, the action becomes:
\begin{equation}
    S = \int d^4x \sqrt{-\tilde{g}} \left[ \frac{1}{2} \tilde{R} - \frac{1}{2} \tilde{g}^{\mu\nu} \partial_\mu \chi \partial_\nu \chi - U(\chi) + \mathcal{L}_m \left( \frac{\tilde{g}_{\mu\nu}}{F(\phi)} \right) \right],
\end{equation}
where \( \chi \) is a redefined scalar field in the Einstein frame, and \( U(\chi) \) is the potential in the Einstein frame. The matter fields \( \mathcal{L}_m \) now couple to the conformally transformed metric \( \tilde{g}_{\mu\nu} \).

In this type of theories, the effective gravitational constant $G_{eff}$ can be expressed as:
\be
G_{eff}\propto 1/F(\phi).
\ee
Given a potential $U(\phi)$ we can derive an expression, as we will see, for the quantity $\Delta G/G_N$ and search for its percentage change. The values of $\Delta G/G_N$ can be connected with observational results and give constraints.

Scalar-tensor theories provide a rich framework for exploring various cosmological phenomena.

\begin{itemize}
    \item \textbf{Cosmic Inflation:} The scalar field can drive inflationary expansion in the early universe \cite{Quiros:2020bcg,Avdeev:2023ycw}. The form of \( U(\phi) \) determines the dynamics of inflation.
    \item \textbf{Dark Energy:} The scalar field can act as a dynamical dark energy component, explaining the accelerated expansion of the universe \cite{Bartolo:1999sq,Agarwal:2007wn,Ji:2024gdc,Wang:2006ze,BuenoSanchez:2011zz}.
    \item\textbf{Dark Matter:}  Symmetron fifth forces have recently been discussed as possible alternatives to particle dark matter \cite{Burrage:2016yjm,OHare:2018ayv,Burrage:2018zuj,Kading:2023hdb,Fischer:2024eic}.
    \item \textbf{Modified Gravity:} These theories modify gravitational interactions on large scales, potentially addressing issues in galaxy rotation curves and structure formation \cite{Cervantes-Cota:2007gst, Gessner:1992flm,Sharma:2022fiw,Ntahompagaze:2022zfx}.
\end{itemize}

Observational constraints on scalar-tensor theories come from various sources.

\begin{itemize}
    \item \textbf{Solar System Tests:} Precision measurements within the solar system, such as the perihelion precession of Mercury and the Shapiro time delay, place stringent limits on deviations from GR \cite{Esposito-Farese:2004azw,Dyadina:2021paa,Gonzalez:2020vzl}.
    \item \textbf{Cosmological Observations:} Data from the Cosmic Microwave Background (CMB), large-scale structure, and supernovae can constrain the dynamics of the scalar field and its effects on cosmic expansion \cite{Ballardini:2021eox,Ballardini:2021evv,Galli:2009pr}.
    \item \textbf{Gravitational Waves:} The observation of gravitational waves from binary mergers provides a new avenue to test the predictions of scalar-tensor theories, especially in strong-field regimes \cite{Will:1994fb, Scharre:2001hn,Paraskevas:2023aae}.
\end{itemize}

 Modifications to gravity, particularly through non-minimal coupling of a scalar field to gravity could be an alternative approach to resolving the cosmological tensions. A recent study \citep{Ferrari:2025egk} analyzed the fit of scalar-tensor gravity models to the latest BAO data from the DESI 2024 \cite{DESI:2024mwx,DESI:2024uvr} release. The authors considered a selection of models from previous works and showed that non-minimally coupled theories provide an improved fit to the new DESI measurements compared to \(\Lambda\)CDM. Notably, their results suggest that a larger non-minimal coupling parameter is favored by the latest BAO observations, reinforcing the role of modified gravity in addressing late-time cosmological tensions. Among the models explored, the Brans-Dicke Galileon framework was identified as the best-fitting scenario, as it can naturally produce a phantom equation of state for dark energy. These insights further motivate investigations into non-minimal couplings as a viable explanation for the observed deviations in cosmic expansion.  
 
 Furthermore, \citep{Ye:2024ywg} highlighted the role of such non-minimal couplings in explaining the latest BAO measurements from the DESI collaboration \cite{DESI:2024mwx,DESI:2024uvr}. These observations suggest a dynamical dark energy evolution crossing the phantom divide, which poses theoretical challenges for standard quintessence models. By analyzing the space of Horndeski gravity theories \cite{Horndeski:1974wa}, the authors identified non-minimal coupling as a crucial mechanism to achieve a stable phantom divide crossing while maintaining consistency with DESI constraints. Extending this idea, a subsequent study \citep{Ye:2024zpk} introduced the \textit{Thawing Gravity} (TG) model, characterized by a non-minimal coupling function of the form \( F(\phi) = 1 - \xi \phi^2 \). This model offers a compelling solution to multiple cosmological tensions, including those related to the Hubble constant (\(H_0\)), the structure growth parameter (\(S_8\)), and the DESI BAO results. Remarkably, TG provides an improved fit to observational data compared to \(\Lambda\)CDM and predicts a \(\sim 5\%\) variation in the effective gravitational constant \(G_{\rm eff}\) on cosmological scales, which could be further tested by upcoming observations.

Overall, scalar-tensor theories offer a versatile and powerful framework for extending GR and addressing some of the most pressing questions in modern cosmology and theoretical physics.

 The goal of the present analysis is to address the following questions:
\begin{enumerate}
    \item Given the scales discussed of an ultra-late time transition, what is the typical scale (size) of the produced true vacuum bubbles assuming that they form at the present cosmological time?
    \item What is the required range of the scalar field mass so that the produced true vacuum bubbles expand and not collapse?
    \item How do these results change if the scalar field is non-minimally coupled to gravity?
    \item What is the spatial extent of a bubble after the first-order phase transition via a theoretical formula and what are the corresponding observational effects?
    \item Can the true vacuum field mass be constrained in order for the percentage of \( \Delta G/G_N \) to be in agreement with observational results?
\end{enumerate}

In this paper, we introduce a theoretical framework in which false vacuum decay can induce ultra-late-time transitions in the effective gravitational constant, \( G_{eff} \), potentially offering a resolution to the Hubble tension. Our analysis explores the dynamics of first-order phase transitions occurring at recent cosmological times (\( z \sim 0.01 \)), which lead to localized variations in \( G_{eff} \). While we focus on false vacuum decay as the underlying mechanism, other alternative scenarios—such as an abrupt change in \( G_{eff} \) modeled by a step-like function—may also generate ultra-late-time transitions \cite{Alestas:2022xxm}. Understanding these transitions and their potential observational signatures is crucial for assessing whether new physics at low redshifts can explain the discrepancies between early- and late-time measurements of \( H_0 \).

This paper has the following structure: In \hyperref[sec2]{Section II}, we investigate the decay of the cosmological constant-tunneling bubble formation and its anticipated scales. Also, some numerical solutions of the equation of motion of such a bubble are presented, going beyond the thin-wall approximation model. In \hyperref[sec3]{Section III}, the evolution of the true vacuum bubbles is discussed, where the dependence on the scalar field mass plays a crucial role in the bubble's life. In \hyperref[sec4]{Section IV}, we study the effects of gravity, presenting the Coleman–De Luccia solutions. We further examine how significant gravitational corrections can be at the scales considered in our analysis. In \hyperref[sec5]{Section V}, we examine the non-minimal coupling case, where the physical mechanism behind the metastable dark energy described by a field is the effective gravitational constant $G_{eff}$. So, we can talk about gravitational constant bubbles. Then, the theoretical modeling of such an ultra-late time transition is discussed in this section. In \hyperref[sec6]{Section VI}, we discuss post-transition bubble scales and attempt to parametrize the transition duration $\delta \tau$ in terms of the mass scale of the scalar field. Our analysis builds upon classical literature, providing a theoretical framework that connects the dynamics of the phase transition to the underlying mass parameters of the field. In the framework of modified gravity theory, in \hyperref[sec7]{Section VII}, we calculate the effective gravitational constant as $G_{eff}\propto 1/F(\phi)$ and check the field mass $m$ and the coupling parameter $\xi$ for the quantity $\Delta G/G_N$ to be $1\%-7\%$ for different coupling functions $F(\phi)$. Observational hints of a first-order phase transition, the potential resolution of the Hubble tension, and some highlights of scientific research in this direction are presented in \hyperref[sec8]{Section VIII}. Finally, in \hyperref[sec9]{Section IX} we conclude, make a summary, and discuss possible future extensions of the present investigation.

\section{Ultra-Late time formation of true vacuum bubbles and their typical scale.}\label{sec2}

\subsection{Quantum Tunneling and Bubble Nucleation: Coleman Formalism and Beyond}\label{subsec2a}

First-order phase transitions have been extensively discussed in the literature in the context of the early universe, electroweak symmetry breaking, baryogenesis, and inflation \cite{Guth:1981uk,Hawking:1982ga,Hawking:1981fz,Albrecht:1982wi} . In first-order phase transitions bubbles of true vacuum form within the false vacuum background, expanding and colliding to complete the decay of the false vacuum. Recently, there has been a resurgence of interest in theories with first-order phase transitions due to the possibility of observing a gravitational wave signal emerging from the collisions of bubbles \cite{Caprini:2019egz, Ellis:2018mja,Hindmarsh:2013xza,Jinno:2016vai,Hindmarsh:2015qta,Hindmarsh:2017gnf,Vaskonen:2016yiu,Apreda:2001us}.

A complete treatment of quantum mechanical tunneling from a metastable false vacuum state to the true vacuum was given in \cite{Coleman:1977py,Callan:1977pt}. More recently, there have been rigorous re-derivations \cite{Andreassen:2016cvx,Ai:2019fri} or novel approaches \cite{Braden:2018tky} that support these results. There have also been attempts to generalize the expressions to the case of finite temperature $T>0$ \cite{Linde:1981zj,Affleck:1980ac,Gould:2021ccf}.
In general, there are two approaches to dealing with tunneling rates; in the first case, one expresses the complete Lagrangian with a potential with a metastable vacuum and using the bounce instanton solution solves for the decay rate to the true vacuum. This will result in a time-independent rate, but time dependence can be inserted if the potential depends on the temperature (e.g. through particle interactions in the early universe) or evolves according to the motion of another field (e.g. as in multifield decay). In the second case, the exact model is suppressed and a phenomenological route leads to different assumptions on the decay rate and the distinct forms of its time dependence.

\begin{figure}
\begin{centering}
\includegraphics[width=0.4\textwidth]{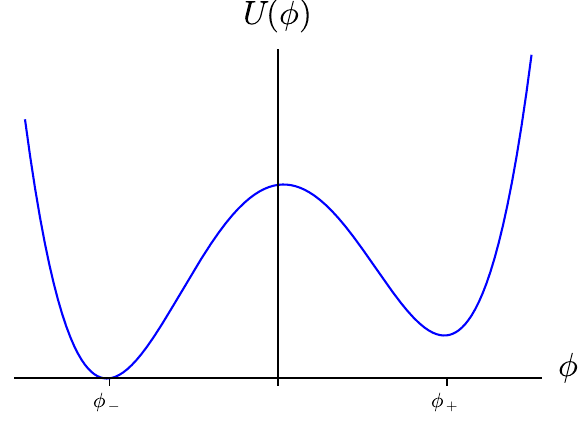}
\par\end{centering}
\caption{A theory with a metastable potential.} 
\label{fig1} 
\end{figure}

We start by presenting a review of the basic formalism that uses the bounce solution to the classical Euclidean equations of motion to estimate decay rates in the thin wall limit \cite{Coleman:1980aw}. The action for a scalar field in the absence of gravity is
\begin{equation}
    S=\int d^4x\, \left(\frac{1}{2}\partial_{\mu}\phi\partial^{\mu}\phi - U(\phi)\right).
\end{equation}

The potential $U(\phi)$ has two minima (check Fig.\ref{fig1}); a metastable minimum at $\phi_+$ and a stable minimum at $\phi_-$. Classically, the field is trapped in the local minimum of its potential, the false vacuum. In quantum field theory though, there is a non-zero transition amplitude $\langle\phi_+(x)|\phi_-(x)\rangle$ from the false vacuum configuration to the true vacuum configuration. Summing over all paths in the Feynman path integral prescription we can associate a decay rate to this quantum mechanical process. In the semi-classical picture quantum mechanical fluctuations of the field at a point `crossover' to the true vacuum, forming nucleation events, or bubbles of a small volume and zero energy that expand at approximately the speed of light. The decay rate gets interpreted as the rate of bubble formation and can be estimated using the bounce formalism as

\begin{equation}\label{Gamma}
    \frac{\Gamma}{V} = A e^{-S_B},
\end{equation}
where \( S_B = S_E(\phi) - S_E(\phi_+) \) \cite{Coleman:1980aw}, represents the bounce action, defined as the difference between the Euclidean action evaluated at the bounce solution \( \phi \) and its value at the false vacuum \( \phi_+ \). Here, \( \phi \) denotes the bounce solution, which satisfies the Euclidean equations of motion.
The $A$ factor includes next-to leading order terms, is expressed in terms of functional determinants and carries the dimensionful part of the rate \cite{Callan:1977pt}. The exact form of the $A$ term will not concern us since the exponent is the dominant contribution to the rate. As $A$ has dimensions of inverse 4-Volume and we can set it equal to a characteristic scale of the system to the inverse fourth power.

We can estimate the number of bubbles formed in a 4-Volume $V^4$ as $\mathcal{N}= (\Gamma/V) V^4$ assuming a constant rate. Then setting $V^4 = H_0^{-4}$, we get the number of bubbles in our Hubble horizon. 

In the absence of gravity, it has been shown that the bounce solution must be $\mathcal {O}(4)$ symmetric \cite{Coleman:1977th}, i.e. $\phi=\phi(\rho)$ where $\rho^2=\tau^2+x^2+y^2+z^2$ and $\tau,x,y,z$ are the Euclidean coordinates. Then the Euclidean action can be expressed in a simplified form

\begin{equation}
    S_E=2\pi^2\int_0^{\infty}\rho^3 d\rho\,\left(\frac{1}{2}\phi'^2 + U(\phi)\right)
\end{equation}
as well as the equation of motion;

\begin{equation}\label{eom}
    \phi''+\frac{3}{\rho}\phi'=\frac{dU}{d\phi},
\end{equation}
with the prime denoting $d\phi/d\rho$.

\begin{figure}
\begin{centering}
\includegraphics[width=0.4\textwidth]{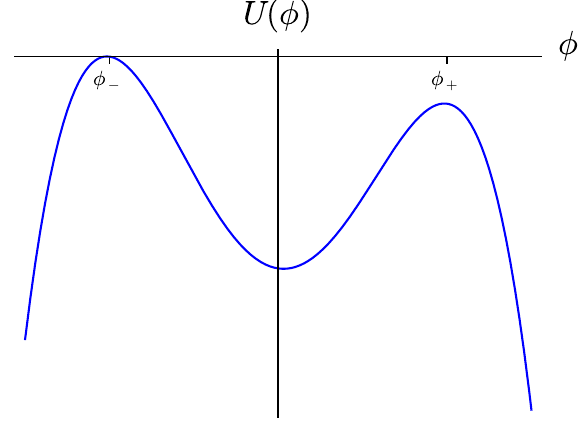}
\par\end{centering}
\caption{The reversed potential of  Fig.\ref{fig1}.} 
\label{fig2} 
\end{figure}

This equation can be interpreted as the equation of a particle moving in the inverted potential under the influence of an extra damping term (Fig.\ref{fig2}).
The bounce solution is constructed assuming the field starts at the false vacuum at time $\tau=-\infty$, reaches the true vacuum at time $\tau=0$ and bounces to the false vacuum once more at time $\tau=\infty$. Thus, in our symmetric coordinates, we have the boundary conditions: $\phi(\infty)=\phi_+$, $\phi(0)=\phi_-$ and $\phi'(0)=0$ to keep our equation of motion results smooth. 
An exact solution can be obtained in the thin wall approximation. In this approximation the two minima are nearly degenerate with an energy difference of $U(\phi_+)-U(\phi_-)=\epsilon$. We can assume that the bounce solution in Euclidean space has the form of a bubble of true vacuum $\phi_-$ up to a radius $\bar{\rho}$, with a thin wall, inside the false vacuum $\phi_+$. We can evaluate the difference of the two integrals in $S_B$ in these three regions from $\rho=0$ to $\rho=\infty$: $S_B=S_{B,inside}+S_{B,wall}+S_{B,out}$. Thus, one can show that

\begin{equation}\label{Action}
    S_B=-\frac{\pi^2\bar{\rho}^4\epsilon}{2}+2\pi^2\bar{\rho}^3S_1
\end{equation}
with $S_1=\int_{\phi_-}^{\phi_+}d\phi\,[2(U(\phi)-U(\phi_+)]^{1/2}$. The first term on the right-hand side is the energy gained from the decay of the false vacuum inside the bubble, while the second term can be interpreted as the energy we expand in forming the bubble wall with surface tension $S_1$. Since the bounce is an extremum of the action, we must find the value of $\rho$ that extremizes $S_B$, thus solving for a critical bubble where the energy gained exactly cancels the surface tension contribution. One finds

\begin{equation}\label{Radius}
    \bar{\rho}_0=\frac{3S_1}{\epsilon},
\end{equation}
while $S_B$ becomes

\begin{equation}\label{Minimized action}
   S_{B,0}=\frac{27\pi^2S_1^4}{2\epsilon^3}.
\end{equation}

After a bubble nucleates, we can model its evolution using the analytic continuation of the bounce solution. For the radius of the bounce we have $\bar{\rho}^2=\tau^2+x^2+y^2+z^2$ in Euclidean space, which after the analytic continuation becomes $\bar{\rho}^2=-t^2+x^2+y^2+z^2$ in Minkowski. The bubble nucleates at $t=0$ and its expansion is described by the hyperboloid above for $t>0$. Thus, the bounce radius models the radius of the bubble at the moment of its materialization. Furthermore, we can deduce that the bubble expansion velocity will asymptotically approach the speed of light.

Someone can go a step further, beyond the thin-wall approximation \cite{Dunne:2005rt}, and solve Eq.\eqref{eom} numerically. For a given potential, let us say of the form
\begin{equation}
    U(\phi)=\frac{\lambda}{8}(\phi^2-a^2)^2-\frac{\epsilon}{2a}(\phi-a),
    \label{pot}
\end{equation}
with $a>0$, $\lambda>0$, and $\epsilon>0$ the external cause which breaks the symmetry of the double well. After some mathematical manipulation, presented in \hyperref[AppI]{APPENDIX A} we expand the scalar field about $\phi_+$, its false vacuum, in the following way
\begin{equation}
    \phi=\phi_+ +\varphi.
\end{equation}
In this way, we get rid of the linear terms of the potential, if we Taylor expand around $\phi_+$, because $dU(\phi)/d\phi|_{\phi=\phi_+}=0$.

Then, the potential up to dimension four is given by \cite{Baacke:2003uw}
\begin{equation}
    U(\varphi)=\frac{m}{2}\varphi^2-\eta \varphi^3 +\frac{\lambda}{8}\varphi^4,
\end{equation}
where we have defined
\begin{equation}
    m^2=\frac{\lambda}{2}(3\phi_+^2-a^2) , \qquad \eta=\frac{\lambda}{2}|\phi_+|.
\end{equation}
If we rescale the field $\varphi$ as well as the coordinates of space-time as
\begin{equation}
    \varphi=\frac{m^2}{2\eta}\Phi, \qquad \Tilde{x}=mx,
\end{equation}
the classical Euclidean action after some trivial algebra will simplify to
\begin{equation}
    S_E[\Phi]=\bigg(\frac{m^2}{4\eta^2}\bigg)\int d^4\Tilde{x}\bigg[\frac12(\Tilde{\partial_\mu} \Phi)^2+\frac12\Phi^2-\frac13\Phi^3+\frac{k}{8}\Phi^4\bigg],
    \label{actI}
\end{equation}
where the dimensionless $k$ next to the quartic coupling strength is defined as
\begin{equation}
    k=\frac{\lambda m^2}{4\eta^2}=1-\frac{\epsilon}{2\lambda a^4}+\dots.
\end{equation}
It is obvious from the above that $\alpha$ tends to one in the ``thin-wall'' limit. From Eq.\eqref{actI} we get the dimensionless potential
\begin{equation}\label{dimlesspot}
    U(\Phi)=\frac12\Phi^2-\frac12\Phi^3+\frac{k}{8}\Phi^4.
\end{equation}

The shape of the potential (Fig.\hyperref[I1]{20}) is determined by $k$, which is dimensionless and $0<k<1$, its divergence from the unity represents a measurement of the vacuum energy difference in relation to the height of the barrier.

The next step in the process is to find the classical bounce solution $\Phi_{cl}$. With a little more algebra shown in our appendix, one can get a rescaled nonlinear ordinary differential equation which the classical bounce $\Phi_{cl}$ solves. This will be:
\begin{equation}
    -\Phi_{cl}''-\frac{3}{r}\Phi_{cl}+\Phi_{cl}-\frac32\Phi_{cl}^2+\frac{k}{2}\Phi_{cl}^3=0,
    \label{neweom}
\end{equation}
with the primes indicating differentiation with respect to $r$ and the boundary conditions given by
\begin{equation}
    \Phi'_{cl}(0)=0,\qquad \Phi_{cl}(\infty)=\Phi_+.
    \label{boundc}
\end{equation}

\begin{figure}[h]
\begin{center}
\includegraphics[scale=0.5]{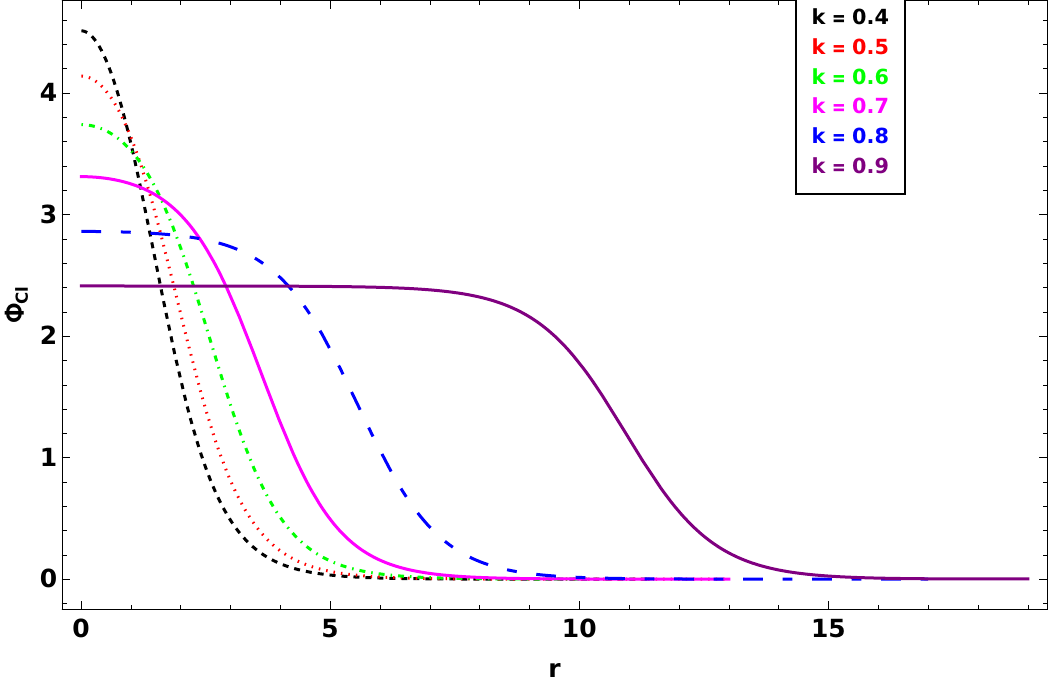}
 \caption{Various bounce solutions $\Phi_{cl}(r)$ of \eqref{neweom} plotted for $k=0.4$, $0.5$ $0.6$, $0.7$, $0.8$, $0.9$.}
    \end{center}
  \label{numerical1}
\end{figure}

In Fig.\hyperref[numerical1]{3} some bubble profiles are depicted for different values of $k$. An observation that can be made is that as $k$ increases, the scalar field fulfills the second boundary condition at a larger bubble radius. Also, a second observation is that the bounce profiles have a form that can be described by $\Phi_{cl}\propto \tanh(-r)$ exactly as expected from the soliton-instanton solutions.

Moreover, while $k\rightarrow 1$, the potential reaches the double-well potential. The vacua become degenerate. As $k$ gets lower, $\epsilon$ (the energy density between the vacua, playing the role of the cosmological constant) increases, and from the figure above we can see that the bubble radius drops too. The physical reason for this is that bigger $\epsilon$ means more energy available for the false vacuum to true vacuum conversion. Hence, less volume is necessary to compensate the wall energy cost and the radius is getting lower. Also, as $\epsilon$ decreases, we can say that the validity of the thin-wall approach becomes weaker.

\subsection{Ultra-Late Cosmological Transitions: Bubble Formation and Characteristic Scales}\label{subsec2b}
Given a potential we can use equation (\ref{Minimized action}) to calculate the decay rate (\ref{Gamma}). We would like to estimate the scales involved in a typical decay rate calculation in the thin-wall approximation and for decay times relevant in cosmology. Substituting equation (\ref{Radius}), this time for $S_1$ into (\ref{Action}) we find for the decay rate

\begin{equation}
    \frac{\Gamma}{V} = A e^{-\frac{\pi^2 \bar{\rho}^4\epsilon}{6}}.
\end{equation}

To study a present-time decay, we must establish certain assumptions. First, we recall that in the current $\Lambda$CDM model, the present age of the Universe is given in a good approximation by the inverse of the Hubble parameter, $H_0^{-1}$ \cite{Lima:2007kr}. The total number of bubble nucleation events, integrated over a Hubble volume $\left( d_H^3 \approx H_0^{-3} \right)$ and the age of the Universe $\left( t_U \approx H_0^{-1} \right)$  can be approximated as given in \cite{Turner:1992tz,Krauss:2013vya}. It is
\begin{equation}
    \mathcal{N}=(\Gamma/V)H_0^{-4}.
\end{equation}
If we set the decay rate to an upper bound for a slow, time-dependent, uncompleted transition\cite{Lima:2020nwt, Landim:2016isc},

\begin{equation}
    \frac{\Gamma}{V} = \mathcal{N}H_0^4
\end{equation}
then the decay time $t_d \sim H_0^{-1}$ will be comparable to the age of the universe.
 As we have already written, the energy density between the vacua is playing the role of the cosmological constant $\Lambda$. So we can write that
\begin{equation}\label{epsilonzeta}
    \epsilon = \frac{3}{8\pi} M_{Pl}^2 H_0^2\Omega_{\Lambda,0}\sim M_{Pl}^2H_0^2
\end{equation}
where $M_{Pl}=1/\sqrt{G_N}$ is the Planck mass.

We can further posit that on dimensional grounds $A\sim m^4$ \cite{Linde:1981zj}, in order to solve for the bubble radius as

\begin{equation}\label{ultratimerho}
 \bar{\rho}=\left(\frac{24}{\pi^2\epsilon}\text{ln}\frac{ m}{\mathcal{N}^{1/4}H_0}\right)^{1/4}.
\end{equation}
Using (\ref{Radius}) we infer that for the potential barrier

\begin{equation}\label{ultratimeS1}
    S_1=\left(\frac{8\epsilon^3}{27\pi^2}\text{ln}\frac{ m}{\mathcal{N}^{1/4}H_0}\right)^{1/4}.
\end{equation}

For an order of magnitude estimate we can ignore the logarithmic terms, since for a wide range of values of $m>H_0$, and $\mathcal{N}$, as we will see later in this article, $[\text{ln}( m/\mathcal{N}^{1/4}H_0)]^{1/4} \sim \mathcal {O}(1)$. Thus

\begin{equation}
    \bar{\rho}\sim \epsilon^{-1/4},
    \label{barro}
\end{equation}

\begin{equation}\label{s1z}
    S_1\sim \epsilon^{3/4}.
\end{equation}

The above order of magnitude analysis is equivalent to setting $S_B\sim \mathcal {O}(1)$, which is not a cause for concern as $S_B$ needs to be at least $\mathcal {O}(10^3)$ to change our estimates. Similarly, in a time-dependent transition, it is usually the case that, as the false vacuum starts decaying, the decay rate is highly suppressed with $S_B\sim \mathcal {O}(10^4)$ and as the transition proceeds $S_B$ evolves to smaller values till $S_B\sim \mathcal {O}(1)-\mathcal {O}(10^3)$ (depending on the exact model \cite{Caprini:2019egz}) and the transition can proceed rapidly. Thus, our estimates are valid for the relevant range where the transition is taking place with non-negligible probabilities. For a cosmological constant transition, we need to pick $\epsilon$ small enough for the thin-wall approximation to give accurate results \cite{Samuel:1991mz}. For an upper bound $\epsilon\sim 10^{-2}\rho_{\Lambda}\sim 10^{-2} H_0^2 M_{Pl}^2 $ we have $\epsilon\sim 10^{-48}\, GeV^4$, $\bar{\rho}\sim 200\, \mu m $, $S_1\sim  10^{-36}\, \text{Gev}^3 $. These estimates for the radius agree with \cite{Coleman:1980aw} in that the bubbles form with a negligible size compared to cosmological distances. Since the bubble wall traces out the hyperboloid $\bar{\rho}^2=-t^2+x^2+y^2+z^2$ we can, as has usually been the case in the literature, assume that the bubble is created with essentially zero radius and propagates outward at the speed of light. 

As an example of a time-dependent, pre-recombination transition, we might take the NEDE model \cite{Niedermann:2019olb,Niedermann:2023ssr}, assuming a thin wall. For $\Gamma/V\sim M^4 e^{-S_B} \,eV^4$, we have $M\sim eV$ and an ultra-light mass scale $m\sim H(z=5000)\sim 10^{-27}\,eV$. Tunneling becomes efficient when the decay rate outperforms the Hubble expansion. Setting $\Gamma/VH^{-4}=M^4/m^4 e^{-S_B}\gtrsim 1$ we get $S_B\lesssim  250$ so that $\bar{\rho}\sim 3.5\times\epsilon^{-1/4}$, $S_1\sim 1.2\times \epsilon^{3/4}$ for when the transition becomes relevant.

\begin{figure}
    \centering
    \includegraphics[width=1\linewidth]{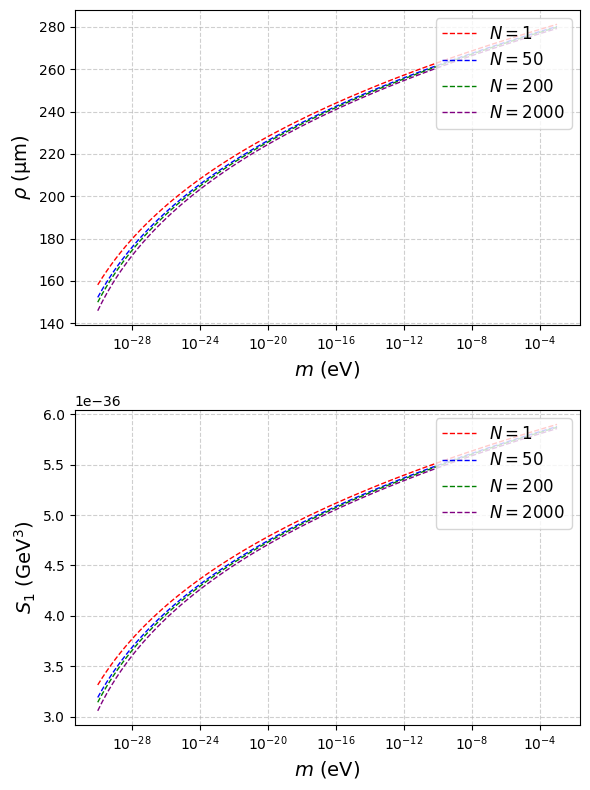}
    \caption{Equations \eqref{ultratimerho} and \eqref{ultratimeS1} versus the mass scale of the transition $m$ for different values of the number of the nucleation events $\mathcal{N}$.}
    \label{rhoS1vsm}
\end{figure}

However, let us take a closer and more precise look at the previous results. In Fig.\ref{rhoS1vsm}, equations \eqref{ultratimerho} and \eqref{ultratimeS1} are plotted versus the transition mass scale $m$. The radius at the moment of nucleation ranges from $\sim 150\, \mu m$ to $\sim 280\,\mu m$ for four different scenarios of the number of nucleation events. The logarithmic dependence of $\rho$ on the nucleation-events number $\mathcal{N}$ and the mass $m$ naturally leads to results that do not change significantly, despite the large range of masses and $\mathcal{N}$ values we have chosen. Similarly, the surface tension $S_1$ exhibits the same logarithmic dependence and is found to be approximately $10^{-36} \, GeV^3$, as we previously estimated through a rough order-of-magnitude calculation.

\section{Evolution of true vacuum bubbles: Expansion vs Collapse.}\label{sec3}
The fate of vacuum bubbles—whether they expand or collapse—is determined by the scalar field critical mass $m_{crit}$, which represents the maximum mass allowing for bubble expansion \cite{Coleman:1980aw,Callan:1977pt}. To determine this critical mass, we begin by analyzing the total energy of a thin-walled bubble at the moment of its formation, in the absence of gravitational effects \cite{Coleman:1980aw}:  

\begin{equation}
    E=-\frac{4}{3}\pi\bar{\rho}^3\epsilon+4\pi\bar{\rho}^2m^3,
    \label{ttlenergy}
\end{equation}  

This energy consists of two competing terms: a negative volume term, $-\frac{4}{3}\pi\bar{\rho}^3\epsilon$, representing the energy gained from converting false vacuum to true vacuum, and a positive surface term, $4\pi\bar{\rho}^2m^3$, corresponding to the energy cost of maintaining the bubble wall. In this expression, we use the relation $S_1 \sim m^3$ \cite{Coleman:1977py,Linde:1981zj}, which emerges naturally from a theory with the potential  

\begin{equation}
    U(\phi)=\frac{\lambda}{8}\left(\phi^2-\alpha^2\right)^2 +\mathcal{O}(\epsilon),
    \label{potential}
\end{equation}  
where $\alpha=m^2/\lambda$ and $\mathcal{O}(\epsilon)$ represents small symmetry-breaking terms \cite{Weinberg:2012pjx}.  

Following the nucleation theory developed by Fletcher \cite{1958JChPh..29..572F} and extended by others \cite{Langer:1969bc,Affleck:1980ac}, the critical radius $\bar{\rho}_c$ represents the minimum radius at which a bubble becomes viable and begins to expand. This occurs at the maximum of the total energy, found by setting $dE/d\bar{\rho}=0$:  

\begin{equation}
  \bar{\rho}_c=\frac{2m^3}{\epsilon} =\frac{2m^3}{M_{Pl}^2H_0^2},
    \label{critradius}
\end{equation}  
where we have used the relation $\epsilon \sim M_{Pl}^2H_0^2$ for a cosmological constant transition \cite{Weinberg:2008zzc}. Comparing this critical radius with the actual nucleation radius given by Eq.~(\ref{barro}), we obtain  

\begin{equation}
    \frac{\bar{\rho}}{\bar{\rho}_c}\approx\frac{(M_{Pl}H_0)^{-1/2}}{(M_{Pl}H_0)^{-2}m^3}
      \geq 1.
\end{equation}  
This inequality leads to a crucial constraint on the scalar field mass:  

\begin{equation}\label{mcritgr}
 m\leq 0.0017\,eV \equiv m_{crit}.
\end{equation}  

\begin{figure}
    \centering
    \includegraphics[width=1\linewidth]{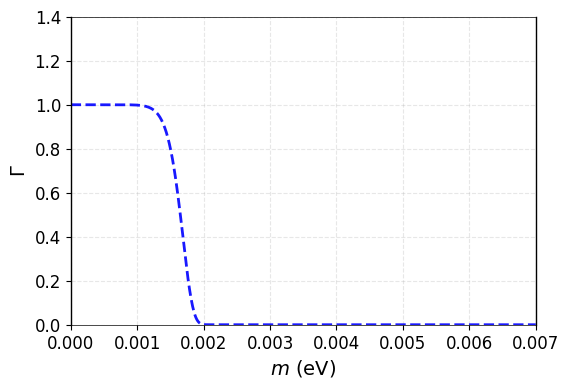}
    \caption{ The decay probability as a function of the scalar field mass. In an ultra-late time transition, for $m<m_{crit}=0.0017\,eV$ the bubble nucleates and expands, for $m \sim 0.002 \,eV$ the false vacuum becomes effectively stable and no bubble nucleates. In the regime $0.0017 \,eV < m < 0.002\,eV$ the bubble nucleates but collapses.}
    \label{fig3}
\end{figure}

The behavior of vacuum bubbles is governed by the value of $m$. For $m \leq m_{crit}$, bubbles expand after formation, while for $0.0017\,\text{eV} < m < 0.002\,\text{eV}$, bubbles nucleate but subsequently collapse due to surface tension dominance. When $m \gtrsim 0.002\,\text{eV}$, the decay rate becomes effectively zero,   $ \Gamma \sim e^{-S_{B}} \sim e^{-S_1^4/\epsilon^3} \sim e^{-m^{12}\times 10^{36}} \rightarrow 0,$ resulting in the stability of the false vacuum. At the critical mass, $m = m_{crit}$, the decay probability takes the value $\Gamma = 0.4$. The complete behavior of the decay probability, $\Gamma \sim e^{m^{12}\epsilon^{-3}}$, as a function of mass is illustrated in Fig.~\ref{fig3}. This analysis demonstrates how the mass parameter crucially determines whether vacuum decay proceeds through expanding bubbles or remains effectively suppressed.  

\section{The effects of gravity.}\label{sec4}

The decay rate calculation is strictly only valid for flat spacetime but using the same formalism, the effects of gravity on the above calculations were estimated by Coleman and De Luccia \cite{Coleman:1980aw}. The relevant gravitational action for a minimally coupled scalar field is:

\begin{equation}
     S=\int d^4x\sqrt{-g}\, \left[\frac{R}{2\kappa}+\frac{1}{2}\partial_{\mu}\phi\partial^{\mu}\phi - U(\phi)\right],
\end{equation}
where we have introduced a metric $g_{\mu\nu}$, the Ricci scalar $R$ and the coupling $\kappa = 8\pi G_N$. Enforcing the (reasonable yet unproved in the presence of gravity \cite{Coleman:1980aw}) assumption of $\mathcal {O}(4)$ symmetry in the metric 

\begin{equation}
    ds^2=d\eta^2 + \rho^2(\eta)[d\chi^2+\text{sin}^2\chi(d\theta^2+\text{sin}^2\theta d\phi^2)]
    \label{metric}
\end{equation}
the Euclidean action becomes

\begin{equation}
    S_E=2\pi^2\int d\eta\,\left[\rho^3\left(\frac{1}{2}\phi'^2+U(\phi)\right)+\frac{3}{\kappa}(\rho^2\rho''+\rho\rho'^2-\rho)\right],
\end{equation}
with the following equations of motion 

\begin{equation}
    \rho'^2=1+\frac{\kappa\rho^2}{3}\left(\frac{1}{2}\phi'^2-U(\phi)\right),
    \label{cdl1}
\end{equation}

\begin{equation}
    \phi''+3\frac{\rho'}{\rho}\phi'=\frac{dU(\phi)}{d\phi},
    \label{cdl2}
\end{equation}
where '$'$` indicates $d/d\eta$.

Coleman-De Luccia considered transitions with potentials with an arbitrarily small ($\epsilon$) cosmological constant. For transitions from de-Sitter ($U(\phi_+)=\epsilon$) to flat space ($U(\phi_-)=0$) and from flat space ($U(\phi_+)=0$) to Anti de-Sitter ($U(\phi_-)=-\epsilon$), using the thin wall formalism and extremizing the action, they found that

\begin{equation}
    \bar{\rho}=\frac{\bar{\rho}_0}{1\pm\left(\frac{\bar{\rho}_0}{2D}\right)^2},
\end{equation}
\begin{equation}
    S_B=\frac{S_{B,0}}{\left[1\pm\left(\frac{\bar{\rho}_0}{2D}\right)^2\right]^2}
\end{equation}
where $\bar{\rho}_0$, $S_{B,0}$ are the results in the absence of gravity equations \eqref{Radius},\eqref{Minimized action}, $D=(\kappa\epsilon/3)^{-1/2}$ and the $+(-)$ denotes the former (latter) case.

The appropriate generalization to arbitrary vacuum potential values ($U(\phi_+)=U_+$ and $U(\phi_-)=U_-$) was given by \cite{Parke:1982pm} as

\begin{equation}
    \bar{\rho}=\frac{\bar{\rho}_0}{\sqrt{1+2\left(\frac{\bar{\rho}_0}{2d}\right)^2+\left(\frac{\bar{\rho}_0}{2D}\right)^4}}
\end{equation}

\begin{equation}
    S_B=2S_{B,0}\frac{\left\{\left[1+\left(\frac{\bar{\rho}_0}{2d}\right)^2\right]-\left[1+2\left(\frac{\bar{\rho}_0}{2d}\right)^2+\left(\frac{\bar{\rho}_0}{2D}\right)^4\right]^{1/2}\right\}}{\left(\frac{\bar{\rho}_0}{2D}\right)^4\left[\left(\frac{D}{d}\right)^4-1\right]\left[1+2\left(\frac{\bar{\rho}_0}{2d}\right)^2+\left(\frac{\bar{\rho}_0}{2D}\right)^4\right]^{1/2}}
\end{equation}
where $d=[\kappa(U(\phi_+)+U(\phi_-))/3]^{-1/2}$.
Of course, they contain the Coleman-De Luccia case as the appropriate limit $(D/d)^2$ goes to $\pm1$.

The gravitational corrections become important close to the Planck scale (as $\bar{\rho}_0\sim D)$ where we expect the semi-classical approximation scheme for the decay rate to break down. In fact, for these scales $S_B\sim M_{Pl}^{12}/\epsilon^3$ and the decay rate gets a huge exponential suppression unless the potentials are comparable to the Planck mass. Thus, when gravity is strong enough to significantly modify the transition rates, the bubble nucleation events will have become irrelevant for times less than or comparable to $H_0^{-1}$. Estimating the gravitational corrections as $\bar{\rho}_0\ll D$ we can infer that they are insignificant in the case of a cosmological constant transition as described above. More specifically,

\begin{equation}\label{graveffect}
     \frac{\bar{\rho}^2}{(2D)^2}=\frac{\bar{\rho}^2\epsilon}{12M_{Pl}^2} \sim 10^{-63}.
     \end{equation}

\section{Non-minimal coupling: Bubbles of Gravitational Constant.}\label{sec5}

\subsection{Theoretical Framework and Numerical Analysis of Gravitational Constant Bubbles}\label{subsec5a}

One can further extrapolate the above results in the case of a non-minimal coupling of the scalar field to gravity \cite{Lee:2005ki,Czerwinska:2017pgb}. Such coupling appears naturally in various theoretical frameworks, including scalar-tensor theories and modified gravity models \cite{Faraoni:2023hwu}. The action describing this system takes the form:

\begin{equation}
     S=\int d^4x\sqrt{-g}\, \left[F(\phi)\frac{R}{2}+\frac{1}{2}\partial_{\mu}\phi\partial^{\mu}\phi - U(\phi)\right]
\end{equation}
where the function $F(\phi)$ describes the non-minimal coupling between the scalar field and gravity, encapsulating the effects of a varying gravitational constant in the physical Jordan frame as follows \cite{Nesseris:2006jc}
\begin{equation}
G_{eff}=\frac{1}{F(\phi)} \frac{F(\phi)+2 \left( \frac{dF}{d \phi} \right)^2}{F(\phi)+\frac{3}{2} \left( \frac{dF}{d \phi} \right)^2} \propto \frac{1}{F(\phi)}.
\label{5.55}
\end{equation}

With a proper redefinition of the field $\phi$ the above action is equivalent to the Brans-Dicke action with a potential term, where the same formalism can be applied \cite{Kim:2010yr,Faraoni:1998qx}. In both of these cases the non-minimally coupled scalar field tunnels to the true vacuum of the potential. Alternatively, a second, minimally coupled, scalar field can tunnel to the true vacuum in the presence of a Brans-Dicke gravitational background \cite{Holman:1989gh,Accetta:1989zn}.

As in the previous case, the mechanism consists in the scalar field slowly rolling towards one of its minima during the expansion history, eventually arriving there at low redshift. If this minimum is metastable, there will be a decay rate associated with the transition, which will then turn on at very late times.

Consider a theory where
\begin{equation}
    F(\phi)=\frac1\kappa-\xi\phi^2,
\end{equation}
with $\kappa=8 \pi G_N$ and $\xi$ the non-minimal coupling between the field and the curvature scalar. 

The Euclidean field equations, derived from varying the action with respect to $\phi$, yield:

\begin{equation}\label{eomnonminimal}
      \phi''+3\frac{\rho'}{\rho}\phi'-\xi R\phi=\frac{dU(\phi)}{d\phi},
\end{equation}
This equation differs from the Coleman-De Luccia result \eqref{cdl2} by the addition of the coupling term $-\xi R\phi$.
  
Following this, we proceed with the variation of the action with respect to the metric. Through this procedure, we obtain the energy-momentum tensor.
\begin{equation}
     T_{\mu\nu}=\frac{1}{1-\xi\phi^2\kappa}\left[\partial_\mu\phi\partial_\nu\phi -g_{\mu\nu} \left(\frac{1}{2}g^{\alpha\beta}\partial_\alpha\phi\partial_\beta\phi + U(\phi)\right) \right].
     \label{eq.3.9}
   \end{equation}
   From the $\eta\eta$ component, we obtain the modified Friedmann equation:
  \begin{equation}\label{ffeq}
      \rho'^2=1+\frac{\kappa \rho^2}{3(1-\xi\phi^2\kappa)}\left(\frac12\phi'^2-U+6\xi\phi'\phi\frac{\rho'}{\rho}\right).
\end{equation}
Finally, the Euclidean action can be expressed in terms of $\rho$ as follows
\begin{multline}
S_E=2\pi^2\int_{0}^{+\infty}d\eta\Bigg[\rho^3\left(\frac12\phi'^2+U\right)+\frac3\kappa(\rho\rho'^2+\rho^2\rho''-\rho)\\-3\xi\phi^2(\rho\rho'^2+\rho^2\rho''-\rho)\Bigg].
\end{multline}

Following the standard thin-wall procedure someone can obtain an analytical expression for the bounce action $S_B$:
\begin{multline}
S_B=2\pi^2\Bigg[\bar\rho^3S_1+\xi\bar\rho C-\frac23 \frac{(1-\bar\rho^2\kappa_+U_+)^{3/2}-1}{\kappa_+^2U_+}\\+\frac23\frac{(1-\bar\rho^2\kappa_-U_-)^{3/2}-1}{\kappa_-^2U_-}\Bigg]
\label{olga}
\end{multline}
where $ C=12\alpha/\sqrt\lambda$ for the potential of \eqref{potential} and $\kappa_\pm=\kappa/(1-\kappa\xi\phi_\pm^2)$. We will use Eq.\eqref{olga} for the physically relevant case of a de Sitter to Minkowski transition, this reduces to:
\begin{equation}
    S_B = 2 \pi^2 \left(\bar{\rho}^3 S_1 + \xi \bar{\rho} C - \frac{2}{3} \frac{(1 - \bar{\rho}^2 \kappa_+ \epsilon)^{3/2} - 1}{\kappa_+^2 \epsilon} - \frac{\bar{\rho}^2}{\kappa_-}\right)
\end{equation}

Yet again, to move beyond the thin-wall approximation, we employ a numerical shooting method technique described in \hyperref[app2]{APPENDIX B } of this paper. The system comprises the scalar field equation \eqref{eomnonminimal} and a second Friedmann-like equation derived from \eqref{ffeq}. If we differentiate the aforementioned equation with respect to the imaginary time parameter we find:
\begin{equation}
    \rho''=\frac{\kappa\rho}{3(1-\kappa\xi\phi^2)}\bigg[-\phi'^2-U+3\xi\bigg(\phi'^2+\phi''\phi +\phi'\phi\frac{\rho'}{\rho}\bigg)\bigg],
    \label{friedman2}
\end{equation}
subject to bounce boundary conditions 
\begin{align}
\lim_{\eta\to\infty}\phi(\eta)=&\phi_+,            &\frac{d\phi}{d\eta}\Bigg|_ {\eta=0}=0.
\label{bc}
\end{align}

For numerical analysis, we employ an instructive toy model potential:
\begin{equation}
    U=-\frac14 a^2(3b-1)\phi^2+\frac12 a(b-1)\phi^3 +\frac14 \phi^4 +a^4 c.
\end{equation}

\begin{figure}[h]
\begin{center}
        \includegraphics[scale=.55]{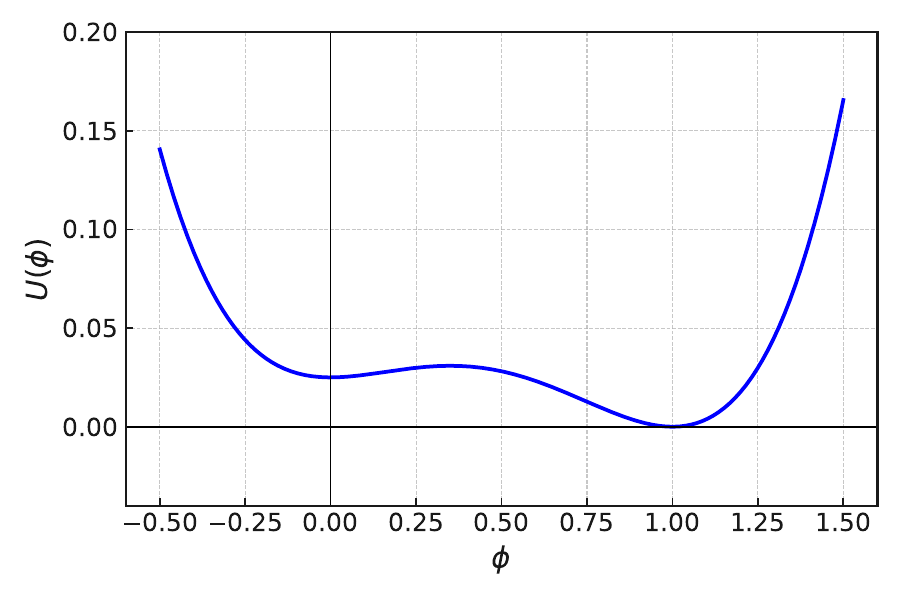}
    \caption{The toy model potential for a $dS\rightarrow M$ transition with $a=1$, $b=0.1$ and $c=0.025$.} 
    \end{center}
  \label{potdsm}
\end{figure}
\begin{figure}[h]
\begin{center}\label{potmads}
        \includegraphics[scale=.55]{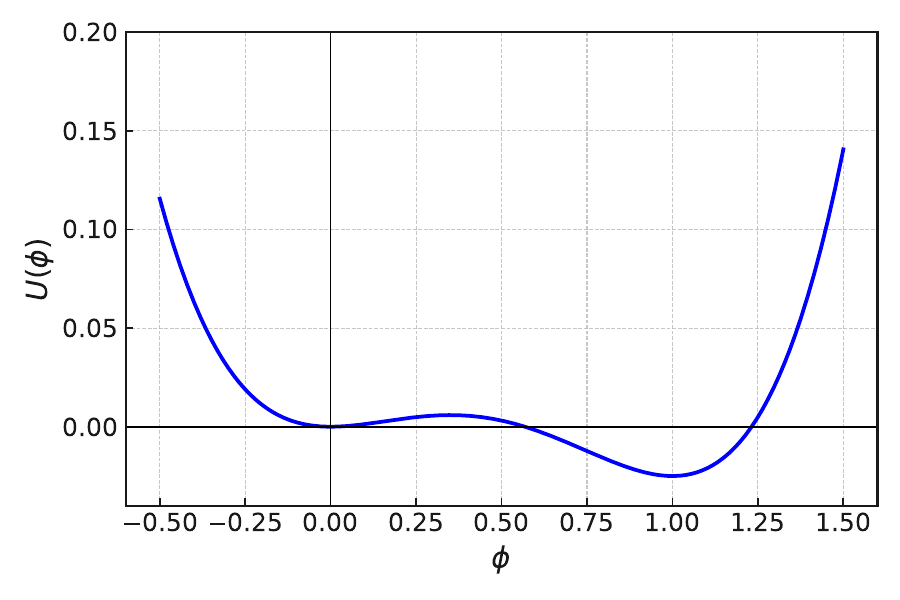}
    \caption{The toy model potential for a $M\rightarrow AdS$ transition with $a=1$, $b=0.1$ and $c=0$.} 
    \end{center}
\end{figure}

This potential features a false vacuum at $\phi_f=0$ and a true vacuum at $\phi_t=a$, which we set to unity in Planck units. The parameter $c$ controls the vacuum energy, with positive values corresponding to a de Sitter background  (Fig.\hyperref[potdsm]{6}) and zero values to a Minkowski background (Fig.\hyperref[potmads]{7}). The parameter $b$, set to 0.1 in our analysis, controls the degree of vacuum degeneracy.

\begin{figure}[htbp]
    \begin{subfigure}[b]{\columnwidth}
        \centering
        \includegraphics[width=0.8\linewidth]{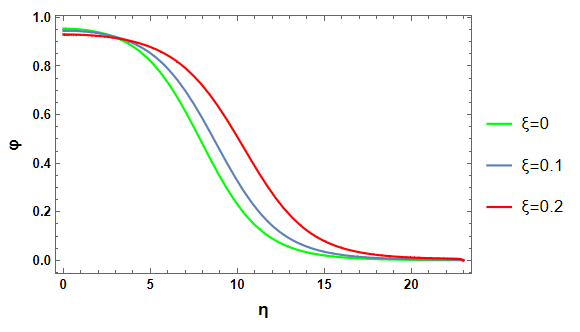}
        \caption{}
        \label{fig:ds1}
    \end{subfigure}
    
    \begin{subfigure}[b]{\columnwidth}
        \centering
        \includegraphics[width=0.8\linewidth]{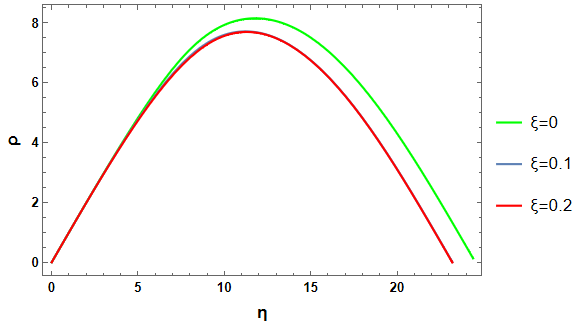}
        \caption{}
        \label{fig:ds2}
    \end{subfigure}
    
    \caption{a) The bubble profiles for a $dS\rightarrow M$ transition for several values of the coupling $\xi$. b) The bubble radii for the same procedure.}
    \label{fig:ds}
\end{figure}

\begin{figure}[htbp]
    \begin{subfigure}[b]{\columnwidth}
        \centering
        \includegraphics[width=0.8\linewidth]{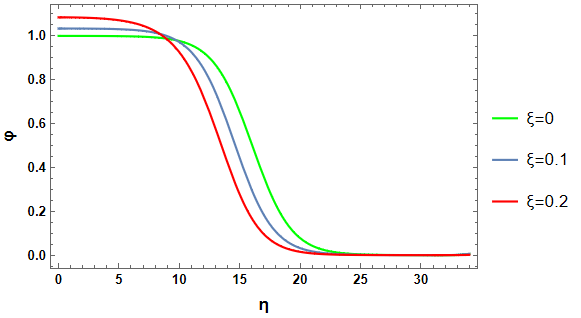}
        \caption{}
        \label{fig:M1}
    \end{subfigure}
    
    \begin{subfigure}[b]{\columnwidth}
        \centering
        \includegraphics[width=0.8\linewidth]{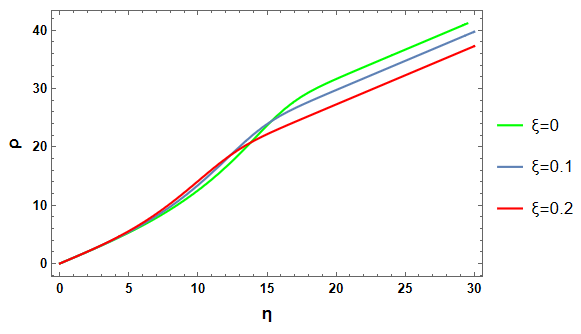}
        \caption{}
        \label{fig:M2}
    \end{subfigure}
    
    \caption{a) The bubble profiles for a $M\rightarrow AdS$ transition for several values of the coupling $\xi$. b) The bubble radii for the same procedure.}
    \label{fig:M}
\end{figure}

In the figures below, bubble profiles and bubble radii are depicted for a $dS\rightarrow M $ transition and a $M \rightarrow AdS$ transition respectively, for different values of the coupling $\xi$. The numerical solutions reveal distinct behaviors for different transitions. In de Sitter to Minkowski transitions ($c=0.05$) (Fig.\hyperref[fig:ds1]{8.a} and Fig.\hyperref[fig:ds2]{8.b}), the scale factor exhibits a second zero, consistent with the findings of Guth and others \cite{Guth:1982pn}, while the bubble grows within the de Sitter hypersphere. For Minkowski to anti-de Sitter transitions ($c=0$) (Fig.\hyperref[fig:M1]{9.a} and Fig.\hyperref[fig:M2]{9.b}), the scale factor approaches $\rho=\eta$ asymptotically, and the resulting anti-de Sitter bubble proves unstable, eventually collapsing as predicted by Coleman and De Luccia \cite{Coleman:1980aw}.

The non-minimal coupling $\xi$ influences these dynamics primarily through the maximum bubble radius, which increases as $\xi$ decreases. However, for the parameter ranges considered in this work, the coupling's influence remains modest, consistent with previous studies \cite{Lee:2005ki,Czerwinska:2017pgb}. These results demonstrate the robustness of vacuum decay processes across different coupling regimes while highlighting the subtle modifications introduced by non-minimal gravitational interactions.

\subsection{Ultra-Late Time Phase Transitions: Bubble Nucleation Radius and Critical Masses}\label{subsec5b}

The requirements for an ultra-late time transition will be the same as in the simple case without gravitation studied previously. That is, $\epsilon \sim 10^{-48}\, GeV^4$ and $\Gamma/V=\mathcal{N}H_0^4 \rightarrow m^4 e^{-S_B}=\mathcal{N}H_0^4$. We will substitute in the equation of $S_B$ dimensionless variables to solve the problem numerically in \textit{Mathematica} for several values of $\xi$ in the interval $[10^{-8},3]$ and the mass scale of the transition $m$ in the regime $[10^{-23}\,eV,10^{-3}\,eV]$. The dimensionless variables are the following
\begin{equation}
 \frac{\lambda\phi^2}{m^2}=\tilde{\phi}^2,\quad
   \frac{\lambda\epsilon}{m^4}=\tilde{\epsilon},\quad
m\rho=\tilde{\rho},\quad
\frac{m^2}{\lambda}\kappa=\tilde{\kappa}.
\end{equation}
We set $\lambda=1$ in our analysis. Before we solve, the dimensionless $\tilde{S}_B$ will be expanded in the small gravity ($\Tilde{\kappa}\rightarrow 0$). The expansion gives
\begin{equation}
    \tilde{S}_B = \frac{1}{2} \pi^2 \left(4    \tilde{ C} \xi \tilde{\bar{\rho}} - \tilde{\bar{\rho}}^4 \epsilon + 4 \tilde{\bar{\rho}}^3 \tilde{S_1}\right) - \frac{1}{12} \tilde\kappa \pi^2 \tilde{\bar{\rho}}^6 \epsilon^2 +\mathcal{O}(\tilde{\kappa}^2).
\end{equation}
Consequently, the equation to solve, in the mass scale we chose is the following
\begin{equation}
    \frac{1}{2} \pi^2 \left(4 \tilde{C} \xi \tilde{\bar{\rho}} - \tilde{\bar{\rho}}^4 \tilde{\epsilon} + 4 \tilde{\bar{\rho}}^3 \tilde{S}_1\right) - \frac{1}{12} \tilde{\kappa} \pi^2 \tilde{\bar{\rho}}^6 \tilde{\epsilon}^2 = \ln{(m^4/H_0^4)}.
    \label{numeqch6}
\end{equation}

In Table \ref{tab:rho_values}, we present the positive solutions for the dimensionless radius $\tilde{\bar{\rho}}$ and the corresponding converted bubble radii $\bar{\rho}$ in $\mu m$, as obtained from equation \eqref{numeqch6} for various values of $\xi$ and different mass scales. The term $\mathcal{O}(\tilde{\kappa})$ has been neglected, given that it is on the order of $\tilde{\kappa} \geq 10^{-62}$. As can be observed, in the context of modified gravity with a non-minimal coupling, the nucleation radius in a present-day phase transition remains of the order of micrometers, similar to the Coleman case where $\bar{\rho}_0 \sim 250\,\mu m$. This holds for values of $\xi$ ranging from $\xi \ll 1$ to $\xi \sim \mathcal{O}(1)$, across a wide range of transition mass scales. Notably, as the mass increases, the solutions become more sensitive to variations in the coupling parameter $\xi$.

Using the solutions for $\bar{\rho}$ (Table \ref{tab:rho_values}), we impose the condition $\bar{\rho}/\rho_c \geq 1$ to determine the critical radius. This allows us to derive an expression for the scalar field critical mass, given by  
\begin{equation}\label{mcritcoupling}
    m_{crit} = \left(\frac{1}{2} \epsilon \bar{\rho} \right)^{1/3} = \left(\frac{1}{2} M_{\text{pl}}^2 H_0^2 \Omega_{\Lambda,0} \bar{\rho} \right)^{1/3}.
\end{equation}  

\begin{figure}
        \centering
\includegraphics[width=1\linewidth]{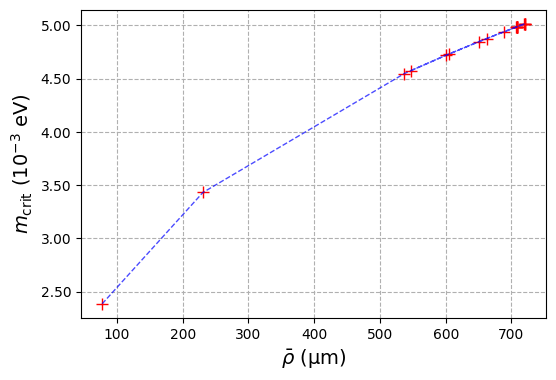}
       \caption{The critical mass plotted as a function of the bubble radius \eqref{mcritcoupling} at the moment of the nucleation. The numerical solutions of Table \ref{tab:rho_values} are indicated with red crosses.}
        \label{mcritvsrho}
\end{figure}

In Fig.\ref{mcritvsrho}, we illustrate equation \eqref{mcritcoupling}, where crosses indicate the numerical solutions for $\bar{\rho}$ obtained from Table \ref{tab:rho_values}. As expected, the order of magnitude of the critical mass remains consistent with the non-gravitational case described by equation \eqref{mcritgr}.

\begin{table}[h]
    \centering
    \renewcommand{\arraystretch}{1.3}
    \begin{tabular}{c|c|c|c}
        \hline
        \hline
        $m$ (eV) & $\xi$ & $\tilde{\bar{\rho}}$ & $\bar\rho\,(\mu m)$ \\
        \hline
        \hline
        $10^{-23}$ & $10^{-8} - 3$ & 2735.47 & 547.094 \\
        \hline
        $10^{-18}$ & $10^{-8} - 3$ & 3027.29 & 605.459 \\
        \hline
        $10^{-13}$ & $10^{-8} - 3$ & 3253.04 & 650.608 \\
        \hline
        $10^{-8}$ & $10^{-8} - 3$ & 3439.67 & 687.934 \\
        \hline
        \hline
        $10^{-5}$ & $10^{-8}$ & 3538.52 & 707.703 \\
        \cline{2-4}
        & $10^{-5}$ & 3538.52 & 707.703 \\
        \cline{2-4}
        & $10^{-3}$ & 3538.52 & 707.703 \\
        \cline{2-4}
        & $10^{-1}$ & 3538.49 & 707.698 \\
        \cline{2-4}
        & $0.2$ & 3538.46 & 707.692 \\
        \cline{2-4}
        & $0.3$ & 3538.43 & 707.686 \\
        \cline{2-4}
        & $1$ & 3538.23 & 707.646 \\
        \cline{2-4}
        & $3$ & 3537.08 & 720.531 \\
        \hline
        \hline
        $10^{-3}$ & $10^{-8}$ & 3600.08 & 720.016 \\
        \cline{2-4}
        & $10^{-5}$ & 3600.05 & 720.01 \\
        \cline{2-4}
        & $10^{-3}$ & 3597.3 & 719.46 \\
        \cline{2-4}
        & $10^{-1}$ & 3311.21 & 662.346 \\
        \cline{2-4}
        & $0.2$ & 3004.21 & 600.841 \\
        \cline{2-4}
        & $0.3$ & 2685.1 & 537.019 \\
        \cline{2-4}
        & $1$ & 1154.18 & 230.836 \\
        \cline{2-4}
        & $3$ &388.782 & 77.7563\\
        \hline
        \hline
    \end{tabular}
    \caption{Bubble nucleation dimensionless radius $\tilde{\bar{\rho}}$ as solution of \eqref{numeqch6} and its conversion to $\bar\rho\, (\mu m)$ for different mass scales $m$ and $\xi$.}
    \label{tab:rho_values}
\end{table}

\section{Post-Transition Bubble Scale.}\label{sec6}

\subsection{Bubble Nucleation and Expansion Dynamics}\label{subsec6a}

Late-time first-order phase transitions offer a potential mechanism to break the cosmological principle by introducing non-trivial local effects in the universe. During such a transition, bubbles of true vacuum nucleate and expand within the false vacuum background. At a given time \( t \), the fraction of the spatial volume occupied by these bubbles is expressed as \( F = 1 - p \), where \( p \) represents the fraction of the volume that remains in the false vacuum phase throughout the transition process \cite{Guth:1981uk}. Then

\begin{equation}\label{hoganpi}
    p=e^{-I(t)}
\end{equation}
where
\begin{align}\label{hoganI}
    I(t) &= \left[\int_{t_0}^{t} dt'\,\Gamma(t')a^3(t') V(t,t')\right] \nonumber \\
         &= \frac{4\pi}{3}\left[\int_{t_0}^{t} dt'\,\Gamma(t')a^3(t') r^3(t,t')\right].
\end{align}

The function $V(t,t')$ represents the comoving volume of a bubble at time $t$ that nucleated at time $t'$ with an initially negligible radius and subsequently expanded at approximately the speed of light. In equation \eqref{hoganI}, $\Gamma(t)$ denotes the rate of bubble production per unit proper volume, with the explicit volume element in the denominator suppressed for simplicity. The term $a(t)$ corresponds to the standard FLRW scale factor. 

In the context of a time-dependent transition, three distinct regimes can be identified, characterized by the ratio $\Gamma/H$. Specifically, $p$ transitions from $p \gg 1$ to $p \sim \mathcal{O}(1)$ and eventually to $p \ll 1$. Considering the first extreme, where $p \gg 1$, an observer placed in a universe undergoing a slow and incomplete phase transition—while not being inside any bubble, which is highly probable in this regime—would observe a standard FLRW universe that appears homogeneous and isotropic. Conversely, in the unlikely case where the observer resides within a bubble, their observations would reveal an inhomogeneous and generally anisotropic universe. On the other extreme, for $p\ll 1$ the phase transition has already been completed and the observer would perceive a universe with a distinct expansion history from the base $\Lambda CDM$ model. 

The function $I(t)$ remains approximately valid until the time $t_1$, at which the fraction of space occupied by the new phase, $F$, approaches unity, that is, $F(t_1) \approx 1$. At this stage, the universe is almost entirely in its true vacuum state, corresponding to $p \ll 1$. An appropriate approximation for $F$ at times $t \leq t_1$ is given by
\begin{equation}
     F \approx \int_{0}^{t} dt' \, \Gamma(t') a^3(t') V(t,t').
     \label{eq.5.32}
\end{equation}
 We define the characteristic time of the transition as $\delta \tau = (F/\dot{F})_{t_1}$, representing the period during which most of the matter transitions from the false vacuum phase to the true vacuum phase. Since a bubble expands nearly at the speed of light ($c=1$), the characteristic time $\delta \tau$ also corresponds to the typical radius of the bubbles at the end of the transition, i.e., $R_b = c \delta \tau = \delta \tau$. As $\dot{F}$ increases, the rate at which bubbles occupy space accelerates, leading to a decrease in the characteristic time $\delta \tau$. In general, $\delta \tau$ represents the time available for bubbles to nucleate and expand at the speed of light and satisfies the condition $\delta \tau \ll t$, where $t$ is the cosmic time. 

For a monotonically and rapidly increasing nucleation rate $\Gamma$, typically described by $\Gamma \sim e^{-S}$, it can be shown \cite{Hogan:1983ixn} that
\begin{equation}
 \delta \tau \approx \left( \frac{\Gamma}{\dot{\Gamma}} \right)_{t_1} = -\frac{1}{\dot{S}(t_1)}.
   \label{eq.5.36}
\end{equation}
In this context, we adopt the following general functional form for the exponential term in the nucleation rate as a function of temperature, suitable for both thermal and quantum tunneling scenarios \cite{Hogan:1983ixn}:
\begin{equation}
    S = s(T)\left(\frac{T_c}{T} - 1\right)^{-\sigma},
    \label{eq.5.29}
\end{equation}
where $\sigma > 0$, $s > 0$, and $d \ln s / d \ln T > 0$, with this last condition typically being of $\mathcal{O}(1)$ if the second factor is not significantly greater than one. Although $\sigma$ is assumed to be constant, the model remains effective if $\sigma$ is allowed to vary slowly under certain restrictions. The critical temperature $T_c$ marks the point at which the universe's potential becomes metastable, initiating the phase transition. As the temperature approaches $T_c$, the action $S$ diverges, i.e., $S \rightarrow \infty$ as $T \rightarrow T_c$.

To compute the value of the action $S(t_1) \equiv S_1$ at the time $t_1$, when the first catastrophic bubble forms, we can approximate the integral $F_1$ as follows:
\begin{equation}
    F(t_1) = C \left(\frac{4\pi}{3}\right) (T \delta \tau)^4 e^{-S_1} \sim 1,
    \label{eq.5.38}
\end{equation}
where $C$ is a numerical parameter. From equation \eqref{eq.5.38}, we obtain
\begin{equation}
     S_1 \approx 4 \ln \left[ T \delta \tau \left( \frac{4\pi C}{3} \right)^{1/4} \right].
     \label{eq.5.39}
\end{equation}

A theoretical prediction for the spatial extent of the gravitational constant bubbles at the end of the phase transition can be derived using the formula provided in \cite{Hogan:1983ixn}:
\begin{equation}
    \Delta \approx \left[ 4 B_1 \ln \left( \frac{M_{Pl}}{T} \right) \right]^{-1},
    \label{eqhogan}
\end{equation}
where $B_1$ represents the logarithmic derivative of the action in units of the cosmological time at the time $t_1$ of catastrophic bubble formation. It can be shown that $B_1$ is typically of $\mathcal{O}(1)$ \cite{Hogan:1983ixn}. The parameter $B_1$ is defined as
\begin{equation}\label{logderiv}
    B = -t \left( \frac{\dot{S}}{S} \right) = -\frac{d \ln S}{d \ln t}.
\end{equation}
Finally, $\Delta$ corresponds to the ratio of the typical bubble size to the cosmic time at the end of the phase transition:
\begin{equation}\label{deltascale}
    \Delta = \frac{\delta \tau}{t} \equiv \frac{1}{S_1 B_1}.
\end{equation}

\subsection{Post-Predicted Bubble Scales and Observational Implications}\label{subsec6b}

Although equation \eqref{eqhogan} is strictly valid only during radiation-dominated epochs, where $t \propto T^{-2}$, the discrepancies between this approximation and a more precise expression are minimal. These differences appear in the logarithm argument and become negligible because $M_{Pl} \gg T$ \cite{Patwardhan:2014iha}. In the case of a very recent false vacuum decay—occurring, for example, between $50$ and $300\, Myrs$ ago—with vacuum energy comparable to the cosmological constant, the characteristic scale of the produced bubbles can be approximated by
\begin{equation}
    R_b \approx \frac{\Delta}{H_0},
    \label{eq.5.52}
\end{equation}
where $T = 2.7^\text{o}\,K \sim 2 \times 10^{-4}\,eV$ is used as the temperature of the photon background at the onset of the phase transition. Additionally, $B_1$ is assumed to be of $\mathcal{O}(1)$. Under these conditions, we have $\Delta \sim 1/(300 B_1)$, and by employing $H_0 = 70\, km\,s^{-1}\,Mpc^{-1}$ in equation \eqref{eqhogan}, we obtain the following estimate for the bubble radius:
\begin{equation}\label{scale}
    R_b \sim 20 \times B_1^{-1} \, Mpc.
\end{equation}

For instance, considering a phase transition lasting between $25$ and $500\, Myrs$, one can determine, with the aid of the analysis presented in \cite{Hogan:1983ixn}, that $B_1$ lies within the range $0.1 \lesssim B_1 \lesssim 2.23$. Substituting these values into equation \eqref{scale} yields a range for $R_b$ between $8.5\, Mpc$ and $120\, Mpc$, corresponding to redshifts in the interval $z \in [0.0055, 0.027]$ at the conclusion of the transition.

An important question arises: how can we be confident that $B_1$ is indeed of $\mathcal{O}(1)$ for our computations within a specific theoretical framework? While we have numerically computed its value for certain choices of the characteristic time $\delta \tau$ using equation \eqref{deltascale}, it is also instructive to explore the analytical behavior of $B_1$ within a general class of theories. This analytical investigation will help establish the consistency of our numerical estimates and provide deeper insights into the mathematical structure underlying these cosmological phase transitions.

In his classical, non-gravitational analysis \cite{Coleman:1977py}, Coleman computed the action
\begin{equation} \label{colemanaction}
S_{Col} = 6 \pi^2 m^{12} \lambda^{-4} \epsilon^{-3},
\end{equation}
where $m$ and $\lambda$ are the mass and coupling constant, respectively, in a symmetry-breaking model of the Higgs field, and $\epsilon$ denotes the usual energy difference between the two vacuum states. If the parameters defining the potential are assumed to vary with the background temperature, with exact degeneracy occurring at the critical temperature $T_c$, we can write \cite{Hogan:1983ixn}
\begin{equation}\label{hoganepsilon}
    \epsilon \propto (T_c - T)^{-\beta},
\end{equation}
where $\beta$ is a positive critical exponent. By employing the definition of $B$ as a logarithmic derivative, given by equation \eqref{logderiv}, we find
\begin{equation}\label{Bcol}
B \equiv \frac{3\beta}{2} \left( \frac{T_c}{T} - 1 \right)^{-1}.
\end{equation}

It is evident from this expression that for temperatures close to the critical temperature $T_c$, the value of $B$ becomes significantly larger than unity, $B \gg 1$. As the temperature decreases, $B$ gradually reduces to values of order unity. Consequently, as the phase transition proceeds, time advances towards $t_1$, and $B$ approaches $B_1$. 

The natural question that arises is: how does $B$ behave in a modified theory of gravity? To explore this, we compute $B$ in the context of the theory discussed in the previous section. Neglecting gravitational effects, as calculated in equation \eqref{graveffect}, one can express the action—after expanding to first order in $\xi$—as \cite{Lee:2005ki}
\begin{equation}\label{llaction}
    S = S_{Col} + \mathcal{O}(\xi) = S_{Col} - 100 \xi m^4 \lambda^{-1} \epsilon^{-1}.
\end{equation}

In this case, using the definition of $B$, the relation $t \propto T^{-2}$, and the temperature dependence $\epsilon \propto (T_c - T)^{-\beta}$, we obtain
\begin{equation}
   B_{mod} = \left( \frac{T_c}{T} - 1 \right)^{-1} \frac{\beta}{2} \left[ \frac{18 \pi^2 (T_c - T)^{2\beta} - 100 \xi}{6 \pi^2 (T_c - T)^{2\beta} - 100 \xi} \right],
\end{equation}
where, as expected, the result reduces to equation \eqref{Bcol} when $\xi = 0$.

\begin{figure}[h]
    \centering
    \includegraphics[width=0.4\textwidth]{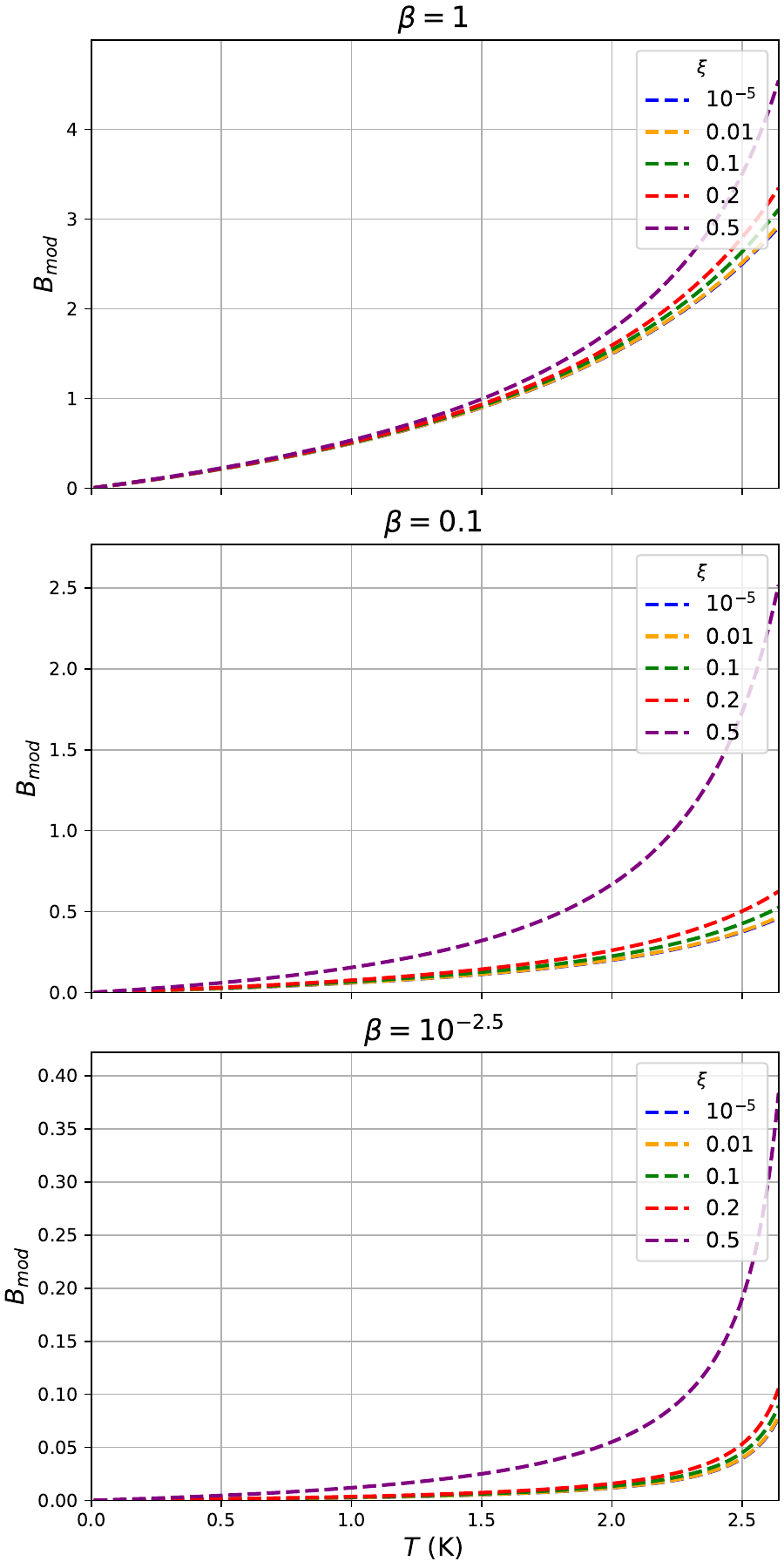}
    \caption{Logarithmic derivative $B_{mod}$ for the modified gravity theory in eq.\eqref{llaction}. The different graphs are plotted for $T_c=2.8\,K$ and diffrenet values of $\xi$ and $\beta$.  The transition is set around $300\,Myrs$ ago and finished in 150 million years $(T=2.64\,K)$, with $B_1$ to be $\mathcal{O}(1)$.}\label{fig:modifiedgravity}
    \end{figure}

In Fig.\ref{fig:modifiedgravity} we can observe the behavior of $B_{mod}$ illustrated for different values of $\beta$ and $\xi$. For $\xi\ll 1$ the behavior is the same as that of \eqref{Bcol}. The graphs are plotted for $T_C=2.8\, K$, a transition that was triggered about 300 million years ago and will be finished in 150 million years, that is, $T=2.64\,K$. $B_{mod}$ is of the order of one at temperature $T=2.64\,K$ where the transition is (or almost) complete. 

\begin{figure}[h]
    \centering
\begin{tikzpicture}
    \draw[fill=cyan, thick] (0,0) circle [radius=3cm];

    \draw[fill=green, thick] (0,0) circle [radius=1.5cm];

    \draw[fill=blue] (0,0) circle [radius=0.1cm];
    \node at (0, -0.3) {Solar System};

    \draw[red, thick, ->] (0,0) -- (1.5,0); 
 \node[above, black] at (0.75, 0) {\fontsize{8}{8}\selectfont 50 Mpc};

   \draw[red, thick, ->] (0,0) -- (0,3); 
\node[left, black,rotate=90] at (-0.2, 2.2) {100 Mpc}; 
\end{tikzpicture}
 \caption{We could potentially live inside a bubble of size $100\,Mpc$, nucleated 300 million years ago. But we could potentially observe of bubble signs at its half of expansion, when the bubble was around $50 \,Mpc$, 100 milion years ago, because of the finite speed of light.}
    \label{bubblecircles}
\end{figure}
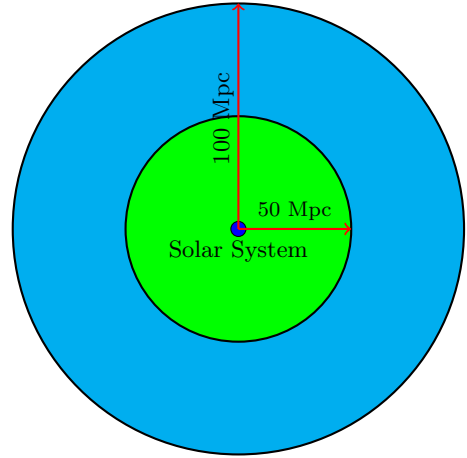

An interesting assumption, drawn from the existing literature and directly related to our analysis, is the possibility that we could be living near the center—or in the region surrounding the center—of a bubble characterized by a different gravitational constant (or cosmological constant). This hypothetical bubble may have nucleated with a negligible initial radius approximately $300\,Myrs$ ago, give or take, and after expanding at the speed of light, could have reached a radius of $90$–$100\,Mpc$ today. However, due to the nature of cosmic observations, we would only be able to observe its spatial extent up to $40$–$50\,Mpc$. This observational limitation arises because, as we look deeper into the Universe, we are effectively looking further back in time (see Fig.\ref{bubblecircles}).

For instance, if we conduct observations at redshift $z_{obs}=0.01$ (corresponding to a distance of approximately $50\,Mpc$), we would be observing the state of the bubble as it was about $150\,Myrs$ ago, when its radius was around $50\,Mpc$. On the other hand, if we extend our observations out to $100\,Mpc$, we would detect no anomalies and observe an isotropic Universe. This occurs because, at that greater distance, we are essentially seeing the Universe as it was $300\,Myrs$ ago, when the bubble had just nucleated at a negligible scale—likely within the vicinity of our solar system—and had not yet expanded to affect that distant region. This phenomenon beautifully illustrates the principles of relativity, highlighting how the finite speed of light shapes our perception of cosmic events.

An alternative scenario to this gradual expansion model could involve a bubble that nucleated at a much larger scale—on the order of $50\,Mpc$—relatively recently, perhaps $100\,Myrs$ ago or less. However, our theoretical calculations consistently predict a nucleation radius on the scale of micrometers ($\mu m$), rendering such a large initial nucleation scale unlikely. 

In a subsequent section (\ref{sec8}), we will explore observational data and recent literature in search of evidence for such bubbles that could signal new physics. The confirmation of a first-order phase transition scenario of this nature could provide a compelling solution to the ongoing Hubble tension problem.

Let us extend our previous hypothesis a step further. Suppose that we are currently located at the center (or around the center) of a bubble that nucleated $x$ million years ago with an initial radius on the scale of micrometers ($\mu m$), as demonstrated in a previous section. After $x$ million years of expansion at nearly the speed of light, this bubble would now have reached its actual spatial extent, $R_{actual}$, and may still be expanding. 

The only observational evidence we have for the existence of such a bubble comes from detected anisotropies in cosmological observations, such as the Hubble tension, observed at a specific redshift $z_{obs}$. The crucial question that arises is: can we, based solely on this observational clue, determine or constrain the current size of the bubble, $R_{actual}$? 

This question is significant because it connects theoretical predictions about late-time phase transitions and bubble nucleation with observable cosmological signatures. If a relationship can be established between the observed anisotropies at $z_{obs}$ and the actual spatial extent $R_{actual}$, it would provide a powerful tool for testing scenarios involving first-order phase transitions in the late Universe.

The radius \( R_b \) of a bubble after a time interval \( \delta t \) is given by: 
\begin{equation}
R_b = c \, \delta t = \delta t,
\end{equation}
where \( c = 1 \). This implies that the comoving radius \( R_b \) is directly proportional to the elapsed time \( \delta t \).

The comoving distance \( D_c(z) \) to an object at redshift \( z \) in a flat universe is:
\begin{equation}
D_c(z) = \int_0^z \frac{c \, dz'}{H(z')}.
\end{equation}
For \( c = 1 \), this simplifies to:
\begin{equation}
D_c(z) = \int_0^z \frac{dz'}{H(z')}.
\end{equation}
This comoving distance \( D_c(z_{obs}) \) represents the distance from us to a redshift \( z_{obs} \).

Assuming that the radius \( R_{obs} \) of the bubble is equal to the comoving distance \( D_c(z_{obs}) \), we can write:
\begin{equation}
R_{obs} \approx \delta t_{obs} \approx D_c(z_{obs}),
\end{equation}
which implies that \( \delta t_{obs} \) can be calculated from the comoving distance to the observed redshift \( z_{obs} \):
\begin{equation}
\delta t_{obs} = \int_0^{z_{obs}} \frac{dz'}{H(z')}.
\end{equation}

The Hubble parameter \( H(z) \) in a flat universe, consisting of matter and dark energy, is given by
\begin{equation}
    H^2(z)=H_0^2[\Omega_{m,0} (1+z)^3 + \Omega_{\Lambda,0}].
\end{equation}
Evaluated today:
\begin{equation}
    \Omega_{m,0}+ \Omega_{\Lambda,0} = 1.
\end{equation}
Therefore substituting for $\Omega_{\Lambda,0}$:
\begin{equation}
     H^2(z)=H_0^2[\Omega_{m,0} (1+z)^3 + 1-\Omega_{m,0}]
\end{equation}
Plugging this expression into the integral, we find:
\begin{equation}
\delta t_{obs} = \int_0^{z_{obs}} \frac{dz'}{H_0 \sqrt{\Omega_{m,0}(1 + z')^3 + 1 - \Omega_{m,0}}}.
\end{equation}

This integral provides the time \( \delta t_{obs} \) associated with the observed redshift \( z_{obs} \). For specific values of \( z_{obs} \), \( \delta t_{obs} \) can be numerically evaluated. For example, for \( z_{obs} = 0.01 \), we find \( \delta t_{obs} \sim 139 \, Myrs \) and \( R_{obs} \sim 45 \, Mpc \). Consequently, the actual size of the bubble today, after approximately \( 280 \) million years of nucleation, would be \( R_{actual} = 2 \delta t_{obs} \sim 90 \, Mpc \).

The nucleation of a bubble within a single Hubble volume, especially one positioned near our solar system, may initially appear to be an extremely improbable event. However, consider an alternative scenario where \( \mathcal{N} \) nucleation events occur throughout space. In such a case, neighboring bubbles could be located at a distance \( d \) from our own. Naturally, we would not expect to detect any observational signals from these distant bubbles by simply observing those regions of the Universe. In this simplified framework, the nucleation of neighboring bubbles is assumed to occur at roughly the same time as that of our own bubble. 

As we observe deeper into the cosmos, we look back in time, potentially before the nucleation of these bubbles. However, since the walls of these bubbles propagate at the speed of light, there will come a point when neighboring bubbles collide and merge. This collision process will produce gravitational waves \cite{Hawking:1982ga,Turner:1992tz,Kosowsky:1992rz,Kosowsky:1992vn,Kosowsky:2001xp,Giblin:2010bd,Gogoberidze:2007an,Caprini:2006jb,Caprini:2009yp}, eventually filling the space with the new vacuum state. The completion of this transition occurs after a characteristic time interval equal to \( \delta \tau \).

According to \cite{Hawking:1982ga,Turner:1992tz,Kosowsky:1992rz,Kosowsky:1992vn,Kosowsky:2001xp}, the duration of the phase transition \( \delta \tau \) (and the distance \( d \) between the centers of adjacent bubbles) can be parametrized in terms of the mass scale associated with the field responsible for the transition.

To begin, the action is expanded around the time \( t_* \), which marks the completion of the phase transition:
\begin{equation}
 S(t) = S_* - \zeta (t - t_*),
\end{equation}
where the parameter \( \zeta \) is defined as:
\begin{equation}
    \zeta = -\left( \frac{d \ln{\Gamma}}{dt} \right)\Bigg|_{t_*} = \left( \frac{\partial S}{\partial t} \right)\Bigg|_{t_*}.
\end{equation}
The parameter \( \zeta \) plays a central role, as it sets the characteristic duration and length scale of the phase transition.

Previously, we defined the probability \( p \) that a given point remains in the false vacuum, as shown in equation \eqref{hoganpi}, where \( I(t) \) is the integral in the exponent representing the expected fraction of space occupied by true-vacuum bubbles at time \( t \).

For simplicity and consistency with earlier assumptions, we neglect the expansion of the Universe by approximating \( \alpha r(t, t') = t - t' \) in the expression for \( I(t) \). This approximation remains valid as long as the transition duration is significantly shorter than \( H^{-1} \). Additionally, it is assumed that bubbles expand to sizes much larger than their initial nucleation radius, a condition that generally holds for the scenarios considered here. Allowing \( t_0 \to -\infty \) without introducing significant error, we obtain:
\[
I(t) \approx \frac{8 \pi}{\zeta^4} \Gamma(t).
\]

Thus, the fraction of space remaining in the false vacuum can be written as:
\[
p(t) = e^{-I(t)} \approx e^{- \frac{8 \pi \Gamma(t)}{\zeta^4}},
\]
where the exponential factor accounts for bubble overlap. From this expression, the duration of the phase transition can be estimated. Although the precise definitions of the start and end of the transition are somewhat arbitrary, this uncertainty does not significantly affect the overall results.

To provide a working definition, we define the start of the transition at time \( t_{c_2} \), where \( p(t_{c_2}) = e^{-c_2} \approx 1 \), with \( c_2 \sim 0.01-1 \). Similarly, the end of the transition is defined at time \( t_* \), where \( p(t_*) = e^{-c_1} \approx 0 \), with \( c_1 \sim 10-30 \) \cite{Turner:1992tz,Kosowsky:1992rz,Kosowsky:1992vn,Kosowsky:2001xp}. Consequently, the total duration of the phase transition is given by:
\begin{equation}
    \delta \tau = t_* - t_{c_2} = \ln\left( \frac{c_1}{c_2} \right) \zeta^{-1}.
\end{equation}
This expression shows that \( \delta \tau \) depends logarithmically on the definitions chosen for the beginning and end of the transition.

To understand the relationship between the key parameter \( \zeta \), the Hubble parameter \( H \), and the mass scale \( m \), we note that \( H^{-1} \) sets the characteristic timescale for all cosmological dynamics. On general grounds, \( \zeta = -\left( \partial S / \partial t \right)_{t_*} \) is expected to be of the order of \( S(t_*) / H^{-1} \), or equivalently \( \zeta^{-1} \sim H^{-1} / S_* \).

For the phase transition to proceed via vacuum bubble nucleation, \( S_* \) must satisfy \( S_* \gg 1 \). This ensures that the transition occurs "rapidly," meaning \( \zeta^{-1} \ll H^{-1} \), a condition typically satisfied in these scenarios. Additionally, we can estimate \( S_* \) using the relation \( \Gamma(t_*) \sim m^4 e^{-S_*} \sim H^4 \sim m^8 / M_{Pl}^4 \). From this, we find that \( S_* \) should be approximately equal to \( \ln(M_{Pl} / m) \). Substituting this into the expression for \( \delta \tau \), we obtain:
\begin{equation}\label{deltatafparam}
    \delta \tau = \ln{\left( \frac{c_1}{c_2} \right)} \frac{H(z_{obs})^{-1}}{\ln{\left( \frac{M_{Pl}}{m} \right)}}.
\end{equation}

\begin{figure}
    \centering
    \includegraphics[width=1\linewidth]{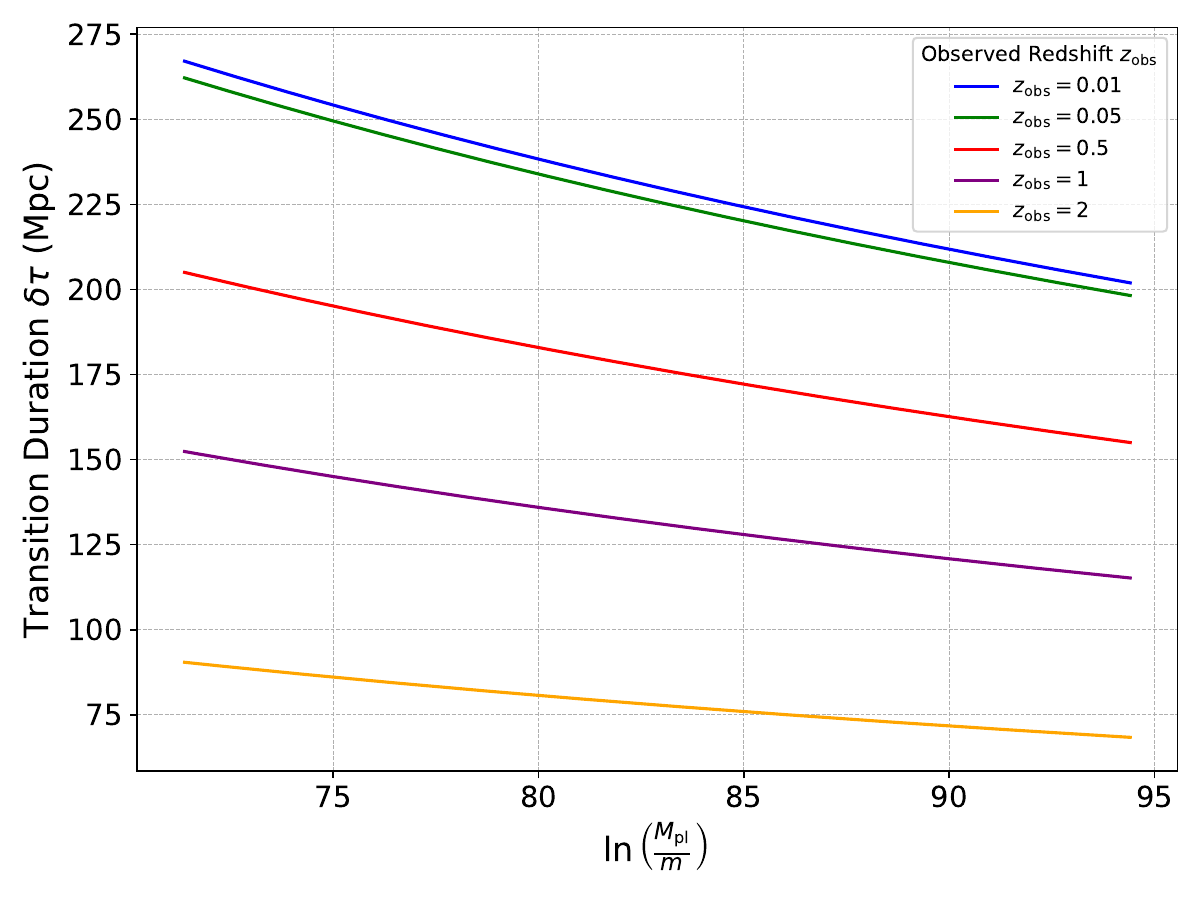}
   \caption{Transition Duration $\delta \tau$ as a Function of $\ln{(M_{Pl}/m)}$, for different values of the observation redshift $z_{obs}$. The mass ranges from $10^{-13}\,eV$ to $10^{-3}\,eV$. The figure is plotted for $c_1=12$ and $c_2=0.1$.}
    \label{deltatauparam}
\end{figure}

In Figure \ref{deltatauparam}, equation \eqref{deltatafparam} is illustrated for \( c_1 = 12 \) and \( c_2 = 0.1 \), across five different values of the observational redshift \( z_{obs} \) and mass ranges \( m = 10^{-13} - 10^{-3} \, eV \). The results indicate that the expected timescale of the phase transition, in the observational region of interest (\( z_{obs} \sim 0.01 \)), is on the order of hundreds of millions of years (\( \sim 600-810 \, Myrs \)). As \( z_{obs} \) increases, the transition duration decreases by a few tens of millions of years. Due to the logarithmic dependence on \( c_1 \) and \( c_2 \), the order of magnitude of \( \delta \tau \) remains unchanged, and its value does not shift significantly when these parameters are varied within reasonable ranges. For example, for \( c_1 = 10 \) and \( c_2 = 0.001 \), we obtain \( \delta \tau = 280-350 \, Myrs \) for \( z_{obs} = 0.01 \) within our considered mass range. For \( c_1 = 30 \) and \( c_2 = 0.001 \), the transition duration is found to be \( \delta \tau = 450-550 \, Myrs \).

In \cite{Kosowsky:1991ua}, Kosowsky, following the previously discussed analysis, provides an expression for the expected scale of the phase transition duration \( \delta \tau \) and the separation distance \( d \) between adjacent bubbles:
\begin{multline}\label{sepdist}
    \delta \tau \sim \frac{d}{2} \sim \mathcal{A} \times 10^{-2} H^{-1} \sim 10^{-2} \frac{M_{Pl}}{\sqrt{\lambda} \phi_t^2} \\
\sim  10^{-2} \left( \frac{M_{Pl}}{\phi_t} \right) \rho_{nucl} \, .
\end{multline}

\begin{figure}
    \centering
    \includegraphics[width=.75\linewidth]{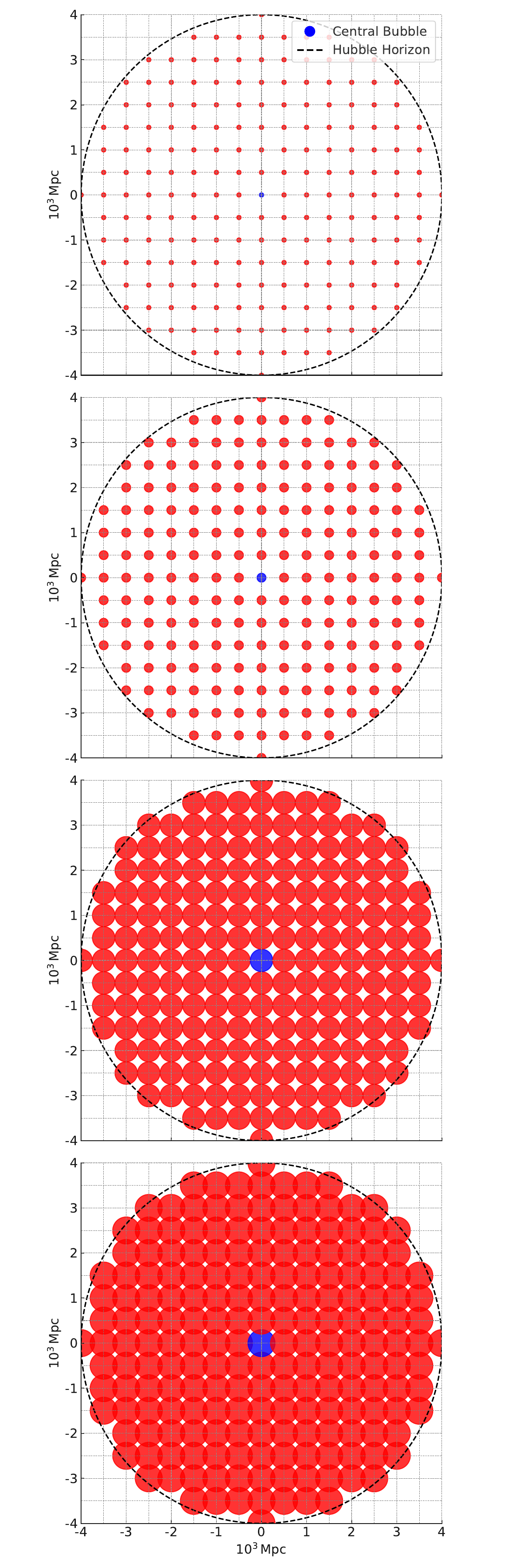}
     \caption{Two-dimensional visualization of a hypothetical phase transition beginning approximately $300 \, Myrs$ ago, with bubbles separated by $d = 500 \,Mpc$. (Top) Bubbles observed at $z_{obs} = 0.01$, with observed radii $R_{obs} \sim 50 \, Mpc$. (Second) Bubbles at their actual sizes today, $R_{actual} \sim 100 \, Mpc$, occupying a significant portion of the Hubble plane. (Third) Bubbles at the stage where their walls first collide, $R_b \sim 250\,Mpc$, consistent with the Hogan approximation, where $F_1 \sim 1$ after $\delta \tau \sim 750\,Myrs$. (Bottom) Final snapshot shows bubbles nearly filling the entire Hubble plane ($F \cong 1$). For $R_b = 0.6d$, the total number of nucleation events in this 2D depiction is $\mathcal{N} = 197$ bubbles.}
    \label{snapshots}
\end{figure}

Here, \( \mathcal{A} \) is a parameter that can be of order unity or larger and can be appropriately constrained. For example, assuming \( \mathcal{A} \sim 6 \), with \( \lambda = 0.07 \), \( \rho_{nucl} \sim 220 \, \mu m \), and \( \phi_t \sim m_{crit} \sim 10^{-3}\,eV \), we obtain \( \delta \tau \sim 250\,Mpc \). This corresponds to a separation distance of approximately \( 500\,Mpc \)\footnote{We assume \( \delta \tau \sim d/2 \) under the approximation of a square grid distribution of the bubbles, as illustrated in Fig.\ref{snapshots}.}.

Next, we justify the choice of \( \beta \sim 10^{-2.5} \), which leads to \( B_1 \sim 0.08 \) in Fig.\ref{fig:modifiedgravity}. Computing \( B_1 \) from equation \eqref{deltascale} for \( \delta t = 750\,Myrs \) yields \( B_1 \sim 0.07 \), resulting in \( R_b \sim 285\,Mpc \) using equation \eqref{scale}. This estimate is in good agreement with the value \( R_b = 250\,Mpc \), obtained using \( B_1 = 0.08 \) from Fig.\ref{fig:modifiedgravity} and equation \eqref{sepdist}. Thus, by employing a theoretical calculation of \( \rho_{nucl} \) within a specific framework, we achieve consistent approximations from three independent approaches: equation \eqref{scale}, equation \eqref{sepdist}, and Fig.\ref{fig:modifiedgravity}, following the model proposed in \cite{Hogan:1983ixn}.

Figure \ref{snapshots} provides a two-dimensional visualization of a hypothetical phase transition that began approximately \( 300 \, Myrs \) ago, with bubbles distributed at a separation distance \( d = 500 \, Mpc \). The first snapshot (top) displays the bubbles at the moment of observation corresponding to \( z_{obs} = 0.01 \), where their observed radii are approximately \( R_{obs} \sim 50 \, Mpc \). In the second snapshot, the bubbles are shown at their actual current sizes, with radii nearing \( 100 \, Mpc \), at which point they occupy a significant portion of the Hubble plane. The third snapshot shows the bubbles at the stage where their walls first come into contact, occurring at \( R_b \sim 250\,Mpc \), which validates the Hogan approximation during this phase of the transition process. At this point, \( F_1 \sim 1 \) after \( \delta \tau \sim 750\,Myrs \) of expansion. Finally, the last snapshot shows the bubbles almost filling the entirety of the Hubble plane, with \( F \cong 1 \) ($R_b=0.6d$). Given the initial separation distance \( d = 500 \, Mpc \) and the relationship \( R_b = d/2 \), the total number of nucleation events in this two-dimensional scenario is \( \mathcal{N} = 197 \) bubbles.

It is worth noting that the nucleation times of these bubbles are nearly identical. Even if these times differ by a few million years, this small discrepancy does not significantly affect the validity of the approximation. Over the course of hundreds of millions of years of expansion, the radii of the bubbles would remain nearly uniform.

For simplicity, we assume a square grid configuration for the placement of the bubbles, ensuring a uniform filling of the Hubble plane. The distances between neighboring bubbles were set to \( d \) and \( \sqrt{2} \, d \). Naturally, our own bubble is located at the center of the Hubble plane in this depiction.

In conclusion, if our bubble nucleated approximately \( 300\,Myrs \) ago, along with the other 200 bubbles, they will complete the transition after about \( 450\,Myrs \). This suggests that we are currently situated roughly at the midpoint of the entire transition process.

\section{Variation of the Effective Gravitational Constant in Modified Gravity: Theory and Observational Constraints.}\label{sec7}

So far, we have determined the bubble radius both at the moment of nucleation and after its expansion within the framework of a modified theory of gravity. In addition, we have computed the critical mass of the bubbles and derived a relation of the transition duration to the mass scale of the catastrophic event. Now, it is crucial to examine how the effective gravitational constant evolves in these theories and how this aligns with observational constraints.

We define the quantity $\Delta G$ as follows:
\begin{eqnarray}\label{deltaG}
    \Delta G= G_{eff,f}-G_{eff,t}=X G_N.
\end{eqnarray}
This represents the difference between the effective gravitational constant in the false vacuum and its value in the true vacuum. This difference corresponds to a fraction of the local Newtonian gravitational constant \( G_N \). Living in a true vacuum bubble means $G_{eff,t}=G_N$. 

To analyze this behavior, we consider a general class of modified gravity theories where the coupling function takes the form 
\begin{equation}
    F(\phi)=\frac{1}{\kappa}-\xi \phi^n.
\end{equation}
Previously, we have examined the specific case where \( n=2 \). Within this family of theories, the parameter \( \xi \) carries a dimensionality given by  
\begin{eqnarray}
    [\xi]=[eV]^{2-n}.
\end{eqnarray}
So, in the case $n=2$, $\xi$ is dimensionless.

From \eqref{5.55} the effective gravitational constant is going to be:
\begin{equation}
    G_{eff} \approx \frac{1}{F(\phi)}=\frac{\kappa}{1 - \kappa \cdot \xi \cdot \phi^n}
\end{equation}
The change in the gravitational constant \( G_{eff} \) due to the influence of the scalar fields is given by:
\begin{equation}
\Delta G = \frac{\kappa^2 \xi \left( \phi_{f}^n - \phi_{t}^n \right)}{(1 - \kappa \xi \phi_{f}^n) \left( 1 - \kappa \xi \phi_{t}^n \right)}
\end{equation}
where \eqref{deltaG} was used. First order to $\xi$, $\Delta G$ becomes
\begin{eqnarray}\label{deltaG2}
    \Delta G=\kappa^2 \xi\left( \phi_{f}^n - \phi_{t}^n \right).
\end{eqnarray}
If we take a potential of the form below
\begin{equation}
    U(\phi) = \frac{1}{8} \left( \phi^2 - \phi_t^2 \right) \left( \phi^2 - \phi_{f}^2 \right) \pm \frac{\epsilon}{2m} (\phi - \phi_t)
\end{equation}
with $\phi_t=m$ and $\phi_f=m(1+\epsilon/m^4)$. Then plugging the values of the false and true vacuum in \eqref{deltaG2} we derive after some manipulation the quantity of interest:
\begin{equation}\label{deltaG_GN}
    \frac{\Delta G}{G_N} = 64 \pi^2 \frac{m^n}{M_{Pl}^2} \, \xi \left( \left(1 + \frac{H_0^2 M_{Pl}^2\Omega_{\Lambda,0}}{m^4}\right)^n - 1 \right)
\end{equation}

\begin{figure}
    \centering
    \includegraphics[width=1\linewidth]{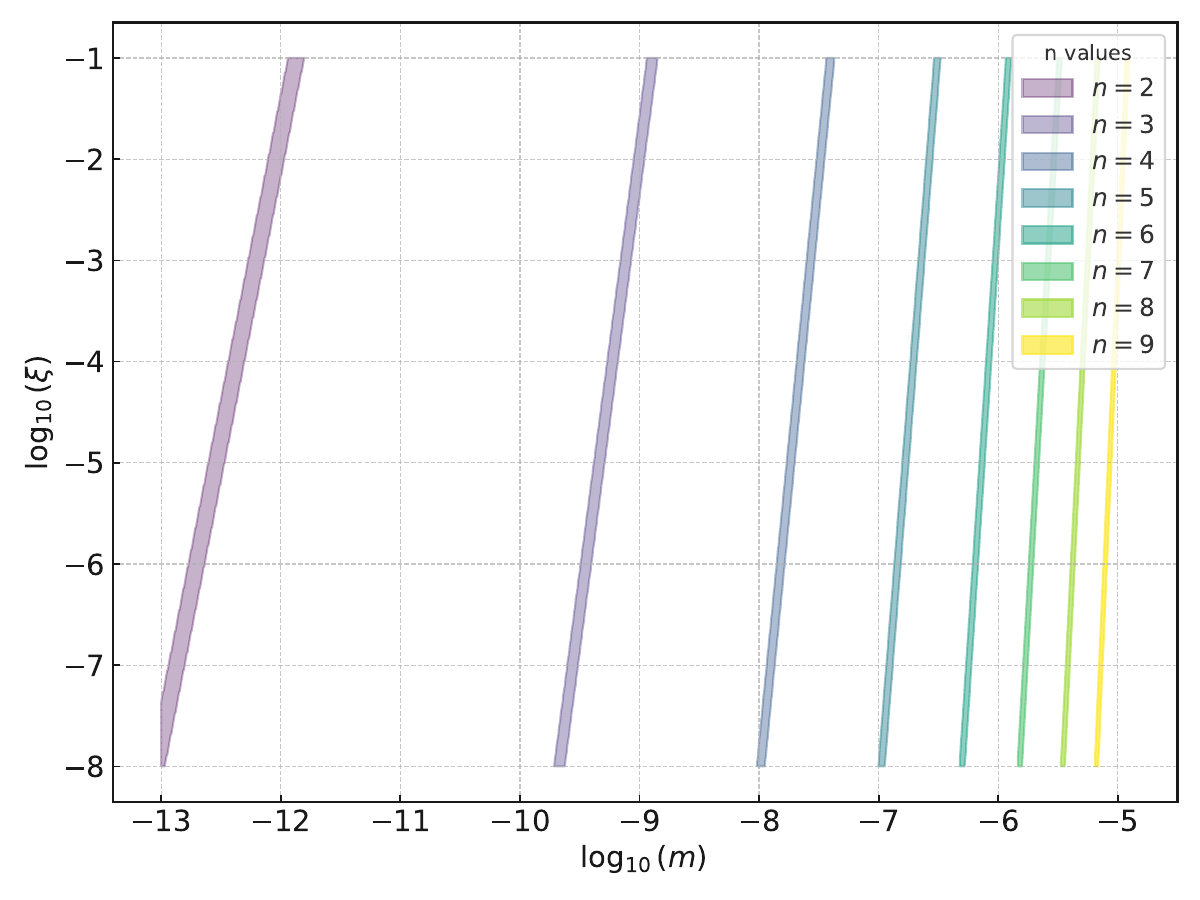}
    \caption{Areas with $\Delta G/G_N$ fixed to $1\%-7\%$ with $n$ running from 2 to 9. The axes are logarithmical. The coupling term $\xi$ ranges from $10^{-8}\, eV^{2-n}$ to $10^{-1}\, eV^{2-n}$ and the mass $m$ ranges from $10^{-13}\, eV$ to $10^{-3}\,eV$.}
    \label{deltaGGNfixed}
\end{figure}

\begin{figure}
    \centering
    \includegraphics[width=1.20\linewidth]{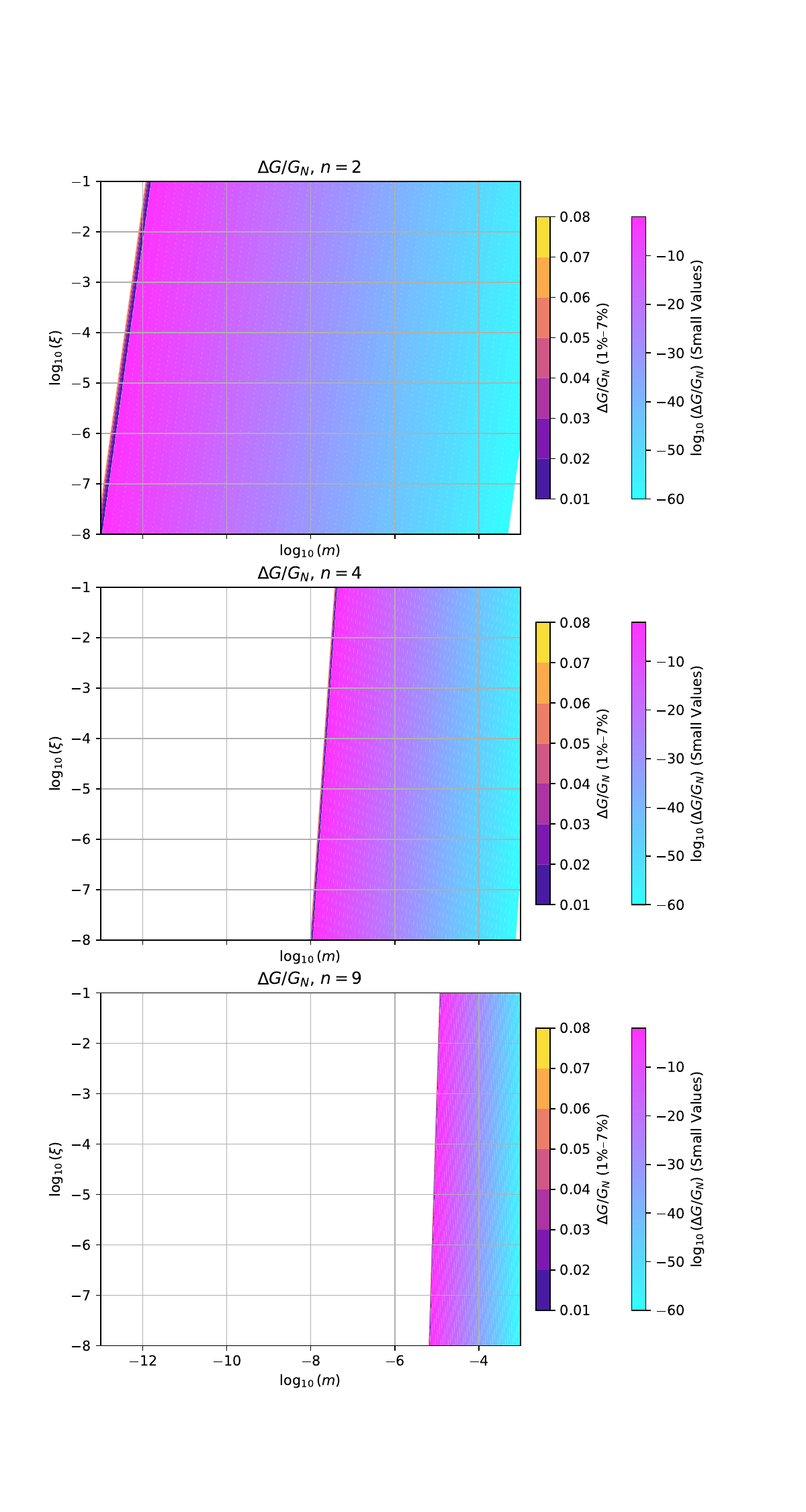}
   \caption{Filled contour plots for $\Delta G/G_N$ for $n=2$, $n=4$, and $n=9$. The dark thin contour lines indicate the parameter space where $\Delta G/G_N$ falls within the $1\%-7\%$ range. The pink-to-blue shading represents regions where $\Delta G/G_N$ drops below 0.01, extending down to $10^{-60}$. The white regions correspond to parameter values where $\Delta G/G_N$ reaches $\mathcal{O}(1)$ or larger. Both axes are logarithmic. The coupling parameter $\xi$ ranges from $10^{-8}\, eV^{2-n}$ to $10^{-1}\, eV^{2-n}$, while the mass $m$ spans from $10^{-13}\, eV$ to $10^{-3}\,eV$.}
\label{contourplot}
\end{figure}

In Fig.~\ref{deltaGGNfixed}, we depict the parameter space corresponding to \(\Delta G/G_N = 1\%-7\%\) for values ranging from \( n=2 \) to \( n=9 \). As \( n \) increases, the mass required to achieve the anticipated \(\Delta G/G_N\) also increases. Furthermore, in Fig.~\ref{contourplot}, the thin dark lines correspond to the contours where \(\Delta G/G_N\) falls within the \( 1\%-7\% \) range, while the pink-blue shaded regions represent values of \(\Delta G/G_N \) that drop below 0.01, extending down to \( 10^{-60} \). The white region, on the other hand, signifies parameter values where \(\Delta G/G_N\) takes values from \(\mathcal{O}(1)\) up to \(\gg \mathcal{O}(1)\).  

For the case of \( n=1 \), the required mass scale for achieving \(\Delta G/G_N = 1\%-7\%\) is approximately \( m \sim 10^{-24} \) eV. This value is significantly lower than the mass scales for higher \( n \), leading to a concentration of all contour regions in a narrow area on the right side of the graph. To enhance visual clarity and aesthetics, we have chosen not to include the \( n=1 \) case in the graphical representation.  

For the coupling parameter \(\xi\) within the range of \( 10^{-8} \, \text{eV}^{2-n} \) to \( 10^{-1} \, \text{eV}^{2-n} \), the corresponding mass range required to achieve \(\Delta G/G_N = 1\%-7\%\) varies as follows:  
 \( n=2 \): \( m \sim 10^{-13} \) eV to \( 10^{-11.8} \) eV,  
 \( n=4 \): \( m \sim 10^{-8} \) eV to \( 10^{-7.4} \) eV, and
 \( n=9 \): \( m \sim 10^{-5.1} \) eV to \( 10^{-4.9} \) eV.  

The chosen range of values for the coupling parameter \(\xi\) is motivated by existing constraints on \(\xi\) in \( n=2 \) theories \cite{Perrotta:1999am,Hrycyna:2015vvs,Ballardini:2021evv,Arapoglu:2022vbf,Rossi:2019lgt}. In future work, we aim to extend these constraints to theories with \( n>2 \), while maintaining the assumption \(\abs{\xi} \ll 1\) to ensure compatibility with the observed precision of General Relativity. Additionally, our analysis focuses solely on positive values of \(\xi\). Allowing negative values would only lead to a sign change in \(\Delta G/G_N\), implying a stronger gravitational constant in the true vacuum. Since this scenario falls outside the scope of our investigation, we do not consider it further.

In Chapter 6 of \cite{Uzan:2024ded} in Table 24 $\abs{\Delta G/G_N}$ is constrained as $\abs{-0.07^{+0.05}_{-0.04}}$ according to measurements from the large Magelanic cloud \cite{Desmond:2020nde}. From the observation of the binary neutron star GW170817 \cite{Vijaykumar:2020nzc} emerges the constraint $-1\leq\Delta G/G_N\leq 8$. In cosmological observations now, an investigation \cite{Wu:2009zb} suggests that both the "WMAP-5yr data" and the "all CMB data" point toward a slightly non-zero (positive) value for $\dot{G}/G$. However, when including the SDSS power spectrum data, the preferred value returns to zero, leading to the constraint $-0.083 < \Delta G/G_N < 0.095$ over the period between recombination and the present.

Several studies have placed constraints on the variation of \(\Delta G / G_N\) over cosmic timescales. Early analyses, based on the abundances of light elements \cite{Accetta:1990au} such as D, \(^3\text{He}\), \(^4\text{He}\), and \(^7\text{Li}\), suggested that \(-0.3 < \Delta G / G_N < 0.4\). These investigations assumed that other physical constants remained fixed. Further studies incorporating constraints from extra relativistic degrees of freedom translated the limits on the speed-up factor of the early universe into bounds on \(\Delta G / G_N\), concluding \(-0.10 < \Delta G / G_N < 0.13\) \cite{Cyburt:2004yc}. More recent work \cite{Yeh:2022heq} refined this range to \(-0.040 < \Delta G / G_N < 0.006\), using improved observational data and updated cosmological assumptions. Using the PRIMAT code and assuming variations only in \(G\) while keeping all other cosmological parameters fixed \cite{Alvey:2019ctk}, a further constraint was obtained: \(\Delta G / G_N = -0.01^{+0.06}_{-0.05}\). These bounds highlight significant progress in constraining the variation of the gravitational constant across different epochs of cosmic history.
 
\section{ Ultra Late-Time Cosmological Transitions and Alternative Resolutions to the Hubble Tension.}\label{sec8}
\subsection{Observational signatures of a gravitational transition: A review}\label{subsec8a}

In the first part of this section, we review studies proposing an ultra-late-time transition as a resolution to the Hubble tension. In the literature \cite{Banik:2024yzi}, this class of models is commonly referred to as the "G-Step Model" (GSM) \cite{Perivolaropoulos:2021bds,Marra:2021fvf,Ruchika:2024ymt}. The core idea behind GSM is that the effective gravitational constant, \( G_{eff} \), underwent a sudden transition at a recent cosmic time, modifying the inferred expansion rate. This transition could be induced by the false vacuum decay mechanism, that was thoroughly examined in this paper, or a step-like variation in \( G_{eff} \), leading to an abrupt shift in the absolute magnitude of Type Ia Supernovae (SnIa). Such a shift would affect distance calibrations in the cosmic distance ladder, offering a potential explanation for the observed discrepancy in \( H_0 \). In this subsection, we explore the observational implications of this scenario, as discussed in recent works.

In a recent study conducted with Skara \cite{Perivolaropoulos:2022khd}, a transition in one of the four key parameters of the Cepheid+SnIa sample was introduced at a specific comoving distance \( D_c \) or cosmic time \( t_c \). This analysis, which reproduced the baseline SH0ES model results, incorporated a single additional degree of freedom, allowing for a possible transition in any of these parameters. One particularly significant parameter in this context is the fiducial luminosity of Type Ia Supernovae (\( M_B \)).

The results were particularly striking in the case of an \( M_B \) transition occurring at \( D_c \approx 50\,Mpc \) (corresponding to approximately \( 160 \) million years ago). Specifically, the best-fit value of the Hubble parameter was found to decrease from \( H_0 = 73.04 \pm 1.04 \,km \,s^{-1}\, Mpc^{-1} \) to \( H_0 = 67.32 \pm 4.64 \,km\, s^{-1}\, Mpc^{-1} \), bringing it into agreement with the Planck measurement. This suggests a potential transition mechanism capable of resolving the Hubble tension, an effect that remains robust even when incorporating the inverse distance ladder constraint on \( M_B \) \cite{Marra:2021fvf, Camarena:2021jlr, Gomez-Valent:2021hda}. This constraint is given by:
\begin{equation}\label{5.70}
    M_B^{P18}=-19.401 \pm 0.027.
\end{equation}
When applying this constraint, the uncertainty in \( H_0 \) is significantly reduced, yielding \( H_0 = 68.2 \pm 0.8\, km \,s^{-1}\, Mpc^{-1} \). 

At this same critical distance \( D_c \), indications of a transitional behavior are also observed for the other three key parameters of the study. However, in these cases, the best-fit value of the Hubble constant does not undergo a significant shift, making the \( M_B \) transition the primary focus of interest. A transition of this nature could be the result of a sudden variation in a fundamental physical constant over the past \( 300 \, Myrs \), possibly associated with a first-order phase transition, giving rise to a gravitational constant bubble, whose properties were examined in the previous section.

\begin{figure}
\begin{centering}
\includegraphics[width=0.5\textwidth]{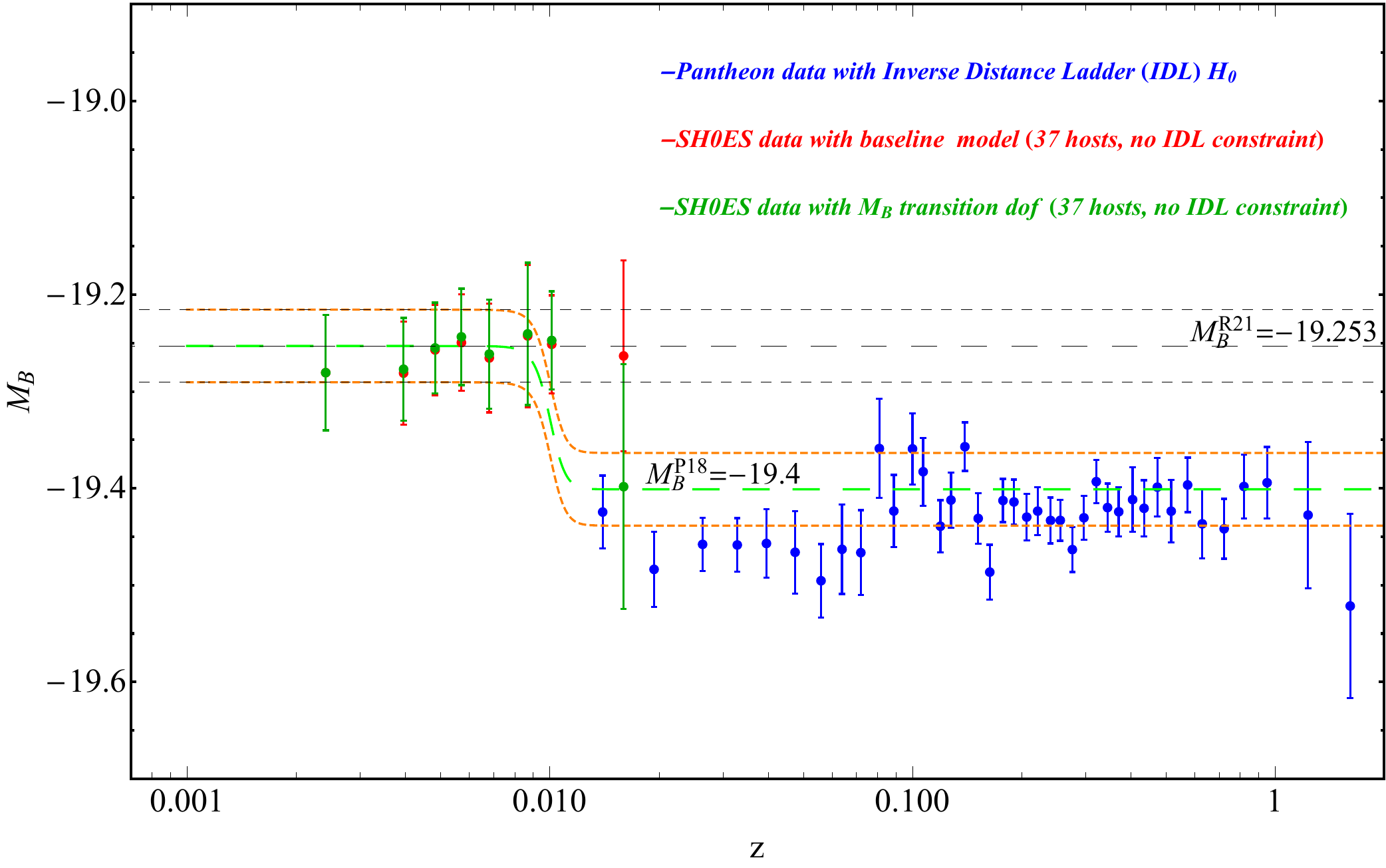}
\par\end{centering}
\caption{The binned (5 host bins with 2 hosts in last bin where $z > 0.01$) Cepheid+SnIa host values of $M_B$ obtained assuming the baseline SH0ES model (red points) and the $M_B$ transition model ($D_c = 50\,Mpc$, green points) are shown along with the inverse distance ladder calibrated binned $M_B$ of the Hubble flow SnIa of the Pantheon dataset (blue points). When the transition degree of freedom is allowed, the data excite it and a hint for a transition appears (the green data point of the transition model is clearly below the red point corresponding to the constant $M_B$ SH0ES baseline model) (reconstructed from \cite{Perivolaropoulos:2022khd}).} 
\label{fig4} 
\end{figure}

In Fig.\ref{fig4}\footnote{All \textit{Mathematica} code used to reproduce Fig.\ref{fig4} from \cite{Marra:2021fvf} can be found in the \href{https://github.com/FOTEINISKARA/A-reanalysis-of-the-SH0ES-data-for-H_0}{A reanalysis of the SH0ES data for $H_0$} \textit{Github} repository of \cite{Perivolaropoulos:2022khd}.}, a clear indication of a transition is visible. The green data point, corresponding to the \( M_B \) transition model, lies below the red data point representing the baseline SH0ES model with a constant \( M_B \). In this figure, red points correspond to binned Cepheid+SnIa host values of \( M_B \) from the SH0ES model, green points represent the results obtained from the transition model, and blue points correspond to the inverse distance ladder-calibrated binned \( M_B \) from the Hubble flow SnIa in the Pantheon dataset. This observed transition could either be a systematic effect or a genuine transition of the effective Newtonian gravitational constant, where \( G_{eff} \) was approximately \( 10\% \) lower in the past, with the change occurring roughly \( 80 \) million years ago or more recently \cite{Marra:2021fvf}.

This observed transition of \( \Delta M_B \sim -0.2 \) can be understood under the assumption of a power-law dependence of the luminosity:
\begin{equation}
    L\sim G_{eff}^{-b}
\end{equation}
where $b>0$ and $\mathcal{O}(1)$. The simplest hypothesis that can be made is $L\sim m_c$ \cite{Marra:2021fvf}, where $m_c$ is the Chandrasekhar mass. This choice leads to $b=3/2$ \cite{Gaztanaga:2001fh}.

The relevance of the Chandrasekhar mass stems from its dependence on the effective gravitational constant and the mass per electron \( m' \) \cite{Amendola:1999vu}:
\begin{equation}
    m_c\simeq \frac{3}{m'}\bigg(\frac{1}{G_{eff}}\bigg)^{3/2}.
\end{equation}
A possible fundamental constant transition would trigger a transition to Chandrasekhar mass and the SnIa peak absolute luminosity too \cite{1982ApJ...253..785A}. A hypothesis that makes sense is that the luminosity is proportional to $m_c$.

The connection between the absolute luminosity and the SnIa absolute magnitude is given by
\begin{equation}
    M_B-M_{0B}=-\frac52\log_{10}\frac{L}{L_0},
\end{equation}
where $L_0$ represents the local values. It is obvious that a $G_{eff}$ decrease leads to an increase of $L$ and a decrease of $M_B$.

A useful phenomenological approximation to the $M_B$ transition is of the following form:
\begin{equation}
  M_B(z)=\begin{cases}
    M_B^{R21}, & \text{if $z\leq z_t$}\\
    M_B^{R21}+\Delta M_B, & \text{if $z>z_t$},
  \end{cases}
\end{equation}
the needed gap in $L$ is approximately related to the corresponding gap in $H_0$ \cite{Marra:2021fvf}:
\begin{equation}
    \Delta M_B\equiv M_B^{P18}-M_B^{R21}\approx 5\log_{10}\frac{H_0^{P18}}{H_0^{R21}}\approx -0.2.
\end{equation}

From all these, can be concluded that the $M_B$ transition and the corresponding $H_0$ crisis could be explained by a transition of the $G_{eff}$ via:
\begin{equation}\label{transition}
  \mu_G(z)\equiv\frac{G_{eff}}{G_{N}} =\begin{cases}
    1\equiv \mu_G^<, & \text{if $z\leq z_t$}\\
    1+\Delta \mu_G \equiv \mu_G^>, & \text{if $z>z_t$},
  \end{cases}
\end{equation}
with $G_{N}$ the local measured Newton's constant. Now, the change of $\mu_G$ is related to the SnIa $M_B$ change via:
\begin{equation}
   \Delta M_B={-\frac52}\log_{10}\frac{L^{P18}}{L^{R21}}=\frac{15}{4}\log_{10}\mu_G^>.
 \end{equation}
 In the above $L^{P18}$ is the CMB-calibrated SnIa luminosities and $L^{R21}$ is the Cepheid-calibrated ones. So we have that
 \begin{equation}\label{deltamg}
     \Delta \mu_G=10^{\frac{4}{15}\Delta M_B}-1 \approx -0.12.
 \end{equation}
Such a transition has the potential to explain the growth tension too \cite{Marra:2021fvf}.

Stellar evolution may impose significant constraints on the \(\mu_G\) transition scenario. The strength of the gravitational interaction, governed by the effective gravitational constant \(G_{eff}\), plays a crucial role in determining the evolutionary path of a star. In particular, any variation in \(G_{eff}\) would affect the star’s hydrostatic equilibrium, influencing its pressure profile and internal temperature. Consequently, these changes would alter nuclear reaction rates, thereby modifying the expected evolutionary trajectories of various stellar populations \cite{Uzan:2010pm}.

In this context, along with Alestas and Antoniou \cite{Alestas:2021nmi} conducted an analysis that provides evidence supporting the possibility of such a gravitational transition. Using a robust dataset consisting of 118 Tully–Fisher datapoints, we applied a specific statistical methodology to investigate potential deviations in the baryonic Tully–Fisher relation (BTFR). By employing Monte Carlo simulations, we compared the real Tully–Fisher dataset with homogenized datasets constructed under the assumption of a standard BTFR. The analysis revealed indications of a transition in the best-fit BTFR parameters at distances \( D_c \approx 9\,Mpc \) and/or \( D_c \approx 17\,Mpc \).

Interestingly, this result aligns well with findings from an earlier study along with Skara \cite{Perivolaropoulos:2021bds}, which reported evidence for a transition in Cepheid calibrator parameters at distances in the range of \(10\) to \(20\,Mpc\). Both studies suggest that the observed transitions could either be attributed to systematic effects in the datasets or, more intriguingly, to a genuine transition in the effective Newtonian constant. In the latter scenario, \(G_{eff}\) would have been approximately \(10\%\) lower at earlier times, with the transition likely occurring about 70 million years ago or more recently.

Moreover, in a recent study, the hemisphere comparison method was employed to assess the isotropy levels of Type Ia Supernovae (SnIa) within the Pantheon+ and SH0ES datasets, focusing on selected distance bins \cite{Perivolaropoulos:2023tdt}. The analysis revealed a particularly notable result: the redshift bin \([0.005, 0.01]\) in the Pantheon+ sample exhibited a significantly higher level of anisotropy compared to the other five redshift bins considered. Additionally, for the SH0ES dataset, a pronounced drop in anisotropy levels was observed at distances exceeding approximately \(30\,Mpc\).

Interestingly, these anisotropy transitions were found to be rare occurrences in simulated isotropic datasets. Specifically, such transitions appeared in only about 2\% of the SH0ES Monte Carlo isotropic simulations and approximately 7\% of the Pantheon+ isotropic simulated samples. This rarity suggests that the observed anisotropy patterns are unlikely to be mere statistical fluctuations.

\begin{figure}
\centering
\includegraphics[width =.45 \textwidth]{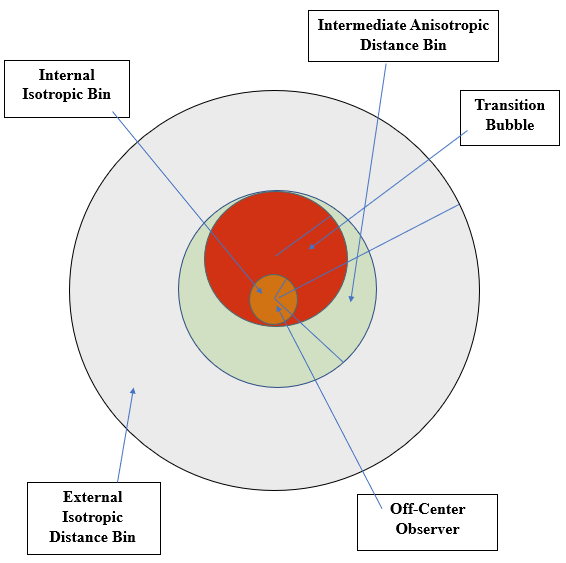}
\caption{Illustration of the hypothetical change of the anisotropy level of SnIa absolute magnitudes in the context of an off-center observer viewpoint (center of the orange circle) located in a bubble (the red circle) of a different gravitational constant by a few percent. The distance bin between the orange inner circle and the green outer circle is predicted to have a low anisotropy, followed by a large anisotropy level and a lesser anisotropy level predicted for the distance bin between the green and grey outer circles (taken from \cite{Perivolaropoulos:2023tdt}).}
\label{fig6}
\end{figure}

As discussed in the study, these intriguing findings could be interpreted as the observational signatures expected from an off-center observer situated within a \(30 - 40\,Mpc\) bubble characterized by unique physical properties or systematics (see Fig.\ref{fig6}). In such a scenario, the off-center observer would detect an abrupt increase in anisotropy in the absolute magnitude \(M_B\) within the intermediate anisotropic region, as illustrated in Fig.\ref{fig6}. Conversely, for distance bins significantly larger or smaller than the bubble’s radius, the universe would appear isotropic to the same observer.

Furthermore, a recent study \cite{perivolaropoulos2022hubblecrisisconnectedextinction} demonstrated that a sudden \( 10\% \) increase in the gravitational constant occurring within the last 100 million years could provide an explanation for the observed doubling or tripling of impact events on Earth and the Moon. This phenomenon may be linked to the Cretaceous-Tertiary (K-T) extinction event, which led to the eradication of approximately \( 75\% \) of life on Earth, including the dinosaurs.

Such a gravitational shift could have significantly amplified the influx of long-period comets (LPCs) into the inner solar system. This effect arises due to velocity perturbations induced by nearby stars or the Galactic tide, which influence the dynamics of the Oort Cloud. A localized change in \( G_{eff} \) could thus lead to an increased number of cometary impact events over geological timescales.

In the context of our model, where a gravitational transition is associated with the expansion of a bubble, the timing of the dinosaurs' extinction event—approximately \( 66 \) million years ago—could imply that our Solar System is positioned around \( 22 \, Mpc \) from the center of the bubble. The impact with the bubble would have occurred after \( 66 \) million years of expansion, triggering the gravitational constant variation responsible for the observed increase in impact events.

\begin{figure}
    \centering
    \includegraphics[width=1.1\linewidth]{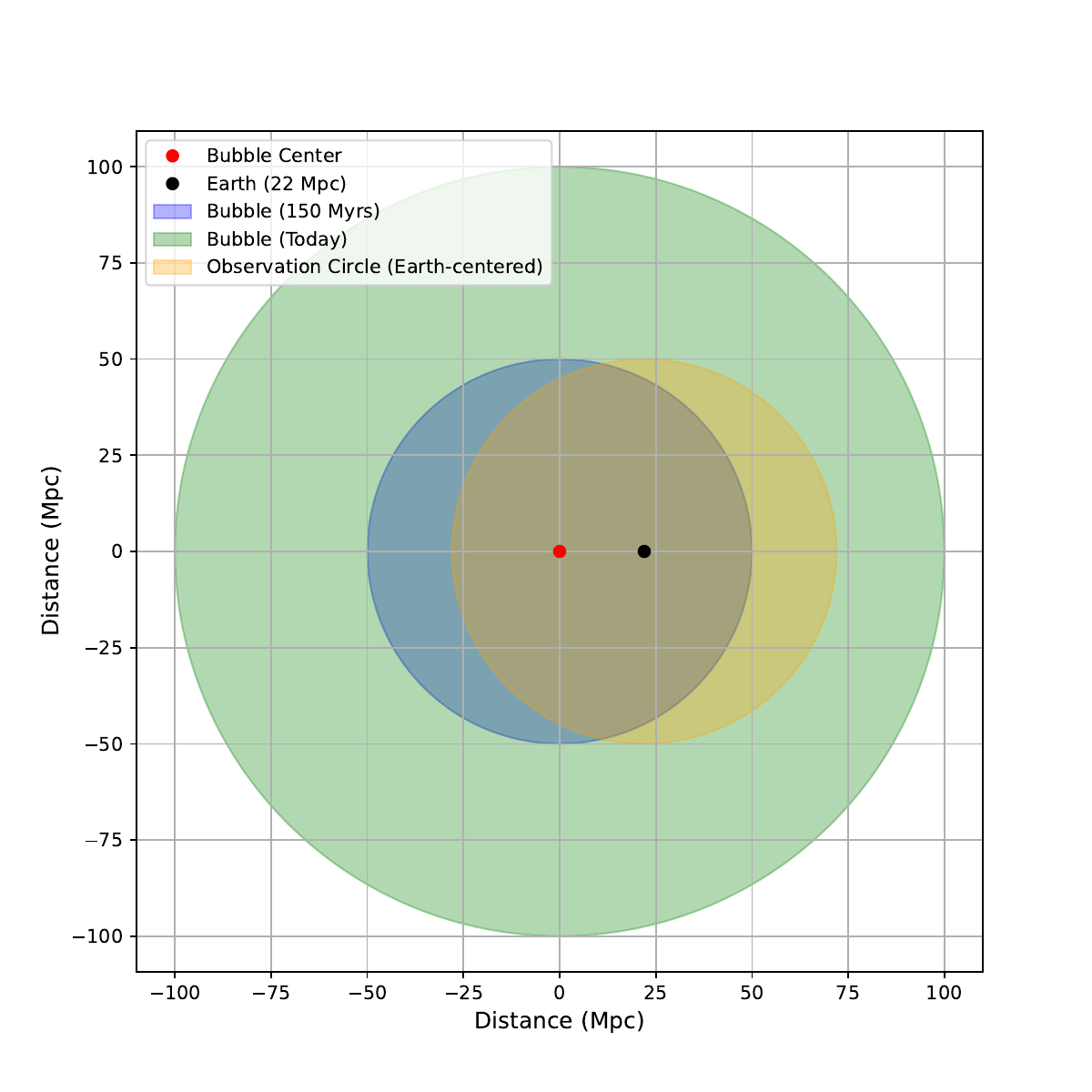}
    \caption{If our Solar System is located $22\,Mpc$ away from the Bubble center, this could probably explain the dinosaurs' extinction occurred 66 million years ago. Observing towards the center of the bubble, regions of new physics can be detected through observational anisotropies.}
    \label{dinosaurs}
\end{figure}

Looking towards the bubble's central regions, the presence of new physics could be inferred through anisotropies in cosmological observations, as indicated by the blue region in Fig.\ref{dinosaurs}. If a fundamental transition in the gravitational constant is indeed linked to such large-scale astrophysical and geological phenomena, this would present a profound intersection between cosmology and planetary science, offering new perspectives on the potential impact of fundamental physics on Earth's history.

An additional noteworthy result was presented in \cite{Perivolaropoulos:2023iqj}, where a variation in the absolute magnitude of Type Ia Supernovae (\( M_B \)) within the Pantheon+ sample was considered. Specifically, the analysis allowed for two different values of \( M_B \): one for lower distances and another for higher distances, with a critical transition occurring at approximately \( 19.95\, Mpc \).

It was found that this transition scenario leads to a significant reduction in the Akaike Information Criterion (AIC) by \( \Delta AIC = -15.5 \). Such a decrease in AIC indicates strong statistical preference for the transitional model over the standard baseline model, which assumes a constant \( M_B \) across all distances. This result suggests that the Pantheon+ dataset may favor a model where the absolute magnitude of SnIa exhibits a transition, rather than remaining fixed.

In \cite{Paraskevas:2023aae}, together with Paraskevas, we examined the evolution of gravitational waves during a sudden transition in the Hubble expansion rate \( H(z) \), potentially linked to a change in the gravitational constant. By applying appropriate boundary conditions, we derived the scale factor and gravitational wave waveform across the singularity. Our analysis included both existing cosmological data and simulated data from future gravitational wave experiments, such as the Einstein Telescope (ET). The findings suggest that mock data from ET could reduce parameter uncertainties by up to a factor of three, depending on the cosmological parameters under consideration.

More specifically, we considered a gravitational transition driven by a phenomenological first-order transition of a scalar field \( \phi(z) \), expressed as:
\begin{equation}\label{gw1}
   \phi(z) = \phi_{0}[1 + \Delta_{1}(\alpha) \Theta(z - z_{s})].
\end{equation}
In the scalar-tensor framework, this scalar field transition induces a gravitational transition and a corresponding Hubble flow:
\begin{equation}\label{gw2}
    H(z) \propto [1 + \sigma \Theta(z - z_s)],
\end{equation}
where \( \sigma \) is a dimensionless parameter, \( z_s \) the transition redshift, and \( \Theta \) the Heaviside step function. Such a transition behaves as a sudden cosmological singularity \cite{Barrow:2004xh,Barrow:2004he,Fernandez-Jambrina:2006qld,Fernandez-Jambrina:2008pwx,Fernandez-Jambrina:2006tkb,Perivolaropoulos:2016nhp}. Since \( H^{2}(z) \sim G_{eff}(z) \), a sudden change in \( G_{eff}(z) \) would produce detectable signatures in cosmological data, potentially observable in the gravitational-wave spectrum.

We highlighted that gravitational-wave observations provide a novel avenue for testing such transitions. Comparing gravitational-wave luminosity distances with those from electromagnetic observations, such as SnIa or galaxy surveys, could reveal deviations from general relativity \cite{Maggiore:2007ulw,Maggiore:2018sht}. In modified gravity theories, the evolution of \( G_{eff}(z) \) leads to distinct propagation patterns for gravitational waves, which can manifest as discrepancies in luminosity distances \cite{Linder:2018jil}. Additionally, in scalar–tensor gravity, gravitational-wave amplitudes after the transition depend on the phase of the wave and the sign of the impulse \( \alpha \), potentially amplifying or diminishing the waveform \cite{Gleyzes:2014rba,Nishizawa:2017nef}.

A plausible mechanism for such a gravitational transition is the formation of a true vacuum bubble within a false vacuum background, with a characteristic scale of approximately \( 20\,Mpc \). This bubble, possessing a distinct value of the effective gravitational constant, could interact with gravitational waves and bound systems in a manner observable as a type II sudden cosmological singularity.

Finally, \cite{Paraskevas:2023aae} explored the so-called sudden-leap model (sLCDM) using real cosmological datasets (Pantheon+, BAO, CMB) and simulated standard siren data from ET observations \cite{Belgacem:2018lbp}. The analysis constrained standard cosmological parameters \( M_B \), \( \Omega_{m,0} \), and \( h \), along with sLCDM parameters \( \alpha \) (transition amplitude) and \( z_s \) (transition redshift). Although the sLCDM model did not outperform the standard \(\Lambda\)CDM in terms of data fit (see Table 4 of the paper), it presents a unique mechanism for addressing the Hubble tension.

Notably, the Pantheon+ dataset yields a best-fit scale factor \( a_s = 0.995 \pm 0.004 \) (\( z_s = 0.005 \)) and a corresponding luminosity distance \( d_L(z_s) \approx 20.96 \,Mpc \). This value closely aligns with previous estimates of a sudden change in SnIa luminosity distance found in \cite{Perivolaropoulos:2023tdt,Perivolaropoulos:2023iqj,Alestas:2021nmi,Perivolaropoulos:2021bds}, which predict a transition scale of approximately \( 20\,Mpc \). This convergence between gravitational and electromagnetic observations warrants further investigation.

 \subsection{Alternative Solutions to the Hubble Tension}\label{subsec8b}

In alternative, non-ultra-late time transition, approaches, recent works have explored the possibility that screened fifth forces may play a crucial role in resolving the Hubble tension—the discrepancy between local measurements of the Hubble constant $H_0$ and those inferred from the CMB. In \cite{Desmond:2019ygn}, the authors propose that a fifth force, acting preferentially in low-density environments due to screening mechanisms such as chameleon or symmetron screening, could affect the calibration of the distance ladder. Their analysis focuses on Cepheid variables used in Type Ia supernova calibration, suggesting that these stars may be less screened compared to those in the host galaxies of supernovae. As a result, the luminosity of Cepheids would be systematically altered, leading to an overestimation of distances and thus an inflated $H_0$. By accounting for this effect, the inferred local value of $H_0$ can be brought into agreement with Planck CMB measurements, providing a potential astrophysical solution to the Hubble tension without invoking new physics in the early universe.

The subsequent study \cite{Desmond:2020wep}, building on the initial findings, extends the framework of screened fifth forces to the Tip of the Red Giant Branch (TRGB) method for $H_0$ calibration. The TRGB approach, considered less susceptible to systematic uncertainties than the Cepheid-based method, relies on the luminosity of red giant stars at the onset of helium burning. The authors demonstrate that a fifth force, if present, would lower the TRGB luminosity in unscreened stars, similarly leading to an overestimated $H_0$. By assuming that the TRGB calibration stars in the Large Magellanic Cloud (LMC) are less screened than those used to calibrate supernovae, the study achieves consistency between TRGB-inferred $H_0$ and CMB-based measurements. The analysis reveals that a fifth force with a strength of approximately $20\%$ of Newtonian gravity in unscreened Red Giant Branch (RGB) stars can reconcile these differences.

In their most recent work, Banik, Desmond, and Samaras \cite{Banik:2024yzi} critically assess the "G-step model" \cite{Perivolaropoulos:2021bds,Marra:2021fvf,Ruchika:2024ymt} as a potential solution to the persistent Hubble tension. The GSM hypothesizes an abrupt decrease in the gravitational constant $G$ approximately 130 Myr ago. This drop would make Type Ia supernovae in the Hubble flow intrinsically brighter compared to those in nearby host galaxies calibrated via Cepheid distances, thereby altering the redshift-distance relation and potentially resolving the observed discrepancy in the Hubble constant $H_0$. Despite the initial appeal of this model, the authors highlight a series of critical inconsistencies. They demonstrate that such a sharp reduction in $G$ would lead to drastic astrophysical and geological consequences, including a substantial decrease in solar luminosity that would have likely triggered a global glaciation episode on Earth, which contradicts the continuous geochronological and cyclostratigraphic records. Furthermore, stellar evolution models under the GSM predict a helioseismic solar age inconsistent with meteoritic data, and a lack of old stellar populations corresponding to the first 3 Gyr of cosmic history, contradicting current galactic observations.

The authors also explore broader astrophysical implications, noting that the GSM would disrupt well-established galaxy scaling relations, such as the radial acceleration relation, and create inconsistencies in stellar luminosities critical for distance measurements. They find that the GSM fails to reconcile these issues while maintaining consistency with CMB and BAO constraints. Concluding their analysis, Banik et al. argue that the severe discrepancies associated with the GSM render it an implausible solution to the Hubble tension. Instead, they propose that more viable solutions might involve environmental screening mechanisms affecting $G$ only in low-density environments, thereby preserving local astrophysical observations while resolving cosmological discrepancies. This comprehensive study significantly narrows the scope of plausible solutions to the Hubble tension and emphasizes the need for models that can coherently integrate local and cosmological observations without invoking drastic astrophysical anomalies.

Even if the tests of Banik et al. \cite{Banik:2024yzi} verified, the possibility of a simultaneous transition of $G$ with other fundamental constants, such as the fine-structure constant $\alpha$ \cite{Bronnikov:2013xh,Bronnikov:2013jua}, would remain viable, thus preserving the broader late physics transition paradigm. Transitions of the type envisaged by this paradigm could be produced by theories such as Dilaton \cite{Damour:1990tw} and Kaluza–Klein theories \cite{Kaluza:1921tu,Klein:1926fj}, TeVeS-like theories \cite{Bekenstein:2004ne,Chaichian:2014dfa}, or varying-$\alpha$ theories \cite{Barrow:2013uza}. Note that the aforementioned constraints do not apply to screening models discussed earlier in this section, as screening naturally suppresses the fifth force locally, precluding a universal change to fundamental constants like $G$.

One of the most compelling alternatives to the standard cosmological model is the $\Lambda_{\rm s}$CDM model \cite{Akarsu:2019hmw,Akarsu:2019pwn,Akarsu:2021fol,Akarsu:2022typ,Akarsu:2023jzv,Akarsu:2023mfb,Akarsu:2024eoo,Akarsu:2024qsi,Yadav:2024duq,Akarsu:2025ijk}, which introduces a sign-switching cosmological constant, $\Lambda_s(z)$, undergoing a rapid anti–de Sitter (AdS) to de Sitter (dS) transition at a characteristic redshift $z^\dagger \sim 2$ \cite{Akarsu:2023mfb}. This transition effectively modifies the expansion history of the Universe post-recombination, allowing for a higher $H_0$ value consistent with local measurements, while simultaneously addressing the $S_8$ tension \cite{Akarsu:2024qsi}. Unlike the traditional $\Lambda$ in $\Lambda$CDM, which remains constant and positive, $\Lambda_s(z)$ adopts negative values at $z > z^\dagger$, transitioning to positive values at lower redshifts. This behavior can be described using a hyperbolic tangent functional form, such as $\Lambda_s(z) = \Lambda_{s0} \tanh[\nu(z^\dagger - z)]/\tanh[\nu z^\dagger]$ \cite{Akarsu:2025gwi}, where $\Lambda_{s0}$ is the present-day value and $\nu$ controls the rapidity of the transition. In the limit $\nu \to \infty$, the model approaches the ``abrupt'' $\Lambda_{\rm s}$CDM scenario, offering a minimal modification that introduces only one additional parameter compared to the standard model \cite{Akarsu:2023mfb}.

Recent studies have provided robust theoretical underpinnings for the $\Lambda_{\rm s}$CDM framework. Notably, Akarsu \textit{et al.} \cite{Akarsu:2023jzv,Yadav:2024duq} propose a scalar field model with a hyperbolic tangent potential, wherein a phantom field with a negative kinetic term drives the AdS-to-dS transition. Remarkably, the step-like form of the potential prevents pathologies such as Big Rip singularities and violations of the weak energy condition (WEC) \cite{Caldwell:2003vq}. The resulting cosmological evolution remains smooth and continuous, with all key kinematical parameters ($H(z)$, $\dot{H}(z)$, and $q(z)$) evolving without discontinuities. The total energy density stays positive, and the effective equation of state remains above $-1$, ensuring physical viability \cite{Yadav:2024duq}. While the energy density of the phantom field crosses zero, introducing a safe singularity in its EoS, the model successfully reconciles the Planck CMB and SH0ES local $H_0$ determinations, addressing the $H_0$ tension without introducing inconsistencies with large-scale structure observations \cite{Riess:2019cxk,Planck:2018vyg}.

Furthermore, extended versions of the $\Lambda_{\rm s}$CDM model incorporate variations in the effective number of neutrino species ($N_\text{eff}$) and the total neutrino mass ($\sum m_\nu$) \cite{Akarsu:2025gwi}. These extensions aim to examine whether late-time physics, specifically the mirror AdS-to-dS transition, suffices to alleviate cosmological tensions without necessitating deviations from the standard model of particle physics \cite{Lesgourgues:2012uu}. Analysis reveals that for datasets including Planck, BAO, and SH0ES, the $\Lambda_{\rm s}$CDM+$N_\text{eff}$+$\sum m_\nu$ model predicts $H_0 \approx 73$ km s$^{-1}$ Mpc$^{-1}$, in excellent agreement with local measurements. Importantly, these results are achieved without requiring significant deviations from $N_\text{eff} = 3.044$, maintaining compatibility with the standard cosmological model. The total neutrino mass upper bounds ($\sum m_\nu \lesssim 0.50$ eV) also align with neutrino oscillation experiments, reinforcing the model's physical plausibility. Consequently, the $\Lambda_{\rm s}$CDM framework emerges as a promising late-time modification to $\Lambda$CDM, capable of addressing major cosmological tensions through minimal yet impactful modifications to the Universe's expansion history.

\section{Conclusion and Discussion.}\label{sec9}

In this work, we have explored the intriguing scenario where the cosmological constant originates from a metastable vacuum state of a scalar field. We examined the formation and evolution of true vacuum bubbles in this framework and extended our analysis to scalar-tensor theories of gravity. These investigations highlight the potential observational signatures of such ultralate transitions, providing a pathway to addressing the current tensions in cosmological data while opening new avenues for probing the fundamental nature of dark energy and gravity.

This study has explored the dynamics of true vacuum bubble formation in the context of first-order phase transitions, a critical phenomenon for understanding early universe processes, including electroweak symmetry breaking and inflation. By employing the thin-wall approximation, we have derived key formulas to estimate the decay rate of the false vacuum and the typical scale of bubble formation.

The thin-wall approximation simplifies the problem by considering nearly degenerate minima and a bubble with a thin wall separating the true and false vacua. Using this approximation, we derived the critical bubble radius, \(\bar{\rho}_0\), and the action \(S_{B,0}\) associated with bubble nucleation. These results indicate that the typical radius of the bubbles is significantly smaller than cosmological scales, and their expansion velocity approaches the speed of light. The estimated decay rates suggest that bubbles form with negligible size relative to the Hubble horizon, aligning with previous theoretical predictions.

Our analysis also extends beyond the thin-wall approximation by solving the field equations numerically for more general potential shapes. This approach confirms that the thin-wall results remain valid for a wide range of parameters, but deviations arise when the potential's shape deviates significantly from the simple double-well form. Specifically, as the parameter \(k\) in the dimensionless potential of Eq.\eqref{dimlesspot} deviates from unity, the bubble radius decreases, reflecting the increased energy available for vacuum conversion.

In the context of cosmology, the implications of these results are profound. For slow, time-dependent transitions, the bubble radius is found to be very small, emphasizing the rapid nature of the phase transition once nucleation occurs. For time-dependent transitions, such as those relevant to early universe scenarios, the decay rate is influenced by the evolving dynamics of the false vacuum, which can lead to faster transitions.

The order-of-magnitude estimates show that the thin-wall approximation provides accurate results as long as the parameter \(\epsilon\) remains small, consistent with the scale of cosmological transitions. For example, in the case of a cosmological constant transition, bubbles are predicted to form with a radius on the order of \(10^{-4}\) meters, significantly smaller than cosmological scales.

Continuing our work, we investigated  the behavior of true vacuum bubbles, focusing on the critical field mass \( m_{crit} \) that determines whether these bubbles will expand or collapse. By analyzing the total energy of a thin-walled bubble and applying the thin-wall approximation, we derived the critical mass and examined its implications for bubble dynamics.

Our findings indicate that the critical mass \( m_{crit} \) is approximately \( 0.0017 \, eV \). Bubbles with a mass \( m \leq m_{crit} \) will expand, whereas those with \( m > m_{crit} \) exhibit different behaviors depending on their mass range. For masses greater than \( 0.002 \, eV \), the decay rate \( \Gamma \) becomes exceedingly small, leading to the effective stability of the false vacuum where bubbles do not nucleate. For intermediate masses between \( 0.0017 \, eV \) and \( 0.002 \, eV \), bubbles nucleate but subsequently collapse.

These results underscore the critical role of bubble mass in determining the fate of vacuum bubbles. The decay rate \( \Gamma \) dramatically decreases for high masses due to an exponential suppression factor, suggesting that the false vacuum becomes increasingly stable as the mass increases beyond \( 0.002 \, eV \). This behavior is consistent with the intuitive expectation that higher masses correspond to a higher energy barrier for bubble nucleation.

The inclusion of gravitational effects in the analysis of true vacuum bubble nucleation reveals critical insights into how curvature influences decay rates and bubble radii during first-order phase transitions. Extending the flat spacetime formalism, the Coleman-De Luccia approach shows that gravitational corrections become relevant only when the characteristic bubble radius \(\bar{\rho}_0\) approaches the gravitational scale \(D\), which is tied to the vacuum energy density \(\epsilon\). Parke’s generalization further expands these results to arbitrary vacuum potential values, offering a robust framework for assessing gravitational influences across various cosmological models.

Importantly, for transitions occurring at energy scales much lower than the Planck scale—such as those associated with the cosmological constant—the gravitational corrections remain negligible. The estimated correction factor on the order of \(10^{-63}\) indicates that curvature effects are effectively suppressed, preserving the predictions of flat spacetime analyses. Only near Planck-scale energies does gravity impose substantial suppression on decay rates, rendering bubble nucleation events cosmologically irrelevant on timescales comparable to \(H_0^{-1}\).

These conclusions carry significant implications for late-time cosmological scenarios. The minimal impact of gravitational corrections ensures that vacuum transitions can still occur with observable consequences, such as gravitational wave signatures or large-scale anisotropies, without invoking Planck-scale physics. Moreover, the analysis confirms the robustness of the semi-classical approach in modeling these transitions, validating its application to cosmological phenomena potentially linked to the late-time acceleration of the Universe.

Overall, the interplay between gravitational effects and vacuum decay processes remains negligible for cosmologically relevant transitions, reinforcing the feasibility of true vacuum bubble nucleation as a mechanism influencing the large-scale dynamics of the Universe.

Following all these, our paper explored the dynamics of true vacuum bubbles in the context of non-minimal coupling between the scalar field and gravity. Our investigation focused on how the non-minimal coupling affects the behavior of these bubbles, particularly their nucleation and stability.

We began by considering the action for a scalar field with non-minimal coupling to gravity:

\begin{equation}
    S=\int d^4x\sqrt{-g}\, \left[F(\phi)\frac{R}{2}+\frac{1}{2}\partial_{\mu}\phi\partial^{\mu}\phi - U(\phi)\right],
\end{equation}
where the function \( F(\phi) \) characterizes the coupling between the scalar field \( \phi \) and the curvature \( R \). We analyzed the effective gravitational constant \( G_{eff} \) in the Jordan frame and found that it is proportional to \( \frac{1}{F(\phi)} \) (see Eq.\eqref{5.55}).

By examining a specific form of \( F(\phi) \) given by \( F(\phi) = \frac{1}{\kappa} - \xi \phi^2 \), with \( \kappa = 8 \pi G_N \) and \( \xi \) being the non-minimal coupling parameter, we derived the Euclidean field equations and expressions for the energy-momentum tensor. This allowed us to determine the impact of the coupling term \( -\xi R \phi \) on bubble dynamics.

The thin-wall approximation provided an analytical expression for the exponential factor \( S_B \), crucial for understanding bubble nucleation rates. The results showed that the transition from de-Sitter space to Minkowski space can be described using:

\begin{equation}
    S_B = 2 \pi^2 \left(\bar{\rho}^3 S_1 + \xi \bar{\rho} C - \frac{2}{3} \frac{(1 - \bar{\rho}^2 \kappa_+ \epsilon)^{3/2} - 1}{\kappa_+^2 \epsilon} - \frac{\bar{\rho}^2}{\kappa_-}\right),
\end{equation}
where \( C \) is related to the potential parameters and \( \kappa_\pm \) are modified gravitational constants. The solutions indicate that the presence of non-minimal coupling affects the bubble radius significantly, particularly for different values of \( \xi \).

We applied a shooting method to solve the field equations numerically, considering a toy model potential for transitions between different vacuum states. The results for bubble profiles and radii demonstrate that the coupling \( \xi \) influences the maximum radius of bubbles but does not fundamentally alter the physics of bubble nucleation. In particular, the radius of bubbles in a de-Sitter false vacuum space is larger for smaller \( \xi \), while in a Minkowski space, the bubbles tend to collapse more quickly.

The analysis of ultra-late time transitions within the framework of non-minimally coupled scalar field theories reveals that the key characteristics of bubble nucleation, such as the nucleation radius and critical scalar field mass, retain similar magnitudes to those found in minimal coupling scenarios. By numerically solving the dimensionless equation \eqref{numeqch6} for various values of the non-minimal coupling parameter \(\xi\) and transition mass scales \(m\), we demonstrated that the nucleation radius \(\bar{\rho}\) remains in the micrometer regime, specifically on the order of a few hundred micrometers.

Table \ref{tab:rho_values} provides a detailed summary of the positive solutions for the dimensionless nucleation radius \(\tilde{\bar{\rho}}\) and the corresponding physical radii \(\bar{\rho}\) in \(\mu m\) across a range of mass scales and coupling strengths. Remarkably, for mass scales spanning \(10^{-23}\,eV\) to \(10^{-3}\,eV\) and coupling values from \(\xi \ll 1\) up to \(\xi \sim \mathcal{O}(1)\), the nucleation radius consistently falls between \(77\,\mu m\) and \(720\,\mu m\). This robustness indicates that the inclusion of a non-minimal coupling does not drastically affect the nucleation size under the conditions considered, maintaining compatibility with General Relativity (\(\bar{\rho}_0 \sim 250\,\mu m\)). However, the sensitivity of the solutions to \(\xi\) increases with larger mass scales, suggesting that the coupling plays a more significant role in higher-energy transitions.

Additionally, the critical mass \(m_{crit}\) was computed using equation \eqref{mcritcoupling}, with results visualized in Fig.\ref{mcritvsrho}. The critical mass defines the threshold above which bubble nucleation can proceed efficiently. Our numerical solutions confirm that \(m_{crit}\) increases as the bubble radius increases, following the expected scaling relation. 

Overall, these findings illustrate that while non-minimal couplings introduce subtle quantitative differences in vacuum decay dynamics, the qualitative behavior remains consistent with GR scenarios. Bubble nucleation at ultra-late times continues to yield radii on the micrometer scale, with critical masses in the \(10^{-3}\,eV\) range. The relative insensitivity of these results to \(\xi\) across broad parameter spaces highlights the robustness of late-time vacuum transition models in modified gravity frameworks.

 The analysis of the scale and properties of vacuum bubbles following a late-time first-order phase transition has illuminated the interplay between cosmological dynamics and observational consequences. We have shown that the growth and spatial extent of these bubbles are intrinsically tied to the parameters governing the transition, including the nucleation rate $\Gamma(t)$, the logarithmic derivative $B$, and the elapsed transition time $\delta t$. These parameters determine the comoving radius of bubbles at the end of the phase transition and how they manifest in our observational data. Notably, under plausible conditions for a recent false vacuum decay, the characteristic size of bubbles today is estimated to be in the range of $20\,Mpc$ to $100\,Mpc$ (coinciding with a gravitational transition taking place around $50-300$ million years ago), with implications for local anisotropies and cosmological tensions.

A significant finding is the logarithmic dependence of the bubble size on the temperature $T$ and the associated nucleation dynamics near the critical temperature $T_c$. The behavior of the logarithmic derivative $B$, which transitions from $B \gg 1$ near $T_c$ to values of $\mathcal{O}(1)$ as the phase transition completes, provides insight into the rate of bubble expansion. This framework is robust across different gravitational theories, with modifications to the action affecting the behavior of $B$ in a manner that remains consistent with observations. Furthermore, the close agreement between analytical estimates, such as $\Delta \sim (S_1 B_1)^{-1}$, and numerical calculations supports the validity of the models employed.

The parametrization of the transition duration $\delta\tau$ was performed, following the work of \cite{Hawking:1982ga,Turner:1992tz,Kosowsky:1992rz,Kosowsky:1992vn,Kosowsky:2001xp}. Plotting \eqref{deltatafparam} for different values of the redshift of observation $z_{obs}$ we found that $\delta\tau$ is in the order of hundreds of Mpc's, and it depends logarithmically on the background mass scale. The plot reveals that for $m \in [10^{-13}, 10^{-3}]\,eV$, transition durations lie within $280-810\,Myrs$. These timescales align with the hypothesis of gravitational bubbles forming recently in cosmic history. The sensitivity of $\delta \tau$ to $c_1$ and $c_2$ variations remains logarithmically suppressed, maintaining robustness in the predicted ranges.

From an observational perspective, our study suggests that a bubble nucleated several hundred million years ago, with an initial radius of $\mu m$, could today extend over scales of up to $\sim 100\,Mpc$. However, due to the finite speed of light, observations at redshifts $z \sim 0.01$ (corresponding to distances $\sim 50\,Mpc$) would capture the bubble at an earlier stage of its evolution, approximately $140\,Myrs$ after nucleation. This highlights the inherent relativistic constraints on probing the full extent of such structures in the universe and raises intriguing questions about the observational signatures of local anisotropies at larger scales.

The scenario involving multiple nucleation events distributed across space introduces gravitational wave production upon bubble collisions. The separation distance $d$ Eq.\eqref{sepdist}, related to $\delta \tau$ through $\delta \tau \sim d/2$, was computed to be approximately $500\,Mpc$ for $\mathcal{A} \sim 6$, $\lambda=0.07$, and $\phi_t \sim 10^{-3}\,eV$. The convergence of results from three independent methods ($B_1$ estimates, $\rho_{nucl}$ calculations, and visual models) suggests $R_b \sim 250-285\,Mpc$ after $750\,Myrs$ of expansion. The 2D simulation, yielding $\mathcal{N} = 197$ bubbles arranged in a square grid, reinforces this picture and shows that these bubbles could complete the transition in $450\,Myrs$—placing our current epoch near the process's midpoint.

In the following section, we explored the variation of the effective gravitational constant, $\Delta G/G_N$, in the context of modified gravity theories characterized by a coupling function of the form $F(\phi)=1/\kappa - \xi \phi^n$. By examining the dependence of $\Delta G/G_N$ on the coupling parameter $\xi$, the mass scale $m$ and the exponent $n$, we quantified the percentage change in the effective gravitational constant due to scalar fields transitioning between false and true vacuum states. This framework allows us to probe the interplay between scalar field dynamics, the coupling parameter $\xi$, and the mass scale $m$, all of which shape the observational signatures of gravitational constant variations.

Our analysis revealed that for specific choices of parameters, such as $n=2$, $m=10^{-12}$ eV, and $\xi=0.01$, the percentage change in $\Delta G/G_N$ reaches $3.25\%$. As illustrated in Fig.~\ref{deltaGGNfixed}, achieving a change of $\Delta G/G_N$ within the observationally motivated range of $1\%-7\%$ depends strongly on the values of $n$, $m$, and $\xi$. Higher values of $n$ demand larger mass scales $m$ to reproduce these changes, while the coupling parameter $\xi$ exhibits sensitivity across several orders of magnitude.

Interestingly, the graphical representation of $\Delta G/G_N$ reveals that most of the viable parameter space occupies values far below observational limits, with only narrow regions reaching the $1\%-7\%$ range. This suggests that while modified gravity theories can introduce variations in $G_{eff}$, these changes are highly constrained and likely subtle. The choice of positive $\xi$ values ensures compatibility with the observed strength of gravity, whereas negative $\xi$ would imply a stronger gravitational constant in the true vacuum, a scenario outside the scope of this investigation.

In light of observational constraints, such as those from the Large Magellanic Cloud \cite{Desmond:2020nde} and neutron star mergers like GW170817 \cite{Vijaykumar:2020nzc}, our results are consistent with current limits on $\Delta G/G_N$. Cosmological data \cite{Wu:2009zb} also support a nearly constant $G$ within bounds of $-0.083 < \Delta G/G_N < 0.095$ since recombination. Constraints from early universe physics, such as primordial nucleosynthesis \cite{Accetta:1990au}, and recent updates using improved cosmological datasets \cite{Yeh:2022heq}, refine the permissible range of $\Delta G/G_N$ even further, emphasizing the remarkable sensitivity of cosmological probes.

Future work can extend this study by incorporating additional modifications to the scalar field potential and coupling functions, exploring their implications across wider redshift ranges. Dilaton-like coupling functions such as 
\be\label{dilatoneq}
F(\phi) = \frac{1}{\kappa} \exp(-\xi \phi^n)
\ee
and logarithmic-form functions like
\be\label{logeq}
F(\phi) = \frac{1}{\kappa} \left( 1 - \ln(\xi \phi^n) \right)
\ee
could be a potentially good path for investigation. Observational data from upcoming missions, including gravitational wave detectors and next-generation cosmological surveys, may provide tighter constraints on $\Delta G/G_N$, enabling deeper insights into the dynamics of modified gravity theories. The interplay between theory and observation remains a fertile ground for advancing our understanding of the gravitational interaction in regimes beyond general relativity.

Future work within the context of scalar-tensor theories characterized by coupling functions of the form \(F(\phi)=1/\kappa - \xi \phi^n\) and \eqref{dilatoneq},\eqref{logeq} includes analytical exploration of the Euclidean field equations using the thin-wall approximation, as well as numerical studies beyond this regime. Such analyses aim to explicitly determine bounce solutions and bubble profiles for various exponents \(n\), mass scales \(m\), and coupling strengths \(\xi\). This systematic approach will offer deeper insights into the detailed dynamics of vacuum decay and the observational signatures associated with transitions in the gravitational constant.

In the final part, we reviewed existing bibliography, searching the evolution in research for transitional signature in cosmological data. Also, we saw that a transition through the false vacuum decay mechanism could potential resolve the Hubble tension, and the growth tension in many cases.

Our focus was on the implications of a transition in the fiducial luminosity of Type Ia supernovae (\(M_B\)) and its effect on the Hubble constant and cosmological growth rates. The analysis revealed that allowing for a transition in \(M_B\) at a critical distance $D_c \approx 50 \, Mpc$ \cite{Perivolaropoulos:2022khd} could effectively reconcile the observed discrepancies between different measurements of \(H_0\). Specifically, a transition leading to a drop in \(H_0\) from $73.04 \pm 1.04 \, \text{km s}^{-1} Mpc^{-1}$ to $67.32 \pm 4.64 \, \text{km s}^{-1} Mpc^{-1}$ aligns with the Planck measurements and provides a substantial resolution to the Hubble tension.

These results also suggest that the observed transition in \(M_B\) can be attributed to a change in the effective Newton's constant (\(G_{eff}\)). This shift, approximately $12\%$ lower in \(G_{eff}\) at higher redshifts, leads to a corresponding increase in the Chandrasekhar mass and a decrease in the absolute magnitude \(M_B\) \cite{Marra:2021fvf}. This hypothesis not only addresses the Hubble tension but also aligns with theoretical predictions of luminosity variations in supernovae. Furthermore, the phase transition model offers a resolution to the growth tension by affecting the growth rate of cosmological matter fluctuations. 

 Recent findings \cite{Perivolaropoulos:2023tdt} using the hemisphere comparison method revealed notable anisotropy variations in the SnIa data, particularly in the redshift bin [0.005, 0.01], which is significantly more anisotropic compared to other redshift bins. These observations suggest that an off-center observer within a $30-40 \, Mpc$ bubble with distinct gravitational properties could perceive abrupt changes in anisotropy levels, consistent with the proposed model of a local transition in \(M_B\). Additionally, the reduction in the Akaike Information Criterion (\(\Delta \text{AIC} = -15.5\)) from allowing a transition in \(M_B\) at a critical distance of approximately $19.95 \, Mpc$ reinforces the preference for a transitional model over a constant \(M_B\) baseline. These results collectively provide compelling evidence for the hypothesis that a phase transition in \(G_{eff}\) could account for observed discrepancies in supernovae data and support the broader implications of such transitions in addressing cosmological tensions.

 The observed anisotropy in supernova data and the implications for stellar evolution provide additional support for the transition model. Specifically, anisotropy transitions observed in data from the Pantheon+ and SH0ES surveys align with the hypothesis of a local bubble with different gravitational properties. This supports the idea that the observed transition in \(M_B\) could be a result of a localized change in the effective Newton constant.

 Moroever, our study explores the implications of a sudden transition in the Hubble expansion rate \(H(z)\), driven by changes in the effective gravitational constant \(G_{eff}(z)\) within a scalar-tensor framework through the work of Paraskevas \cite{Paraskevas:2023aae}. The transition, modeled as \(H(z) \propto [1 + \sigma \Theta(z - z_s)]\), affects gravitational wave propagation, altering GW amplitudes and wavelengths due to modifications in the friction term \( \alpha_{M} \).

A key outcome is the relationship between GW and electromagnetic luminosity distances:
\begin{equation}
    d_{L}^{gw}(z) = d_{L}^{em}(z) \sqrt{\frac{G_{eff}(z)}{G_{eff}(0)}}.
\end{equation}
This relation provides a means to test modified gravity theories through combined GW and electromagnetic observations.

Additionally, a gravitational transition scenario involving a true vacuum bubble with a characteristic scale of $20 \, Mpc$ is investigated. Such a bubble, exhibiting a distinct \(G_{eff}\), could manifest as a type II sudden cosmological singularity. The sudden-leap (sLCDM) model \cite{Paraskevas:2023aae}, though not outperforming $\Lambda$CDM in current data fits, offers a promising approach to addressing the Hubble tension. Notably, the Pantheon+ dataset suggests a best-fit transition scale factor \(a_s = 0.995 \pm 0.004\) ($z_s = 0.005$) and a corresponding luminosity distance \(d_L(z_s) \approx 20.96 \, Mpc\), aligning well with recent theoretical and observational predictions.

In summary, while the sLCDM model requires further observational validation, it remains a compelling framework for exploring gravitational transitions and their potential role in resolving key cosmological tensions.

The possibility that a screened fifth force could be responsible for the observed Hubble tension offers an intriguing astrophysical resolution. Screening mechanisms such as the chameleon and symmetron models suggest that Cepheid variables and TRGB stars may experience a different effective gravitational constant in low-density environments, leading to systematic alterations in their inferred luminosities. The resulting bias in the local distance ladder calibration would naturally account for an inflated measurement of $H_0$, bringing it into better agreement with CMB-based determinations. While these models effectively reconcile local and cosmological constraints on $H_0$, their viability hinges on the precise nature of screening in astrophysical environments. Further observational probes, including independent tests of screening effects in stellar populations and exoplanetary systems, will be critical in determining whether a fifth-force modification to gravity can consistently explain the observed discrepancies without conflicting with other astrophysical constraints.

On the other hand, the $\Lambda_{\rm s}$CDM framework provides a fundamentally distinct approach, modifying the cosmic expansion history itself through a late-time anti-de Sitter to de Sitter transition. By allowing the cosmological constant to undergo a sign-switching transition at redshift $z^\dagger \sim 2$, this model not only resolves the $H_0$ tension but also mitigates the $S_8$ discrepancy, making it one of the few proposals capable of addressing multiple cosmological anomalies simultaneously. Theoretical developments have reinforced the physical plausibility of $\Lambda_{\rm s}$CDM, particularly through scalar field realizations that maintain continuity in cosmic evolution while avoiding singularities. Moreover, the model remains consistent with neutrino physics, large-scale structure observations, and early-time cosmology, positioning it as a compelling alternative to $\Lambda$CDM. Future high-precision cosmological surveys, along with potential constraints from gravitational wave standard sirens, will be instrumental in further testing the predictions of $\Lambda_{\rm s}$CDM and its variants.

Overall, we have introduced a concrete theoretical framework in which false vacuum decay can induce ultra-late-time transitions in the effective gravitational constant, $G_{eff}$, potentially leading to observable cosmological signatures at very low redshifts ($z \sim 0.01$). Our primary motivation for investigating this scenario was to provide a viable theoretical explanation for existing tensions in cosmological data, particularly the Hubble tension. While our analysis has extensively focused on false vacuum decay as the underlying physical mechanism, alternative approaches, such as models involving step-like variations in the gravitational coupling function $F(\phi)$, also exist. Observational signatures, particularly localized anisotropies and gravitational wave signals, may provide definitive tests capable of distinguishing between these theoretical frameworks, thereby advancing our understanding of gravity and dark energy at cosmologically recent epochs.

\raggedleft
\bibliographystyle{apsrev4-1}
\bibliography{bibliography}  

\newpage

\appendix 
\renewcommand{\thesubsection}{\Alph{section}.\arabic{subsection}}
\renewcommand{\theequation}{\Alph{section}.\arabic{equation}}
\numberwithin{equation}{section}

\section{The flat space time case}
\subsection{The rescaled action and the rescaled potential}\label{AppI}

\begin{justify}

  \indent In this appendix, we will follow the methods used in \cite{Dunne:2005rt}. The main quest is to go a step further, beyond our standard thin-wall assumption, and achieve a precise numerical solution for the bounce equation.

Our first goal is to proceed to the Euclidean action and the potential rescaling. consider our familiar potential
\begin{equation}
    U(\phi)=\frac{\lambda}{8}(\phi^2-a^2)^2-\frac{\epsilon}{2a}(\phi-a),
   \end{equation}
with $a>0$, $\lambda>0$, and $\epsilon>0$ the external cause which breaks the symmetry of the double well. The two minima of the potential, in first order with respect to $\epsilon$, will be 
\begin{equation}
    \phi_\pm(\epsilon)=\phi_\pm(0)+\frac12 \dv{\phi_\pm}{\epsilon}\Bigg|_{\epsilon=0}\epsilon +\dots,
    \label{i2}
\end{equation}
with $\phi_\pm(0)=\pm a$, for $\epsilon \ll1$. Then,
\begin{equation}
    \dv{U}{\phi}\Bigg |_{\phi_\pm}=0 \Rightarrow \phi_\pm(\phi_\pm^2-a^2)=\frac{\epsilon}{\lambda a}.
\end{equation}

If we differentiate with respect to $\epsilon$ and calculate the expression in $\epsilon=0$, it becomes
\begin{equation}
    3\phi_\pm^2\dv{\phi_\pm}{\epsilon}\Bigg |_{\epsilon=0}-a^2\dv{\phi_\pm}{\epsilon}\Bigg |_{\epsilon=0}=\frac{1}{\lambda a}\Rightarrow \dv{\phi_\pm}{\epsilon}\Bigg|_{\epsilon=0}=\frac{1}{\lambda a^3}.
    \end{equation}
    
    Finally, \eqref{i2} becomes
    \begin{equation}
        \phi_\pm=\pm a\bigg(1+\frac{\epsilon}{\lambda a^4}+\dots\bigg).
    \end{equation}
    
In the next step, we are going to expand our scalar field about $\phi_+$, its false vacuum, in the following way
\begin{equation}
    \phi=\phi_+ +\varphi.
\end{equation}
In this way, if we Taylor expand the potential we get rid of the linear term as follows
\begin{multline*}
      U(\varphi)=U(\phi_+)+\cancel{U'(\phi_+)(\varphi-\phi_+)}^{0}+\frac{U''(\phi_+)}{2}(\varphi-\phi_+)^2\\+\frac{U'''(\phi_+)}{3!}(\varphi-\phi_+)^3 +\frac{U''''(\phi_+)}{4!}(\varphi-\phi_+)^4+ \dots
 \end{multline*}
 
 \begin{multline}
      \Rightarrow U(\varphi) =U(\phi_+)+\frac{U''(\phi_+)}{2}(\varphi-\phi_+)^2+\frac{U'''(\phi_+)}{6}(\varphi-\phi_+)^3\\+\frac{U''''(\phi_+)}{24}(\varphi-\phi_+)^4+ \dots
\end{multline}

Now, after a great amount of very trivial algebra, which is not shown here due to its size and simplicity, the potential up to dimension four is given by \cite{Baacke:2003uw}
\begin{equation}
    U(\varphi)=\frac{m}{2}\varphi^2-\eta \varphi^3 +\frac{\lambda}{8}\varphi^4,
\end{equation}
where we have defined
\begin{equation}
    m^2=\frac{\lambda}{2}(3\phi_+^2-a^2) , \qquad \eta=\frac{\lambda}{2}|\phi_+|.
\end{equation}
If we rescale the field $\varphi$ as well as the coordinates of space-time as
\begin{equation}
    \varphi=\frac{m^2}{2\eta}\Phi, \qquad \Tilde{x}=mx,
\end{equation}
the classical Euclidean action
\begin{equation}
    S_E[\varphi]=\int d^4x \bigg[\frac12 (\partial_{\mu}\varphi)^2+\frac{m}{2}\varphi^2-\eta\varphi^3+\frac{\lambda}{8}\varphi^4 \bigg]
\end{equation}
after some simple algebra will simplify to
\begin{equation}
    S_E[\Phi]=\bigg(\frac{m^2}{4\eta^2}\bigg)\int d^4\Tilde{x}\bigg[\frac12(\Tilde{\partial_\mu} \Phi)^2+\frac12\Phi^2-\frac13\Phi^3+\frac{k}{8}\Phi^4\bigg],
    \label{act}
\end{equation}
where the dimensionless $k$ next to the quartic coupling strength is defined as
\begin{equation}
    k=\frac{\lambda m^2}{4\eta^2}=1-\frac{\epsilon}{2\lambda a^4}+\dots.
\end{equation}
It is obvious from the above that $\alpha$ tends to one in the ``thin-wall'' limit. From Eq.\eqref{act} we get the dimensionless potential
\begin{equation}
    U(\Phi)=\frac12\Phi^2-\frac12\Phi^3+\frac{k}{8}\Phi^4.
\end{equation}

The shape of the potential is determined by $k$, and its divergence from the unity represents a measurement of the vacuum energy difference in relation to the height of the barrier. Someone can observe this behavior in Fig.\hyperref[I1]{20}.
\begin{figure}[h]
\begin{center}
        \includegraphics[scale=0.40]{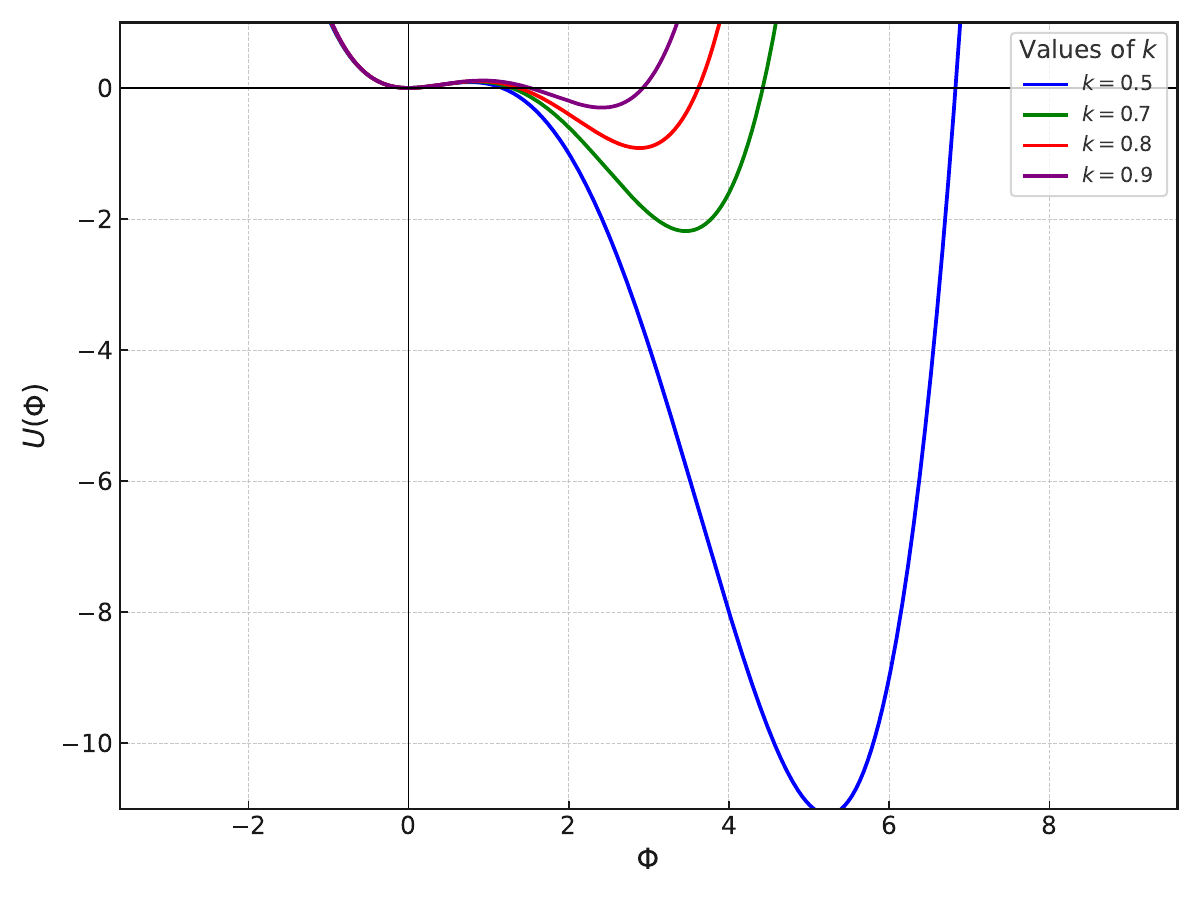}
    \caption{The rescaled potential ploted for $k=0.5$, $0.7$, $0.8$, $0.9$. As this parameter approaches unity, the rescaled potential approaches the double well potential.}
    \end{center}
    \label{I1}
  \end{figure}
  \end{justify}

\subsection{The equation of motion}
\begin{justify}
    
The rescaled equation of motion of the classical bounce $\Phi_{cl}(r)$ can be obtained with the principle of least action of the rescaled Euclidean equation under the usual assumption of the spherical symmetry of the bounce. Let us derive the EoM in a slightly different way from the main text. First, we transform from the Cartesian coordinates $\{x^1,x^2,x^3,x^4\}$ to the spherical polar ones $\{r,\phi_1,\phi_2,\phi_3\}$
\begin{align}
    x_1&=r\cos\phi_1 \nonumber\\
    x_2&=r\sin\phi_2\cos\phi_2 \nonumber\\
    x_3&=r\sin\phi_1\sin\phi_2\cos\phi_3 \nonumber\\
    x_4&=r\sin\phi_2\sin\phi_2\sin\phi_3.
\end{align}
The 4-volume element in Euclidean space can be obtained easily by using the Jacobian:
\begin{align}
    d^4x &= \Bigg|\det \frac{\partial(x_i)}{\partial(r,\phi_j)}\Bigg| drd\phi_1 d\phi_2d\phi_3 \nonumber \\
    &= r^3\sin^2\phi_1\sin\phi_2 drd\phi_1d\phi_2d\phi_3.
\end{align}

Consequently, the Euclidean action can be written as
\begin{equation}
    \begin{split}
        S_E[\Phi]&=\bigg(\frac{m^2}{4\eta^2}\bigg)\int d^4\Tilde{x}\bigg[\frac12(\Tilde{\partial_\mu} \Phi)^2+\frac12\Phi^2-\frac13\Phi^3+\frac{k}{8}\Phi^4\bigg]\\
        &=(2\pi)^2\bigg(\frac{m^2}{4\eta^2}\bigg)\int_{0}^{r_f}r^3dr\bigg(\frac12(\partial_r\Phi)^2+\frac12\Phi^2-\frac13\Phi^3+\frac{k}{8}\Phi^4\bigg).
    \end{split}
\end{equation}
The Lagrangian in the new coordinates will be
\begin{equation}
    \mathcal{L}[\Phi]\propto r^3\bigg(\frac12(\partial_r\Phi)^2+\frac12\Phi^2-\frac13\Phi^3+\frac{k}{8}\Phi^4\bigg).
\end{equation}
The equation of the classical bounce solution can be found by the Euler-Lagrange equation to be the following for our case
\begin{equation}
    -\Phi_{cl}''-\frac{3}{r}\Phi_{cl}+\Phi_{cl}-\frac32\Phi_{cl}^2+\frac{k}{2}\Phi_{cl}^3=0,
    \label{eomapp}
\end{equation}
with the boundary conditions given by
\begin{equation}
    \Phi_{cl}'(0)=0,\qquad \Phi_{cl}(\infty)=\Phi_+.
\end{equation}

The numerical solution to this boundary value problem (BVP) is presented in Fig.\hyperref[numerical1]{3}. The implementation of an intrinsically embedded \textit{Shooting method} command within the \textit{Mathematica} software was employed for the purpose of solution determination.
\end{justify}
\section{The modified gravity case}\label{app2}
\subsection{The shooting method explained}
\begin{justify}
    In this appendix, we will describe the method used to solve numerically the coupled field equations in the modified theory of gravity case. Let us remind the equations of interest, for $\phi$ we have 
\begin{equation}
   \phi''+\frac{3\rho'}{\rho}\phi'-\xi R\phi =\frac{dU(\phi)}{d\phi}
   \label{eqf}
\end{equation}
and for $\rho$ is
 \begin{equation}\label{eqr}
      \rho'^2=1+\frac{\kappa \rho^2}{3(1-\xi\phi^2\kappa)}\left(\frac12\phi'^2-U+6\xi\phi'\phi\frac{\rho'}{\rho}\right).
\end{equation}

In this system, an issue occurs straight away because of the square root of $\dot{\rho}$ in the second equation. A clever way to avoid this ambiguity is to differentiate the equation of $\rho$ with respect to $\eta$ \cite{Rajantie:2016hkj}. Then, someone finds

\begin{equation}
    \rho''=\frac{\kappa\rho}{3(1-\kappa\xi\phi^2)}\bigg[-\phi'^2-U+3\xi\bigg(\phi'^2+\phi''\phi +\phi'\phi\frac{\rho'}{\rho}\bigg)\bigg].
    \label{friedmann2}
\end{equation}

Similar to the previous section we will introduce some dimensionless variables
\begin{equation}
    x=mr,\qquad \Tilde{\phi}(x)=\frac{\phi(\eta)}{m},\qquad \Tilde{U}(\Tilde{\phi})=\frac{U(\phi)}{m^4},\qquad \Tilde{\rho}(x)=\Tilde{\rho}(\eta),
    \end{equation}
where $m$ is the mass, arbitrarily selected. The system of the equations will be expressed in the new dimensionless variables as follows
\begin{equation}
    \Tilde{\phi}''=-3\frac{\Tilde{\rho}'}{\Tilde{\rho}}\Tilde{\phi}'+\dv{\Tilde{U}}{\Tilde{\phi}}+\xi R\phi,
    \label{eqphi}
\end{equation}
\begin{equation}
    \Tilde{\rho}''=-\frac{8\pi}{3(1-\xi\tilde{\phi}^2\kappa)}\Tilde{\rho}\bigg[-\tilde{\phi}'^2-\tilde{U}+3\xi\bigg(\tilde{\phi}'^2+\tilde{\phi}''\tilde{\phi} +\tilde{\phi}'\tilde{\phi}\frac{\tilde{\rho}'}{\tilde{\rho}}\bigg)\bigg],
    \label{eqrho}
\end{equation}
where $M=m/M_{Pl}$.

Contrary to the previous section where we used the \textit{Shooting Method} as a saved command built-in \textit{Mathematica}, here we will construct a shooting algorithm step by step before we begin the coding part.

The boundary conditions for the field $\Tilde{\phi}$ become
\begin{equation}
    \lim_{x\rightarrow\infty}\Tilde{\phi}(x)=\frac{\phi_+}{m}
    \label{bc1}
\end{equation}
\begin{equation}
    \dv{\Tilde{\phi}(x)}{x}\Bigg|_{x=0}=0.
    \label{bc2}
\end{equation}
The boundary conditions for $\Tilde{\rho}$ read
\begin{equation}
    \Tilde{\rho}(0)=0,
    \label{bc3}
\end{equation}
\begin{equation}
    \Tilde{\rho'}(0)=1.
    \label{bc4}
\end{equation}
The first condition means that the bubble at the beginning of its nucleation has a tiny radius, almost zero, and the second condition arises from Eq.\eqref{eqr}.

Finding a solution with numerical methods in the system of the differential equations is not a straightforward process since our boundary conditions define a BVP, not an initial value problem (IVP), but with the \textit{Shooting Method} we can get a solution. We will define an IVP by choosing a new boundary condition for $\Tilde{\phi}(0)$. This condition will have the following form
\begin{equation}
    \Tilde{\phi}(0)=a,
\end{equation}
and with the conditions Eqs.\eqref{bc2}-\eqref{bc4} constitute a Cauchy problem (IVP) which can be solved numerically as a function of $a$. Then, this parameter is adjusted in order for Eq.\eqref{bc1} to be fulfilled.

As before, at $x=0$ a singular point exists at Eq.\eqref{eqphi}; for this reason, the solution must be placed in a range $[x_{min},x_{max}]$ with $x_{min}$ to be positive. Consequently, the initial conditions must be adapted to $x=x_{min}$ from those at $x=0$. To achieve this, we can connect $\Tilde{\phi}(x_{min})$ to $\Tilde{\phi}(0)$ if we Taylor expand as follows
\begin{align}
\Tilde{\phi}(x_{min}) &= a + \cancel{\Tilde{\phi}'(0)x_{min}} + \frac{1}{2} \Tilde{\phi}''(0)x_{min}^2 + \mathcal{O}(x_{min}^3) \nonumber \\
&= a + \frac{1}{2} \Tilde{\phi}''(0)x_{min}^2 + \mathcal{O}(x_{min}^3).
\end{align}
We used Eq.\eqref{bc2} on the above expression. Now, from Eq.\eqref{eqphi} we can get
\begin{equation}
    \Tilde{\phi}''(0)=\Tilde{U}_{\phi}(\Tilde{\phi}(0))-\frac{3\Tilde{\rho}'(x)}{\Tilde{\rho}(x)}\Tilde{\phi}'(x)\Bigg|_{x=0}+\xi\tilde{\phi}(0)R=\Tilde{U}_{\phi}(a)-3\Tilde{\phi}''(0)+\xi a R.
\end{equation}
This leads to
\begin{equation}
    \Tilde{\phi}''(0)=\frac{1}{4}\Bigg(\Tilde{U}_{\phi}(a)+\xi a R\Bigg).
\end{equation}

Therefore, the initial conditions for $\Tilde{\phi}$ become
\begin{equation}
    \Tilde{\phi}(x_{min})=a+\frac18 \Bigg(\Tilde{U}_{\phi}(a)+\xi a R\Bigg) x_{min}^2+\mathcal{O}(x_{min}^3),
    \label{bc5}
\end{equation}
\begin{equation}
    \Tilde{\phi}'(x_{min})=\frac14\Bigg(\Tilde{U}_{\phi}(a)+\xi a R\Bigg) x_{min}+\mathcal{O}(x_{min}^2).
    \label{bc6}
\end{equation}

Acting in a similar way, we Taylor expand for $\Tilde{\rho}(x_{min})$ around $\Tilde{\rho}(0)$ and we get
\begin{equation}
 \Tilde{\rho}(x_{min})=\cancel{\Tilde{\rho}(0)}+ {\Tilde{\rho}'(0)x_{min}}+\frac12  \Tilde{\rho}''(0)x_{min}^2=x_{min} 
\end{equation}
from Eq.\eqref{eqrho} and the conditions Eq.\eqref{bc3} and Eq.\eqref{bc4}. Finally, the initial conditions for $\Tilde{\rho}$ will be
\begin{equation}
    \Tilde{\rho}(x_{min})=x_{min},
    \label{bc7}
\end{equation}
\begin{equation}
    \Tilde{\rho}'(x_{min})=1.
    \label{bc8}
\end{equation}

\end{justify}

\end{document}